\documentclass[ twocolumn,
aps,nofootinbib,showpacs,showkeys, 
tightenlines,preprintnumbers,]{revtex4}

\usepackage{epsf,epsfig,subfigure,graphicx,amsmath,amssymb}
\usepackage{color}

\newcommand*{\nx}{\ensuremath{n_X}}
\newcommand*{\mx}{\ensuremath{m_X}}

\newcommand*{\abund}{\ensuremath{\Omega_X h^2}}
\newcommand*{\abundchi}{\ensuremath{\Omega_{\chi}h^2}}

\newcommand*{\msusy}{\ensuremath{M_{\rm SUSY}}}

\newcommand*{\sigsip}{\ensuremath{\sigma_p^{\rm SI}}}

\newcommand\pb{\,{\rm pb}}

\newcommand{\eq}[1]{Eq.~(\ref{#1})}

\newcommand{\bfrac}[2]{{\left(\frac{#1}{#2} \right)  }}

\newcommand\lsim{\mathrel{\rlap{\lower4pt\hbox{\hskip1pt$\sim$}}
    \raise1pt\hbox{$<$}}}
\newcommand\gsim{\mathrel{\rlap{\lower4pt\hbox{\hskip1pt$\sim$}}
    \raise1pt\hbox{$>$}}}

\newcommand{\SUC}{{SU(3)$_c$}}
\newcommand{\SUW}{{SU(2)$_L$}}
\newcommand{\UY}{{U(1)$_Y$}}
\newcommand{\UPQ}{{U(1)$_{\rm PQ}$}}
\newcommand{\Uone}[1]{${\rm U(1)_{#1}}$}
\newcommand{\etal}{{\it et al.}}

\def\alt{\lesssim}
\def\agt{\gtrsim}
\def\be{\begin{equation}}
\def\ee{\end{equation}}
\def\bea{\begin{eqnarray}}
\def\eea{\end{eqnarray}}
\def\eslt{E_T^{\rm miss}}

\def\to{\rightarrow}

\def\bi{\begin{itemize}}
 \def\ei{\end{itemize}}

\def\c1p{C1^\prime}
\def\msq3{\overline{m}_{\tilde{q}}(3)}
\def\ta{\tilde a}
\def\tG{\tilde G}

\def\ta{\tilde a}

\def\tg{\tilde g}

\def\tq{\tilde q}

\def\tw{{\tilde\chi}}

\def\tz{\tilde\chi^0}
\def\oh2{\Omega_{{\tilde{\chi}}}h^2}
\newcommand{\dis}[1]{\begin{equation}\begin{split}#1\end{split}\end{equation}}
\newcommand{\VEV}[1]{\langle #1 \rangle}
\newcommand{\OVER}[1]{\,\overline{\hskip -0.5mm #1}}

\newcommand\mpc{\,{\rm Mpc}}

\newcommand\cm{\,{\rm cm}}

\newcommand\Tmax{{T_{\rm max}}}
\newcommand\treh{T_{\rm reh}}
\newcommand\Tdec{{T_{\rm dec}}}
\newcommand\mdm{{M_{\rm DM}}}
\newcommand\Gint{{\Gamma_{\rm int}}}

\newcommand\Msun{{M_{\odot}}}
\newcommand\gravitino{\tilde{G}}

\newcommand\gluino{\tilde{g}}

\newcommand\mgluino{m_{\gluino}}   

\newcommand\squark{\widetilde q}

\newcommand\mchi{m_{\chi}}
\newcommand\mnlsp{m_{\rm NLSP}}

\newcommand\omegaantp{\Omega^{\rm NTP}_{\axino}}

\newcommand\dm{ {\rm DM}}
\newcommand\TP{ {\rm TP}}
\newcommand\NTP{ {\rm NTP}}
\newcommand\TD{ T_{D}} 
\newcommand\Tfr{ T_{\rm fr}}
\newcommand\Treh{ T_{\rm  reh}}
\newcommand\Rreh{ R_{\rm  reh}}
\newcommand{\MP}{M_{\rm P}}
\newcommand{\Mp}{\MP}
\newcommand\fa{f_{a}}
\newcommand\axino{{\tilde{a}}}
\newcommand\maxino{{m_{\axino}}}

\newcommand\mgravitino{{m_{3/2}}}
\newcommand\mgravitinoalt{{m_{\gravitino}}}

\newcommand\tev{\,{\rm TeV}}
\newcommand\gev{\,{\rm GeV}}
\newcommand\mev{\,{\rm MeV}}
\newcommand\kev{\,{\rm keV}}
\newcommand\ev{\,{\rm eV}}

\newcommand{\MGEQ}{M_{\rm GUT}}
\newcommand{\MG}{{M_{\rm GUT}}}

\newcommand{\ie}{{\it i.e.}\ }
\newcommand{\Z}{{\bf Z}}

\newcommand{\vew}{{v_{\rm EW}}} 
\newcommand{\LCDM}{{\Lambda{\rm CDM}}}

\def\E6{{\rm E_6}}
\def\EE8{{\rm E_8\times E_8}}

\begin{document}

\title{\Large\bf Dark matter production in the early Universe:
beyond the thermal WIMP paradigm}

\author{Howard Baer}
\affiliation{Department of Physics and Astronomy, University of Oklahoma, Norman, OK 73019, USA }
\author{Ki-Young Choi}
\affiliation{Korea Astronomy and Space Science Institute,  Daejon 305-348, Republic of Korea }
\author{Jihn E. Kim}
\affiliation{Department of Physics,Seoul National University, 1 Gwanakdaero, Gwanak-Gu,  Seoul 151-747, Republic of Korea, and \\
Department of Physics, Kyung Hee University, 26 Gyeongheedaero, Dongdaemun-Gu, Seoul 130-701, Republic of Korea }
\author{Leszek Roszkowski}
\thanks{On leave of absence from University of Sheffield, United Kingdom.}
\affiliation{National Centre for Nuclear Research,  Ho\.za 69, 00-681, Warsaw, Poland}

\begin{abstract}
 Increasingly stringent limits from LHC searches for new physics, coupled with lack of convincing signals of weakly interacting massive particle (WIMP) in dark matter searches, have tightly constrained many realizations of the standard paradigm  of thermally produced WIMPs as cold dark matter. In this article, we review more generally both thermally and non-thermally produced dark matter (DM). One may classify DM models into two broad categories: one involving bosonic coherent motion (BCM) and the other involving WIMPs. BCM and WIMP candidates need, respectively, some approximate global symmetries and almost exact discrete symmetries.  Supersymmetric axion models are highly motivated since they emerge from compelling and elegant solutions to the two fine-tuning problems of the Standard Model: the strong CP problem and the gauge hierarchy problem. We review here non-thermal relics in a general setup, but we also pay particular attention to the rich cosmological properties of various aspects of mixed SUSY/axion dark matter candidates which can involve both WIMPs and BCM in an interwoven manner.  We also review briefly a panoply of alternative thermal and non-thermal DM candidates.
\end{abstract}

\maketitle

\centerline{\bf Contents}
\noindent {\bf 1}.~{\bf Introduction} \hfill 1\\
\indent {\bf A}~A survey of some candidate DM particles \hfill 4\\
\indent {\bf B}~WIMP miracle or nonmiracle? \hfill 5\\
\noindent {\bf 2}.~{\bf Physics in the early Universe} \hfill 11\\
\noindent {\bf 3}.~{\bf Theory of dark matter} \hfill 13\\
\indent {\bf A}~Symmetry considerations for DM physics \hfill 13\\
\indent {\bf B}~WIMP vs. BCM \hfill 13\\
\indent {\bf C}~BCM as exemplified by axions \hfill 14\\
\indent {\bf D}~Weakly interacting massive particles (WIMP)  \hfill 15\\
\indent {\bf E}~Discrete and global symmetries \hfill 16\\
\indent {\bf F}~BCM supersymmetrized \hfill 17\\
\noindent {\bf 4}.~{\bf Thermal production} \hfill 20\\
\indent {\bf A}~Hot relics \hfill 20\\
\indent {\bf B}~Cold relics: case of WIMPs \hfill 20\\
\indent {\bf C}~Cold relics: case of E-WIMPs \hfill 21\\
\indent {\bf D}~Asymmetric dark matter (ADM) \hfill 24\\
\noindent {\bf 5}.~{\bf Non-thermal production of dark matter} \hfill 26\\
\indent {\bf A}~Dark matter from bosonic coherent motion \hfill 26\\
\indent {\bf B}~DM prod. via decay of heavy unstable part. \hfill 31\\
\indent {\bf C}~Cosmological constraints \hfill 33\\
\noindent {\bf 6}.~{\bf Non-thermal SUSY dark matter} \hfill 36\\
\indent {\bf A}~Gravitino dark matter \hfill 36\\
\indent {\bf B}~Non-thermal WIMPs \hfill 37\\
\indent {\bf C}~Axino dark matter \hfill 39\\
\indent {\bf D}~Axino production in the early Universe \hfill 39\\
\indent {\bf E}~Non-thermal axino prod. via sparticle decays \hfill 42\\
\indent {\bf F}~Mixed axion-axino CDM \hfill 44\\
\indent {\bf G}~Mixed axion-neutralino CDM  \hfill 46\\
\noindent {\bf 7}.~{\bf Non-thermal DM: non-SUSY candidates} \hfill 53\\
\indent {\bf A}~Pseudo Nambu-Goldstone boson as DM \hfill 53\\
\indent {\bf B}~Sterile neutrinos as DM \hfill 54\\
\indent {\bf C}~Minimal DM  \hfill 55\\
\indent {\bf D}~Primordial black hole DM \hfill 56\\
\indent {\bf E}~Supermassive DM (Wimpzilla) \hfill 56\\
\indent {\bf F}~Kaluza-Klein DM \hfill 57\\
\indent {\bf G}~Strongly interacting massive particles (SIMPs)  \hfill 57\\
\indent {\bf H}~Dynamical dark matter (DDM) \hfill 58\\
\indent {\bf I}~Chaplygin gas \hfill 58\\
\indent {\bf J}~Mirror-matter DM  \hfill 58\\
\indent {\bf K}~Self-interacting dark matter (SI~DM)  \hfill 59\\
\noindent {\bf 8}.~{\bf Conclusion}  \hfill 60\\
\indent ~Acknowledgments  \hfill 60\\
\indent ~References  \hfill 61

\begin{widetext}
\section{Introduction}
\label{Sec:Introduction}

Some eighty years have elapsed since Zwicky~\cite{ZwickyF33} first speculated that the
Coma cluster, to account for gravitational binding of its constituent galaxies,
 must contain a large amount of non-luminous (dark) matter.
Since that time, the puzzle of exactly what consititutes the dark matter (DM) has become
one of the foremost unresolved questions in particle physics and cosmology.
On the one hand, experimental evidence in favor of its existence has grown over the
years and is currently utterly convincing.\footnote{For a recent review
  of observational evidence see, e.g.~\cite{DelPopolo:2013qba}.  On
the other, since all the evidence is based (directly or
indirectly) solely on gravitational effects, we still don't know what
the DM is actually composed of.}

 Some properties of the Universe provide further evidence as to its structure and history.
The age of our Universe, 13.8 billion years, is very long:
if the matter density were too high, one would expect it to have gravitationally
collapsed in on itself, while if the dark energy (DE) were too high,
then one would expect that accelerated expansion
would have removed all stars and galaxies to beyond our purview~\cite{Weinberg:1988cp}.
Measurements of the cosmic microwave background radiation \cite{Ade:2013zuv} (CMBR)
imply that the Universe on large scales is homogeneous and isotropic to
one part on $10^5$: this is surprisingly smooth for apparently causally-disconnected regions.
Yet on smaller scales the Universe appears quite lumpy: inhomogeneous and anisotropic.
To understand the large scale smoothness, it is hypothesized that the Universe
has gone through an early inflationary epoch of rapid expansion so that the apparently
dis-connected regions were in fact causally connected and
only a tiny matter density existed at the end of inflation \cite{Guth81,Linde82,Stein82,Linde84}.
To understand the small-scale inhomogeneities, it is required that quantum
fluctuations provided the seeds to allow the Universe to have gone through a
gravitational condensation phase so galaxies, stars, planets and ultimately
life forms could have arisen~\cite{Tegmark06}.

Inflationary cosmology predicts that the total mass-energy density of
the Universe is very close to the critical closure density
$\rho_c=3H_0^2/8\pi G_N\simeq 1.88\times 10^{-29}h^2$
g$\cdot$cm$^{-3}$, where $G_N$ stands for the gravitational constant
and $H_0$ denotes the current value of the Hubble parameter which
is parametrized as $H_0\equiv 100\,h$\,km/sec/Mpc and $h\simeq
0.674$~\cite{Ade:2013zuv}. The recent WMAP~\cite{wmap9}
and Planck~\cite{Ade:2013zuv} satellite data confirm
that the energy density of the Universe is nearly $\rho_c$ (spatially
flat) and that the present dark energy is about 68\% of the
critical energy density of the Universe. The simplest
form of DE is the so-called cosmological constant. The
WMAP/Planck data fit to the $\LCDM$ cosmological model (supported by
data from galactic rotation curves, weak lensing measurements, baryon
acoustic oscillations {\it etc.}) imply the matter density in the
Universe lies at the $\sim 32\%$ of closure density level.  Of this
quantity, about $5\%$ lies in baryonic matter, whilst $\sim 27\%$
constitutes cold dark matter (CDM).

On quite general grounds one can expect DM particle candidates to be:
\begin{enumerate}
\item
non-relativistic (and thus massive) since relativistic particles (such as neutrinos) would exceed the escape velocity of clumping baryons and thus couldn't produce the gravitational wells
needed for structure formation,

\item
non-baryonic, \ie carrying neither electric nor (preferably) color charges,\footnote{
 It has been hypothesized that DM comes in clumps of baryons such a black holes,
cold stars or planets. Such clumps are referred to as Machos for MAssive Compact Halo Objects.
Machos have been searched for by the MACHO collaboration \cite{MACHO00}
by scanning for eclipsing stars in the Magellanic clouds. Lack of signal
makes this possibility seem unlikely.
In addition, one must also get around limits on the baryonic content of the
universe from the successful predictions of BBN \cite{Steigman:2007xt}.}

\item
stable (or at least extremely long lived, with the lifetime exceeding the
  age of the Universe by many orders of magnitude).
\end{enumerate}
 While some DM candidates are created just to solve the DM problem,
others emerge quite naturally from solutions to long standing problems
in particle physics. In this latter category, notable candidates include the axion,
which emerges from the Peccei-Quinn (PQ) solution to the strong CP problem
and the neutralino which emerges from a supersymmetric solution to the gauge hierarchy
problem. In cases such as these and others, the relic abundance of DM along with DM detection rates are calculable in terms of fundamental parameters, and thus subject to
experimental searches and tests.

\begin{figure}[t]
  \begin{center}
  \begin{tabular}{c}
   \includegraphics[width=0.65\textwidth]{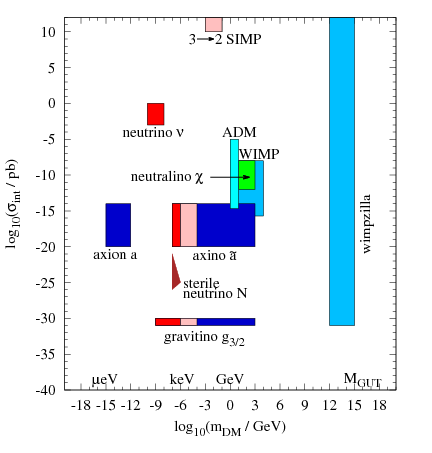}
  \end{tabular}
  \end{center}
  \caption{ (Color online) Several well-motivated candidates of DM are
    shown in the log-log plane of DM relic mass and $\sigma_{\rm int}$
    representing the typical strength of interactions with ordinary
    matter. The red, pink and blue colors represent HDM, WDM and CDM,
    respectively. This plot is an update of the previous
    figures~\cite{RoszIndia,KimRMP10}.
}
\label{fig:DMtype}
\end{figure}

Generally, DM relics are considered to be produced in the early
Universe in (at least) two distinct ways.  One possibility involves DM
particles generated in processes taking place in thermal equilibrium,
which we will generically refer to as {\em thermal production} (TP),
and the relics produced this way will be called {\em thermal
  relics}. On the other hand, {\em non-thermal production} (NTP), will
refer to processes taking place outside of the thermal equilibrium,
and the resulting relics will be called {\em non-thermal relics}. The
first class of processes will include the freeze-out of relics from
thermal equilibrium, or their production in scatterings and decays of
other particles in the plasma. The second will include, for example,
relic production from bosonic field coherent motion or from
out-of-equilibrium decays of heavier states or from bosonic coherent motion.

Working within the Standard Model (SM) of particle physics, it is found
that none of the known particles have the right properties to constitute CDM.
At one time, massive SM(-like) neutrinos were considered a possibility.
Measurements of the number of light neutrinos at LEP combined with
calculations of their relic abundance rule out this possibility \cite{Griest:1989pr}.

Instead, the most often considered theoretical candidate for CDM is
the weakly interacting massive particle, or WIMP.
It is worth stressing, however, that the WIMP is
not a specific elementary particle, but rather a broad class of
possible particles. Lee and Weinberg~\cite{Lee:1977ua}\footnote{For more
  references, see, e.g.,~\cite{Kolb:1990vq}. }  introduced WIMPs in
1977 in the form of stable, massive left-handed neutrinos which could
play the role of CDM. Such weakly-interacting particles were excluded
as CDM long ago due to lack of signal in direct DM detection
experiments. Since then a whole host of various weakly-interacting
particles have been discussed in the literature;  many of these
possibilities have now been excluded by experiment, although many also have
so far survived experimental tests.

\subsection{A survey of some candidate DM particles}

In Fig.~\ref{fig:DMtype}, we present an overview of several well-motivated DM
candidates in the mass {\it vs.} detection cross section plane
~\cite{RoszIndia}. On the vertical axis we show a typical
order of magnitude detection cross section associated with each type of candidate.
For reference, a SM neutrino with mass of order $0.1\ev$ and weak
interaction strength of order $10^{-36}\cm^2=1\pb \simeq
1\gev^{-2}/3.92$ is shown, although such a candidate would constitute hot DM (HDM) and
thus does not meet the need (of its velocity not exceeding the escape velocity
in galaxies) for cold relics. For more details see Sec.~\ref{Subsec:HotRelic}.

The box marked ``WIMP'' represents  ``generic'' weakly interacting massive
particle candidates as thermal relics.
Their mass can lie in the range between a few GeV~\cite{Lee:1977ua}
(below which it would overclose the Universe) and some $\sim100\tev$ from unitarity
constraints~\cite{Griest:1989wd,Profumo:2013yn}.
Their detection cross section is limited from above by direct DM search limits.
Recently, the strongest of these come from the Xenon100~\cite{XENON100:2012}
experiment and the LUX~\cite{Akerib:2013tjd} experiment.
A firm lower limit on the other hand does not
really exist; it can only be estimated on the basis of some kind of
theoretical arguments of ``naturalness''.
A more detailed discussion of thermal WIMPs will be presented in Sec.~\ref{Subsec:WIMP}.

The most highly scrutinized thermal relic is the lightest
neutralino particle of supersymmetric (SUSY) theories~\cite{GoldbergH83,Ellis:1983ew},
hereafter referred to as simply the neutralino.\footnote{For reviews see, e.g.,
~\cite{Jungman96,Bertone:2004pz,Drees:2012ji}.}
The neutralino is particularly well-motivated since, in addition to
solving the DM problem, SUSY extensions of the SM contain a number of
other attractive features both on the particle physics side and in
early Universe cosmology. From below,  the neutralino mass is limited by LEP2 searches
to lie above $\sim 50\,\gev$ in GUT-based SUSY models, but could be significantly lighter
in more general SUSY models~\cite{PDG14}.  
As an upper bound,  the neutralino mass is not expected to significantly exceed the
$\sim 1\tev$ scale based on the theoretical expectation of ``naturalness''.
We will discuss this important candidate in more detail below and in
Sec.~\ref{Subsec:WIMP}.

Another type of dark matter relic is called asymmetric dark matter (ADM).
In this case, in contrast to the standard WIMP scenario, one postulates both
DM and anti-DM particles where an asymmetry can develop between the two, in analogy
to baryonic matter.
The ADM possibility has recently received renewed interest and will be discussed in more
detail in Sec.~\ref{Subsec:AsymDM}.

An alternative possibility consists of strongly interacting massive particles (SIMPs).
Candidate SIMP particles with mass values around the MeV scale have been suggested
as a DM possibility in Ref.~\cite{Hochberg14}.
While usually DM is not expected to interact strongly,
such candidates have been considered in the past
(and for the most part been excluded \cite{Javorsek:2001yv} for instance by searches for
anomalous heavy nuclei or even by collider searches).

Moving down the vertical axis, the axion is a well known example of a non-thermal
relic. Its interaction strength is strongly suppressed relative to the
weak strength by a factor $(m_W/\fa)^2$, where $\fa\sim10^{11}\gev$ is the
PQ breaking scale. Despite being of very light mass ($\sim10^{-5}\ev$),
the axion is nonetheless a CDM candidate since it is produced basically at
rest in the early Universe.
The axion is a highly motivated and interesting candidate for CDM.
It will be discussed in more detail below and in Sec.~\ref{Subsec:ThCollective}.

In SUSY axion models, the axion supermultiplet contains, along with the axion,
the spin-$\frac12$ $R$-parity odd {\it axino} field $\ta$ and the $R$-parity even
spin-$0$ {\it saxion} field $s$.
The axino, as the fermionic partner of the axion, is an example of an
extremely-weakly interacting massive particle (E-WIMPs, or
alternatively super-WIMPs or FIMPs (for feebly interacting massive particles)).
The axino mass is strongly model dependent.
In the case where $\ta$ is the stable LSP, then it could comprise at least
part of the DM. Axinos can be either thermal or non-thermal relics, or both,
since they can be produced in both TP and NTP processes.
Depending on conditions, they may comprise hot, warm or cold relics,
or some combinations of say warm and cold DM.
Axino DM will be examined in Sec.~\ref{Subsec:AxinoDM}.

The gravitino $\gravitino$, the fermionic partner of the graviton, is
another well-motivated example of an E-WIMP.
It shares several properties to the axino.
It is a neutral Majorana fermion whose couplings to ordinary particles (and sparticles) are
strongly suppressed --  this time by the square of the Planck scale, $\sim(m_W/\MP)^2$.
Its mass is likewise strongly particle physics model dependent.
Like the axino, relic gravitinos can have contributions from both thermal and
non-thermal processes, and they can be either hot, warm or cold DM.
The gravitino as a cosmological relic will be further discussed
below and primarily in Sec's.~\ref{Subsec:eWIMP} and~\ref{gravitinodm:sec}.

Finally, a wimpzilla is an example of a nominally non-thermal relic.
While it is not motivated by particle physics, it represents an
alternative type of relic that can be produced in the early Universe by 1. classical
gravitational effects~\cite{Ford:1986sy,Yajnik:1990un}, 2. through
non-perturbative quantum effects during preheating or 3. from vacuum
fluctuations in a first-order phase transition~\cite{Chung:1998zb,Kuzmin:1998uv}.
In addition, 4. under some circumstances it can also have a thermal population~\cite{Chung:1998rq}.
The wimpzilla, along with several other  non-thermal relics, will be discussed in more detail in
Sec.~\ref{Sec:NonSUSY}.
The list of possible DM candidates is much longer than the few cases mentioned here.
In this review, we will focus primarily on non-thermal relics.

\subsection{WIMP miracle or non-miracle?}

As one can see from Fig.~\ref{fig:DMtype}, particle relics with a
correct relic density span a mass range of some thirty-three orders of
magnitude while interaction cross sections range across over forty orders of magnitude.
This is possible because their populations can be
generated by very different production mechanisms in the early Universe.
Of the possible DM candidates, WIMPs as thermal relics remain, however,
the most scrutinized possibility for DM due to a conspicuous connection between the CDM
relic density and the electroweak interaction strength. The argument,
often referred to as the so-called WIMP ``miracle'', goes as follows.  In
the early Universe WIMPs (denoted by $X$) are assumed to be in thermal
equilibrium at temperature $T\agt m_X$.  The WIMP number density $n_X$
as a function of time $t$ is governed by the Boltzmann equation
\be
\frac{d n_X}{dt}=-3H n_X-\langle\sigma_{\rm ann} v\rangle (n_X^2-n_{\rm eq}^2).
\label{eq:Boltzmann}
\ee
Here $n_{\rm eq}$ is the equilibrium density, the Hubble constant for a
radiation-dominated Universe is given by $H^2=\rho_{\rm rad}/3M_P^2$, and
$\langle\sigma_{\rm ann} v\rangle$ denotes the thermally averaged WIMP
annihilation cross section times WIMP relative velocity.  At early
times, the number density tracks the equilibrium density. However, at
some point  --  known as the freeze-out point at the temperature $\Tfr$
 --  the expansion rate outstrips the annihilation rate and the Hubble
term becomes dominant. At that point, the WIMPs freeze-out and their
number density in a co-moving (expanding) volume becomes effectively
constant.

An approximate solution of the Boltzmann equation provides the present-day WIMP
relic density as
\be
\abund \simeq \frac{s_0}{\rho_c/h^2}\left(\frac{45}{\pi^2
    g_*}\right)^{1/2}\frac{1}{x_f \Mp}\frac{1}{\langle\sigma_{\rm ann} v\rangle},
\label{eq:oh2approx}
\ee
where $s_0$ denotes the present day entropy density of the Universe, $g_*$ the number of
relativistic degrees of freedom at freeze-out and $x_f\equiv \Tfr/\mx\sim 1/25$
the freeze-out temperature scaled to  $\mx$  (for a derivation, see Sec.~\ref{Sec:ThPD}).
Plugging in the known values \cite{PDG14} for $s_0$, $\rho_c$ and $\Mp$ and setting
$\Omega_\chi h^2$ to its measured value $\simeq 0.12$, one finds
\be
\frac{\abund}{0.12} \simeq \frac{1} {\VEV{\frac{\sigma_{\rm
        ann}}{10^{-36}\cm^2} \frac{v/c}{0.1}} }.
\label{eq:oh2sigann}
\ee

Thus, an annihilation cross section of weak strength  of order
$\sim10^{-36}\cm^2(=1\pb)$ and typical WIMP velocities at freeze-out give
the correct present day relic density of dark matter.
(A solution of the Boltzmann equation will be discussed in more detail in Sec.~\ref{Sec:ThPD}.)
This remarkable argument has motivated a large number of
investigations into the possibility that DM,  in the form of WIMPs,
may be related to new electroweak physics which is anyway expected to occur near
the Fermi scale (for instance, new physics which is needed to stabilize the Higgs boson mass).

\begin{figure}[t]
  \begin{center}
  \begin{tabular}{c}
   \includegraphics[angle=-90,width=0.65\textwidth]{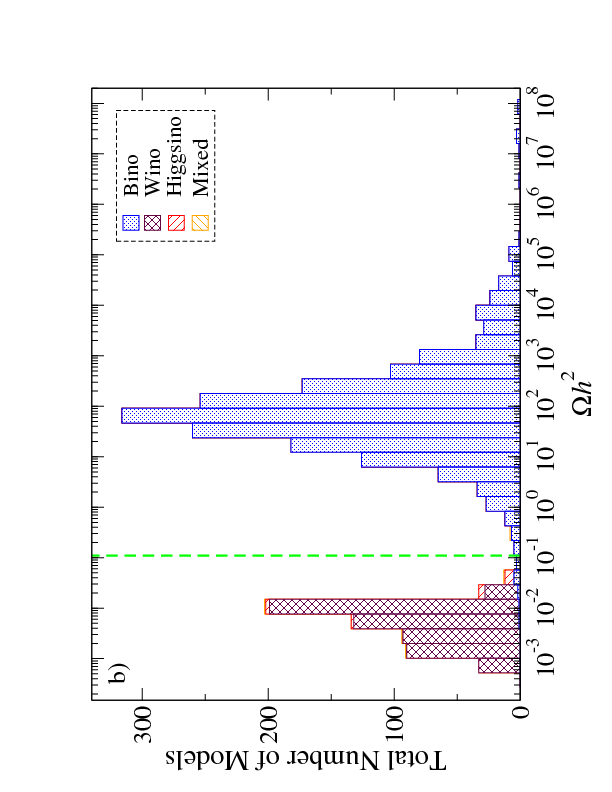}
  \end{tabular}
  \end{center}
  \caption{ (Color online) Neutralino relic density from a scan over 19-dimensional
SUGRA model parameter space, requiring the chargino mass to lie above $103.5$ GeV
(in accord with LEP2 searches)
and $\mchi<500$ GeV (naturalness). The measured value of CDM density is indicated by the
green dashed vertical line (from~\cite{Baer:2010wm}).
}
\label{fig:sug19}
\end{figure}

While the standard paradigm of the thermal production of WIMP dark matter certainly seems both
compelling and simple at face value, under deeper scrutiny the WIMP ``miracle''
scenario faces a series of challenges.

The first issue is whether the thermal WIMP ``miracle'' indeed somehow
{\em implies} that the WIMP mass lies near the electroweak scale.
Already from Fig.~\ref{fig:DMtype} one can see that the answer is negative:
the allowed WIMP mass range can be much larger.
This can be understood as follows.
On dimensional grounds one expects that
\be
\sigma_{\rm ann} \propto \frac{g^4}{m_X^2},
\label{eq:sigmaann}
\ee
where $g$ denotes the coupling of some processes behind WIMP
annihilation. Clearly, only their ratio has to be fixed in order to
obtain the weak interaction strength. A closer examination
~\cite{Feng:2008ya,Profumo:2013yn} shows that very wide ranges of both $g$ and
$m_X$ are consistent with the freeze-out mechanism. However, it is
also true that in some sense the electroweak scale (within roughly an
order of magnitude) can be regarded as being most naturally consistent with the
freeze-out mechanism.

The second issue with the WIMP ``miracle'' entails whether it really
predicts the measured abundance of CDM in realistic models (such as those
based on SUSY).  This problem can be illustrated in the context of
the neutralino $\chi$ of SUSY by computing the neutralino relic density in
a general SUSY model, in this case the 19-parameter supergravity
(gravity-mediation) model, wherein all soft terms are stipulated at
the GUT scale taken to be $\MGEQ\simeq (2\sim 3)\times 10^{16}$
GeV. In Fig.~\ref{fig:sug19}, a thorough scan\footnote{Here, thorough means that if we double
the number of scan points, the resultant histogram hardly changes (aside from normalization).}
of 19-dimensional
parameter space (subject to the chargino mass in excess of 103.5 GeV in accord with LEP2
searches and $\mchi<500$ GeV assuming SUSY naturalness
\footnote{For $m_{\chi}>500$ GeV,
then the superpotential higgsino mass  $\mu\agt 500$ GeV and consequently
there are large, unnatural cancellations in the $Z$ mass since
in SUSY one expects $m_Z^2/2\simeq -\mu^2-m_{H_u}^2$.}) reveals
that for a bino-like lightest supersymmetric particle (LSP), then the
predicted abundance $\abundchi$ lies 2--4 orders of magnitude {\it above}
the measured value while for a wino-like or higgsino-like LSP, then
the predicted abundance lies 1-2 orders of magnitude {\it below} the
measured value.
In fact, the measured CDM abundance lies at exactly the most
improbable value: in the case of SUSY, we would thus expect from
theory either far greater or far lesser values of $\abundchi$ than is
measured. In fact, in the case of SUSY, the WIMP ``miracle'' scenario
works best provided squark and slepton masses lie in the 50-100 GeV
range~\cite{Ellis:1983ew,Baer:1995nc}; such a range of masses has long
ago been excluded by direct collider searches.

In the case of SUSY, in order to match $\abundchi$ to data, then very specific neutralino types or
annihilation mechanisms are required:
well-tempered neutralinos
(just the right mix of bino-wino-higgsino for the LSP)
~\cite{Profumo:2004at,ArkaniHamed:2006mb,Baer:2006te}, co-annihilations (where the
next-to-LSP (NLSP) and LSP have a very small mass gap)~\cite{Griest:1990kh,Ellis:1998kh}, or
resonance annihilation~\cite{Baer:2000jj} ({\it e.g.}  $\chi\chi$ annihilation through
the $s$-channel pseudoscalar Higgs $A$ resonance with $2m_{\chi}\simeq
m_A$).  Much of the parameter space for these enhanced annihilation scenarios
has been excluded by recent limits from sparticle searches and the Higgs mass at LHC8
although they currently do remain as viable possibilities provided one tunes model parameters
to just the right values
~\cite{Fowlie:2011mb,Buchmueller:2011ab,Baer:2012uya,Kowalska:2013hha,
Buchmueller:2013psa,Roszkowski:2014wqa}.\footnote{The possibility of pure wino CDM now seems
in violation of indirect detection limits \cite{MACHO00,Cohen:2013ama}.
Mainly higgsino CDM remains a possibility for higgsino masses above about a TeV.}

The historic discovery of the Higgs boson at the
LHC~\cite{Aad:2012tfa,Chatrchyan:2012ufa} -- apart from being a landmark achievement in
itself -- seems to imply that the scale of SUSY breaking lies in the
TeV range -- consistent with new highly
constraining limits on superpartner masses (reaching in the case of
the gluino and the squarks of the first two generations the scale well
above 1~TeV) and from the lack of sizable deviations from the SM in flavor processes.
In SUSY, the calculated Higgs boson
mass includes radiative corrections~\cite{Carena:2002es} which are
proportional to the scale of SUSY breaking $\msusy$.  The rather large
value of the Higgs mass of about 125\,GeV typically implies $\msusy$ in
the range of 1 to even 12~TeV~\cite{Baer:2011ab,Kowalska:2013hha}.
Interestingly, in the multi-TeV region of superpartner masses
there exists a higgsino-like DM solution with the right amount of relic density and with mass close to
1~TeV~\cite{Kowalska:2014hza}.
The solution is generic since all that is required is the
Higgs/higgsino mass parameter $\mu$ (the higgsino mass) is close to
1~TeV -- no special mechanisms to reduce the relic abundance need
be invoked -- and that it is the LSP. This
can be easily achieved in phenomenological SUSY scenarios, e.g., the Minimal
Supersymmetric Standard Model
(MSSM)~\cite{Profumo:2004at,ArkaniHamed:2006mb,Baer:2006te,Fowlie:2013oua}
where $\mu$ is a free parameter.  In frameworks with grand-unification
assumptions the situation is less obvious since the $\mu$ parameter is
not free there but is determined, via the conditions of electroweak
symmetry breaking, and by the other parameters of the model. The correct
higgsino solution can still be found in large regions of unified MSSSM
models~\cite{Roszkowski:2009sm,Cabrera:2012vu,Kowalska:2013hha}, and
beyond (see, e.g.,~\cite{Kaminska:2013mya}).

Prior to the LHC the higgsino was known as a subdominant relic, while
the bino was favored. The shift of the SUSY breaking scale to the TeV
range, as implied by LHC limits and the large value of the Higgs mass
has led to the emergence of the $\sim1\tev$ higgsino as a motivated
and testable DM candidate. An alternative view, which relies on
naturalness, favors higgsinos in the mass range $\sim 0.1\tev$ but
with too low relic
density~\cite{Baer:2011ec,Baer:2012up,Baer:2013vpa}, in which case one
invokes an additional DM relic, e.g., the axion, to make up for the
rest of the relic density.

A third, and related to the previous, challenge to the thermally
produced WIMP dark matter paradigm is a continuous lack of a confirmed
experimental detection signal. Much experimental activity has been
directed to search for DM using the well-known strategies:
\begin{enumerate}
\item direct WIMP-nucleon scattering in underground detectors,
\item indirect detection of WIMPs in space by measuring the products of their
annihilation into anti-matter or gamma-rays, and
\item detection of WIMP produced at colliding beam experiments such as
  LHC, where the WIMP signal would be revealed as anomalously high
  rates for events containing missing transverse energy $\eslt$.
\end{enumerate}

Current limits from direct search experiments are shown in
Fig.~\ref{fig:dd} where the spin-independent detection cross section
($\sigma$, or $\sigsip$) is plotted against WIMP mass.  Also shown for
comparison are the favored theory regions of the Constrained MSSM
(CMSSM)~\cite{Kane:1993td} for $\mu>0$ following from recent global
analyses of Buchmueller {\it et al.}\,($\chi^2$
approach)~\cite{Buchmueller:2013psa} and Roszkowski  {\it et al.}\,(Bayesian
approach)~\cite{Roszkowski:2014wqa} (updated
from~\cite{Kowalska:2013hha}).

The new LUX limit has excluded the focus point region corresponding to
the mixed (bino-higgsino) neutralino (which had already been excluded
by LHC data on SUSY limits and the Higgs mass~\cite{Kowalska:2013hha})
and it is starting to probe the large $\sim1\tev$ higgsino DM
region. New limits expected later this year from the current runs of
LUX and Xenon100 will reach down to explore a large fraction of
$\sigma$. The rest of the $\sim1\tev$ higgsino region, as well as the
stau co-annihilation and $A$ funnel regions (at lower masses, from
left to right), will have to wait for one-tonne detectors to be at
least partially probed. It is worth stressing here, however, that in
less constrained SUSY models the allowed ranges of both neutralino
mass and spin-independent cross section tend to be wider, and in
phenomenological SUSY scenarios, like the pMSSM, very much wider; for
recent studies, see~\cite{Fowlie:2013oua,Cahill-Rowley:2014boa}. In
such models recent Xenon100 and LUX limits have excluded a wide range
of well-tempered neutralinos~\cite{Baer:2006te,Fowlie:2012im,Kowalska:2013hha} lying at
$\sigma_p^{\rm SI}\sim 10^{-44}\cm^2$.

Claims exist in the experimental community of possible detection of
WIMP signals including seasonal variation of low mass ($\sim 10$ GeV)
WIMP events in sodium iodide crystal in DAMA/Libra
~\cite{Bernabei:2010mq} and direct detection of low mass WIMPs at
CoGeNT~\cite{Aalseth:2010vx}, CREST II~\cite{Angloher:2011uu} and
CDMS/Si~\cite{Agnese:2013rvf}.  These low mass WIMP signals seem at
face value inconsistent amongst themselves
~\cite{Gondolo:2012rs,DelNobile2}, and are now also in strong conflict
with recent null searches by Xenon100~\cite{XENON100:2012},
CDMSLite~\cite{Agnese:2013jaa} and LUX~\cite{Akerib:2013tjd}, even
after taking into account different CDM halo profiles and velocity
distribution~\cite{HuhDel13,DelNobile1,DelNobile2,DelNobile3}.

\begin{figure}[t]
  \begin{center}
  \begin{tabular}{c}
    \includegraphics[width=0.75\textwidth]{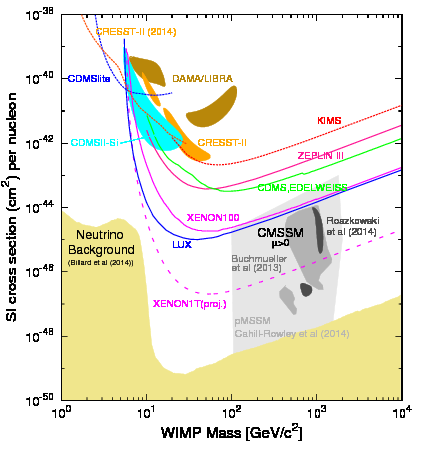}
  \end{tabular}
  \end{center}
  \caption{  (Color online) Direct detection search limits in the $m_X$
    vs. $\sigsip$ plane. For comparison we show the latest theory
    predictions for the usual case of the CMSSM with $\mu>0$ from
    global analyses of Buchmueller \etal\ ($\chi^2$
    approach)~\cite{Buchmueller:2013psa} and Roszkowski \etal~
    (Bayesian approach)~\cite{Roszkowski:2014wqa} (updated
    from~\cite{Kowalska:2013hha}). The pMSSM region of Cahill-Rowley \etal~has a much wider region \cite{Cahill-Rowley:2014boa}. The yellow region below is the neutrino background~\cite{Billard:2013qya}.
}
\label{fig:dd}
\end{figure}

In addition, there exist a variety of claims from indirect detection
experiments.  The recent AMS-02 confirmation of an unexpected rise in
the positron energy distribution confirms previous PAMELA data and may
hint that WIMP-type DM can be around 700 GeV--1 TeV~\cite{AMS13}.
However, an alternative explanation occurs in that positrons may be
created from ordinary pulsar processes
~\cite{Hooper:2008kg,Profumo:2008ms,Barger:2009yt}, so it is unclear if
this signal is really an indication of WIMP dark matter. The Fermi-LAT gamma
ray telescope also sees a possible anomaly in the high energy gamma
ray spectrum~\cite{Weniger:2012tx}. All these claims are weakened by
large and often poorly understood astrophysical backgrounds.

A fourth problem with the thermal WIMP-only CDM scenario is that it
ignores other matter states that may necessarily come along with the
DM particle in a complete theory and which therefore are likely to
also play a role in cosmology.
Such a case is illustrated by the {\it gravitino problem} in SUSY models.
In the case of gravity mediation, one expects the
presence of gravitino $\gravitino$ with mass $m_{3/2}$ (also
denoted by $\mgravitinoalt$) not far
above the weak scale.\footnote{The simplest gauge-mediation and
  anomaly-mediation scenarios seem under pressure by the measured
  value of the Higgs mass at $m_h\simeq 125$ GeV~\cite{Baer:2014ica}.}
Gravitinos, while not in thermal equilibrium in the early Universe,
can still be thermally created via scattering and decay processes at
large rates which are proportional to the reheat temperature $\Treh$
at the end of inflation. If $m_{3/2}>m_X$, then for high enough
$\Treh$, the WIMP abundance will be augmented, possibly in excess of
measured values. Furthermore, late decays of gravitinos into SM
particles may destroy the successful predictions of Big Bang
Nucleosynthesis (BBN). The gravitino problem is illustrated in
Fig.~\ref{fig:gravprob} (taken from Ref.~\cite{Kawasaki:2008qe}).
Large swaths of the $m_{3/2}$
vs. $\Treh$ plane are excluded by BBN/overproduction constraints
unless either $m_{3/2}$ is quite large ($\agt 5$ TeV) or $\Treh$ is
low~\cite{khlopov+linde84} ($\alt 10^5$ GeV).  The latter requirement
disfavors simple thermal baryogenesis-via-leptogenesis scenarios which
seem to require $\Treh\agt 10^9$ GeV~\cite{Buchmuller:2005eh}. The
former seems unlikely under simple naturalness considerations where
$m_{3/2}$ is expected to lie not far beyond the weak scale. However,
such large multi-TeV values of $m_{3/2}$ are allowed by more detailed
scrutinization of naturalness~\cite{Baer:2014ica} and indeed are
favored by the large value of the Higgs boson mass, the stringent lower
limits on superpartner masses from LHC direct searches and by a
decoupling solution to the SUSY flavor and CP problems.\footnote{
On the other hand, when the gravitino is the LSP and stable,
ordinary sparticles can decay into it and an electromagnetic or hadronic shower.
A combination of this and the overclosure argument then puts an upper bound
$\Treh\lsim 10^9$ GeV~\cite{Ellis:1984eq,Ellis:1984er,Moroi:1993mb}.}

\begin{figure}[t]
  \begin{center}
  \begin{tabular}{c}
   \includegraphics[width=0.55\textwidth]{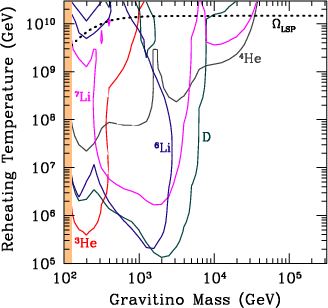}
  \end{tabular}
  \end{center}
  \caption{ (Color online) Regions of the  $m_{3/2}$ vs. $\Treh$ plane which are excluded by
1. overproduction of neutralino dark matter (upper-right) and by
2. BBN constraints on late-decaying gravitinos (from~\cite{Kawasaki:2008qe}).
}
\label{fig:gravprob}
\end{figure}

In addition to these, in the thermal WIMP-only scenario one is compelled to solve the strong CP problem of Quantum Chromodynamics (QCD). If it is solved by a very light axion, axions
contribute to CDM and can change  the thermal WIMP-only scenario.

Briefly, QCD in the limit of two light
quarks possesses an approximate global U$(2)_L\times$U(2)$_R$ symmetry,
which can be recast as U(2)$_V\times$U(2)$_A$. The U(2)$_V$ symmetry
gives rise to the familiar isospin and baryon number symmetries while
the U(2)$_A$ is broken by anomalies: since U$(2)_A$ is rank four, then by Goldstone's
theorem it should give rise to four light pions while in nature we see only three light
pions and one heavy $\eta'$. 't Hooft solved this supposed U(1)$_A$
problem using the QCD $\theta$ vacuum~\cite{tHooft}:
the extra U(1)$_A$ symmetry is badly broken by the anomaly term which
contributes dominantly to the $\eta'$ mass. Two important consequences
of the 't Hooft solution are that:
\begin{enumerate}
\item
$\eta'$ is much heavier than $\pi$'s, and
\item
the anomaly term is real~\cite{KimRMP10}.
\end{enumerate}

Thus, the QCD Lagrangian should contain a term
\be
{\mathcal L}\ni
\frac{\bar{\theta}}{32\pi^2}G_{a\mu\nu}{\tilde G}^{\mu\nu}_a,
\ee
where $\bar{\theta}=\theta+{\rm arg(det} {\mathcal M})$, and
${\mathcal M}$ is the quark mass matrix.  This Lagrangian term gives
contributions to the neutron EDM; comparison with experiment tells us
that $\bar{\theta}\alt 10^{-10}$: \ie the term has to be there, but
it should be somehow minuscule~\cite{KimRMP10}!
This is the strong CP problem of QCD.

After 35 years, perhaps the only compelling solution to the strong CP
problem requires introduction of an additional global PQ symmetry to the Lagrangian ~\cite{Peccei:1977hh,PQprd77} and its
concomitant {\it axion}
~\cite{Weinberg:1977ma,Wilczek:1977pj}.\footnote{One can try to
  address this in specific, calculable models~\cite{KimRMP10}, but the
  model ought to also explain SM phenomena at the level of explaining
  $m_h\simeq 125\,\gev$ (as SUSY models do).  }  The scale of PQ
symmetry breaking $f_a$ is required to be $f_a\agt 10^9$ GeV;
otherwise supernovae cool too quickly
~\cite{dicus,Raffelt:2006cw}. Since axion couplings are proportional
to $f_a^{-1}$, their couplings are of super-weak strength leading to
the so-called {\it invisible} axion
~\cite{Kim79,Shifman80,DFSZ81,Zhitnitsky80}.  The axion mass is
expected to lie in the $\mu \ev$ range.  The introduction of the axion
field allows the offending $G_{a\mu\nu}{\tilde G}^{\mu\nu}_a$ term to
dynamically settle to zero, but in the process fills the Universe with
{\it axions} as CDM via ({\it non-thermal}) bosonic coherent motion (BCM) of
the axion field~\cite{Abbott83,Preskill83,DineFischler83}.
It seems that any theory of DM which neglects the axion stands a
good chance of being incomplete.

For many years, physicists have often considered either WIMP-only \cite{Jungman96}
or axion-only \cite{Sikivie83} theories of CDM.
This appears to us to be a false dichotomy.
In a pure PQ extension of the SM, it is hard to understand
 --  in light of the Higgs boson discovery -- why the running Higgs mass does
not blow up at least to the PQ scale, far beyond the electroweak scale.
In a SUSY extension of the SM, the newly measured Higgs boson mass can
be in accord with theory, but then there is no solution for the strong
CP problem. Nature seems to need both. However, introducing an
axion into any SUSY theory radically changes the calculus  of  the DM relic abundance.
Along with the axion, one must now include its SUSY partners -- the
spin-$\frac12$ axino $\ta$ and the spin-0 saxion $s$.
In such a case,
both thermal and non-thermal production processes of axions, axinos,
saxions, neutralinos and gravitinos can occur and the various
abundances of each can feed into the others via production and decay
processes in the early Universe. Furthermore, embedding the PQ
symmetry into the supersymmetrized SM allows for the possibility that
the $R$-party-odd axino may now be the LSP and is a viable DM
candidate.

The current situation presents an opportunity to review the status of
non-thermal relics in a more complete way and with some generality. In
this report we pay particular attention to SUSY models incorporating
axionic states, but also, for completeness, we review several thermal
as well as non-thermal DM candidates that have been suggested in the
literature.

The remainder of this review article is organized as follows.
In Sec.~\ref{Sec:EarlyUniv}, we present a general overview of physics in the early Universe.
In Sec.~\ref{Sec:DMtheory}, we briefly present the basic reasons for introducing global and discrete symmetries
needed for the BCM and WIMP dark matter scenarios.
After introducing the thermal DM production mechanism in Sec.~\ref{Sec:ThPD},
we continue the non-thermal production  mechanism in  Sec.~\ref{Sec:NTProduction}.
In Sec.~\ref{Sec:SUSY} we present with some detail the case of thermal and non-thermal DM production
in PQ/SUSY extensions of the SM.
In Sec.~\ref{Sec:NonSUSY}, a variety of non-SUSY non-thermal DM candidates are also reviewed.
Sec.~\ref{Sec:Conclusion} contains a brief summary along with a global outlook.

Below we summarize the notation used in this review.  The
superpartners of the SM and some physical constants
are:\footnote{Since the spin-2 graviton is represented by
  $g_{\mu\nu}$, we represent the spin-$\frac32$ and the spin-$\frac12$
  part of the graviton supermultiplet as $g_{3/2}$ and $g_{1/2}$,
  respectively, and the collection of these as $\tilde{G}$ whose mass
  is denoted as $m_{3/2}$ or $\mgravitinoalt$. The goldstino is
  denoted by  $g_{1/2}$ without a tilde.  On the other hand, the gluino is
  represented by $\tilde{g}$ with its mass denoted as $m_{\tilde g}$.
}
\begin{eqnarray}
  && {\tilde G} ={\rm gravitino},
  \nonumber\\
  &&~~g_{3/2}= {\rm  spin~{\tiny\pm 3/2}~components ~of~gravitino},\nonumber\\
  &&~~{g}_{1/2}= {\rm goldstino~ and~ also ~spin~{\tiny\pm 1/2}~components ~of~gravitino},\nonumber\\
  &&\tilde{g} = {\rm gluino},~\tilde{W} ={\rm Wino},~\tilde{Z} = {\rm Zino},~ \tilde{B} = {\rm Bino},~  \tilde{\gamma} = {\rm photino},\nonumber\\
  && \chi = \tilde{\chi} = \tz_1 = \tilde{\chi}_1 = {\rm the~lightest ~neutralino},\nonumber
\end{eqnarray}
\begin{eqnarray}
&&T_{D}= {\rm temperature~ at~the ~time~of}~X ~{\rm decay},\nonumber\\
&&{T}_{\rm fr}= {\rm freezeout~temperature ~of~DM}, \nonumber\\
&&T_{\rm dec}= {\rm decoupling~temperature},\nonumber\\
&&T_{\rm reh}= {\rm reheating~temperature},\nonumber
\end{eqnarray}
\begin{eqnarray}
 &&{\rm GeV\cdot cm}=0.5067689\times 10^{14},  \nonumber\\
 &&{\rm GeV\cdot s}=1.519255\times 10^{24}, \nonumber \\
 &&{\rm GHz }=4.136\times 10^{-6}\,{\rm eV}, \nonumber \\
 &&{\rm ^o K = 0.861735\times 10^{-4}\, eV},   \nonumber
\end{eqnarray}
\begin{eqnarray}
 &&\MP = {\rm Reduced ~Planck~mass}=2.44\times 10^{18}\,{\rm GeV}, \nonumber\\
 && \rho_c =3 H_0^2\MP^2= 1.88\times 10^{-29}h^2\,{\rm g\cdot cm^{-3}}
 =8.1\times 10^{-47}h^2 \,{\rm GeV}^4,  \nonumber \\
 &&t_U=4.3\times 10^{17}\,{\rm s}, \nonumber
 \end{eqnarray}
where $h$ is the present Hubble parameter in units of
100\,km/s/Mpc. The recent Planck satellite data~\cite{Ade:2013zuv}
gives  $h\simeq 0.674$.

\vspace*{2\baselineskip}
\section{Physics in the early Universe}
\label{Sec:EarlyUniv}

In this section, we briefly review  physics in the early Universe,
including some necessary information on particle physics and
cosmology. For a standard textbook treatment, see {\it
  e.g.}~\cite{Kolb:1990vq,RaffeltBook}.

The evolution of the Universe can be described by the Einstein equations which arise in general relativity. Including  matter fields and a cosmological constant (CC) $\Lambda$, then
\dis{
R_{\mu\nu} - \frac12 {\mathcal R} g_{\mu\nu} \equiv G_{\mu\nu} = 8\pi G T_{\mu\nu}+\Lambda g_{\mu\nu},
}
where $G_{\mu\nu}$ is the Einstein tensor and $T_{\mu\nu}$ is the stress-energy tensor which includes
all the fields present -- matter, radiation etc.
For a homogeneous, isotropic Universe, the metric is given by the
Friedmann-Robertson-Walker (FRW) form:
\dis{
ds^2 = dt^2-R^2(t)\left\{\frac{dr^2}{1-kr^2}+r^2
  d\theta^2+r^2\sin^2\theta d\phi^2\right\},
}
where $R(t)$ is the scale factor and $k$ is the curvature $=-1$, 0 or 1 for an open, flat or closed Universe. In this case, the Einstein equations lead to the Friedmann equation
\dis{
H^2+ \frac{k}{R^2} = \frac{8\pi G}{3} \rho, \label{eq:Friedmann}
}
which governs the expansion of the Universe
(here, we introduce the  Hubble parameter $H \equiv \dot{R}/R$ and $\rho$ is the energy density).
Combined with the continuity equation,
\dis{
\dot{\rho} + 3 H (\rho+p) = 0,
}
one can solve the Friedmann equation for each case of radiation, matter or  CC  dominated Universe.
The scale factor then evolves as
\dis{
{\rm Radiation} \qquad&  \rho \propto R^{-4},\\
{\rm Matter} \qquad&  \rho \propto R^{-3},\\
{\rm Cosmological \ Constant} \qquad&  \rho \propto {\rm constant}.
}
The early Universe was dominated early on by relativistic particles
(radiation-dominated, RD) and later on matter became dominating
(matter-dominated, MD).
At present, it is known that the Universe is accelerating and thus is described by
a vacuum- (or CC-) dominated Universe.
The early RD Universe is considered to be preceded
by a different accelerating phase of the Universe known as {\it cosmic inflation}.

An initial inflationary period provides an explanation for the cosmological problems related to the
initial conditions of the standard Big Bang cosmology
(for a review, see {\it e.g.} \cite{Linde:2007fr}).
During inflation the Universe became very flat and homogeneous with only small amounts of fluctuations. After inflation, the oscillating inflaton field $\phi$ briefly makes the Universe matter-dominated until its decay produces relativistic particles: the Universe is then {\it reheated} and thus begins the standard Big Bang Universe. This process is called {\it reheating}.

For simplicity, it is usually assumed that the particles produced from
inflaton decay are thermalized
instantly and the reheating temperature $\treh$ is defined as the temperature
when the energy density of radiation dominates the matter density of the
oscillating inflaton field~\cite{Albrecht:1982mp,Kolb:1990vq},
{\it i.e.}
$\treh$ is the maximum temperature attained {\it during the RD phase}.
That happens around a time
 comparable to the lifetime of the inflaton field,
$t\sim H^{-1} \simeq \tau =\Gamma_{\phi}^{-1}$,
when the inflaton energy density exponentially decreases.
From the Friedmann equation, the reheating temperature can be expressed as
\dis{
\Treh \simeq \left( \frac{90}{4\pi^2g_*}  \right)^{1/4}\sqrt{\Gamma_{\phi} \Mp}.
\label{Treh}
}
However, the maximum temperature $\Tmax$ (the highest temeprature reached
after inflation but which may be attained before the onset of RD)
can be much higher than the reheating temperature~\cite{Kolb:1990vq,Giudice:2000ex}.
If the thermalization is delayed and occurs after RD,
then the reheating temperature can be much lower than that defined by
\eq{Treh}~\cite{Davidson:2000er,Mazumdar:2013gya}.
To maintain the successful predictions for the abundances of light nuclei production
during the standard Big Bang Nucleosynthesis (BBN), it is required that $\treh \gtrsim 4\mev$~\cite{Hannestad:2004px}.

The early Universe after inflation  was filled with relativistic
particles in a plasma that was very hot and dense.
The relativistic particles, collectively referred to as radiation, became thermalized
due to their self-interactions thus reaching local thermodynamic equilibrium.
From the equilibrium distributions, the energy density, number density
and entropy density of radiation are given by
\be
\rho_R =\frac{\pi^2}{30}g_*T^4,  \label{eq:rhoR}
\ee
\be
n_R=\frac{\zeta(3)}{\pi^2}g_{*S}T^3, \label{eq:nR}
\ee
and
\be
s=\frac{2\pi^2}{45}g_{*S}T^3, \label{eq:s}
\ee
where $\zeta(3)=1.20206\ldots$ is the Riemann zeta function of 3,
$g_*$ counts the effective number of relativistic species present in
equilibrium and $g_{*S}$ denotes the effective degrees of freedom of entropy
at the time of decoupling.

For the non-relativistic particles in thermal equilibrium ({\it e.g.}
WIMPs $X$), one finds
\be
\rho_X=m_X n_X,\qquad {\rm with} \qquad n_X=g\bfrac{m_X T}{2\pi}^{3/2} \exp[-(m_X-\mu)/T], \label{eq:nX}
\ee
where $\mu$ is the chemical potential.

Whether any DM particles are thermalized or not is determined by
comparing the interaction rate $\Gint$ to the expansion rate of the
Universe $H$.  When the interaction rate is much faster than the
expansion time scale, \ie,
\dis{
\frac{\Gint}{H} >1,
\label{GammaH}
}
the species are in thermal equilibrium; in the opposite case they never
reach thermal equilibrium.  The temperature $\Tdec$ at the epoch
$\Gint=H$ is called the decoupling temperature of the particles and
is slightly different from the freeze-out temperature $\Tfr$ below which the mass
fraction of the particle stays constant.  Since they are very similar,
we will for the most part not distinguish them in this review.

\vspace*{2\baselineskip}

\section{Theory of dark matter}
\label{Sec:DMtheory}

\subsection{Symmetry considerations for dark matter physics}

To obtain DM candidates theoretically, certain underlying symmetries are required.
The bosonic coherent motion (BCM) such as the axion field oscillation relies on very light bosons.
Light bosons are almost massless. Theoretically, therefore, it is better for them to originate from
Goldstone bosons~\cite{Goldstone60} by feebly breaking the corresponding global symmetries.
However, global symmetries are not
respected by quantum gravity effects~\cite{Giddings88,Gilbert89}.  In
fact, quantum gravity effects were used to argue against the QCD axion
for the case of intermediate scale decay constant $f_a\sim
10^9-10^{13}$ GeV
~\cite{Barr92,Kamionkowski92,Holman92,Ghinga92,Dobrescu97}.  The U(1)s
corresponding to BCM candidates must be explicitly broken so that they
generate a potential for the collective motion to roll down.  In the
QCD axion case, the axion-$G\tilde G$ coupling breaks the global PQ
symmetry \UPQ.  For other coherent motions \cite{BCMreview14}, similar methods can be applied.

The case of WIMPs requires a discrete symmetry, parity (such as
the $R$-parity in SUSY models) or $\Z_2$ symmetry. However, not all
discrete symmetries are safe from quantum gravity effects.  Because of
this reason, only discrete groups which can be subgroups of gauge
groups were suggested for safe discrete groups called {\it discrete
  gauge symmetry}~\cite{Krauss89}.  Because of the gravity dilemma of
global symmetries, global symmetries which arise as approximate global
symmetries from discrete gauge groups were suggested~\cite{KimPLB13},
which has been shown to arise from string compactification
~\cite{KimPRL13,KimPLB13}.  This method was applied even for
generating a dark energy potential if the corresponding \Uone{DE}
is anomaly free~\cite{KimNilles14,KimJKPS14}.

The BCM mechanism based on discrete symmetries works well for
pseudoscalar pseudo-Goldstone bosons.  For scalar Goldstone bosons to
be realized linearly, one notices the difficulty in forbidding the
scalar mass term $m^2\phi^*\phi$ from a discrete symmetry.  For the
pseudoscalar case, even if $m^2\phi^*\phi$ is present as in the soft
terms in supergravity, the phase field in $\phi$ may not carry mass.
The dilaton idea was suggested and treated usually as a non-linearly
realized case~\cite{Peccei88},\footnote{For a recent discussion,
  see~\cite{Csaki14}.}  but it does not enjoy the merit of the
pseudoscalar case mainly because quantum loops generate $d\ne 4$ terms
also.

\subsection{WIMPs vs. BCM}

There are two well known candidates for CDM: 1. the bosonic coherent
motion (as occurs for production of cold axions) and 2. the WIMP
possibility.  The BCM case involving the
axion~\cite{Preskill83,Abbott83,DineFischler83} requires a very light
boson whose lifetime needs to be larger than the age of the Universe.
If DM is the QCD axion, its mass should be smaller than $\lesssim 24$
eV~\cite{KimRMP10} to live long enough until present.

An alternative example of BCM is the case of the inflaton  field $\Phi_{\rm inf}$ whose  oscillation dominates the
energy content of the Universe during inflation. As $\Phi_{\rm inf}$ oscillates and then decays,
it may produce WIMPs and/or other visible sector particles depending on the $\Phi_{\rm inf}$ couplings.
Since the $\Phi_{\rm inf}$ lifetime is of the order the reheating time, the $\Phi_{\rm inf}$ oscillation cannot account for the CDM density.

The WIMP particle, as mentioned in the discussion of the WIMP
``miracle'' scenario, is most naturally realized with both mass and
interaction strength typical of the weak scale.
For example, if the lightest neutralino of SUSY theories is also the  LSP,
then it interacts with other superpartners of the SM particles which also have mass in the TeV region.

The underlying theories of BCM and WIMP DM require symmetries:
some global symmetry for the case of BCM and the discrete symmetry for the case of WIMP.
Asymmetric dark matter (ADM) can be produced via mechanisms similar to those which generate the baryon asymmetry.

There exists another possibility for DM: introducing new forces called {\it dark forces}.
This has been suggested in Ref.~\cite{Arkani-Hamed08} to account
for data from satellite measurements of positron, photon
 and proton spectra~\cite{PAMELA09,Heat97,Beatty04,Kane02,Baltz02,Hooper04,ChangJ08,Marciano13}.
If there exists (baryon)$'$s under the new forces, its stability is guaranteed just as is the proton in the SM.
The kinetic mixing of U(1) forces~\cite{Holdom86} between the SM and new forces may be a portal to the dark force.
For this kind of DM, its stability belongs to one of the above categories and here we will not discuss it separately.

Whether the CDM is a coherent motion or a WIMP, its lifetime must be larger than the age of the Universe,
$t_U\simeq 4.3\times 10^{17}$\,s$\simeq 1/6.45\times 10^{42}\gev$.
If the $X$ decay interaction to lighter particles is given by
\dis{
\frac{1}{\tilde M^n} \phi_1\cdots \phi_\ell X
}
where the $\phi$'s are bosons or fermions, then the decay width to massless $\phi$'s is
\dis{
\Gamma\approx \frac{M_X^{2n+1}}{\tilde M^{2n}} (\textrm{phase space  factor})
=({\rm phase~space~factor}) \left(\frac{M_X}{\tilde M}\right)^{2n} M_X.\label{eq:DecayofX}
}

For the axion, we can take $n=1, \tilde M\sim 10^{13}\gev$ and
$M_X=10^{-5}\ev$ (the phase space factor is $\sim 10^{-2}$) so that
$\Gamma\approx 10^{-61}\ev\approx 1/10^{41}{\rm years}\ll 1/t_U$.  For
$n=1$, a keV particle needs $\tilde M>2.5\times 10^{11}\gev$ for it to
live longer than the age of the Universe.

\subsection{Bosonic coherent motion  as exemplified by axions}
\label{Subsec:ThCollective}

The possibility of CDM via a BCM was proposed
in~\cite{Preskill83,Abbott83,DineFischler83} in connection with the
axion coupling.  String theory, the most popular ultraviolet-completed
theory, houses plenty of pseudoscalar particles.  They come from the
antisymmetric field $B_{MN}$~\cite{Witten84,Witten85} and matter
representation~\cite{KimPRL13}.  Some of these pseudoscalar particles
behave like axions in that they have no couplings except to the
anomalies: these are called `axion-like particles' or ALPs
(for a recent review on ALPs, see~\cite{Jaeckel:2010ni}).
  ALPs are defined for the possibility of detecting them via
  axion-type search experiments, \ie phenomenologically they are
  defined by their coupling to the electromagnetic field $(a_{\rm
    ALP}/f_a)F_{\mu\nu}\tilde{F}^{\mu\nu}$.
Unlike the axion, its mass is a free  parameter, not related to $f_a^{-1}$.

The basic idea of any coherent motion as CDM is not much different from the case of axion CDM.
To be specific, here we discuss the axion case but with the proviso that one may replace the PQ scale
$f_a$ by some other effective mass scale in other bosonic cases if needed,
which means that it does not couple to the QCD anomaly.

The axion case begins with the axion coupling to the gluon anomaly
\dis{
\mathcal{L} = \frac{\alpha_s }{8\pi \fa}\,a\,G^a_{\mu\nu}\widetilde{G}^{a\,\mu\nu},
\label{aGG}
}
where $\alpha_s=g_s^2/4\pi$ is the strong coupling constant and
$\widetilde{G}^{a\,\mu\nu} = \frac12 \epsilon^{\mu\nu\rho\sigma }
G^a_{\rho\sigma}$ is the dual of the field strength $G^{a\,\mu\nu}$ for eight gluons $G^a_\mu\,(a=1,2,\cdots,8)$.
This interaction term can be obtained after integrating out colored heavy fields below the
PQ symmetry breaking scale $\fa$ but above the electroweak scale $\vew$.
The lower bound on the axion decay constant is obtained from beam dump experiments and
from astrophysical studies.
In particular, the currently accepted lower bound comes from the SN1987A energy loss rate,
$f_a\gtrsim 10^{10}\gev$, ~\cite{Raffelt88,Raffelt90,Turner88,Turner90}.
For the coupling to electron,
the red-giant branch gives a bound $g_{ae}<4.3\times 10^{-13}$ \cite{RGB13}.
An upper bound of $\fa \lsim 10^{12}\gev$ is frequently mentioned in the literature to
avoid overproduction of axions; this limit is highly dependent on assumptions
regarding the initial misalignment angle~\cite{BaeHuhKim09,Visinelli:2009zm,Turner86}
and entropy dilution in the early Universe and hence will not be used here.
For string axions from $B_{MN}$, the decay constant is bigger than $10^{16}\,\gev$~\cite{ChoiKimFa,Svrcek06}
(alternative string-based axion models with $f_a$ within the ``axion window'' $\sim 10^{11}$ GeV
are shown in~\cite{Honecker:2013mya}).
It was pointed out that if there is no inflation after the PQ phase transition,
the energy density of hot axions generated by axionic string oscillation lowers the upper bound of the
decay constant $f_a$ to $10^{10}-10^{11}\,\gev$~\cite{YamaguchiM99,Shellard10}.
For an ALP, the ALP mass and the coupling to the QCD anomaly are not related.
A recent summary of ALPs has been presented in~\cite{AriasP12}.

A general low-energy axion interaction Lagrangian can be written in
terms of the effective couplings $c_1$, $c_2,$ and
$c_3$ with the SM fields that arise after integrating out all heavy PQ-charge carrying
fields.  The resulting effective axion interaction Lagrangian terms are~\cite{KimRMP10}
\dis{
\mathcal{L}&^{\rm eff}_{\rm int} \,=\,c_1 \frac{(\partial_\mu a)}{\fa}
\sum_q \bar{q}  \gamma^\mu\gamma_5q \\
& - \sum_q  (\bar{q}_L m q_R e^{ic_2 a/\fa }+ \textrm{h.c.} )
+\frac{c_3}{32\pi^2 \fa}\, a\, G\widetilde{G}\\
 & + \frac{C_{aWW}}{32\pi^2 \fa}
 a W\widetilde{W} +\frac{C_{aYY}}{32\pi^2 \fa}
 a Y\widetilde{Y}  + \mathcal{L}_{\rm leptons},
 \label{eq:efflagr}
}
where $c_3$ can be set to one by rescaling $\fa$.
The axion decay constant $f_S$, $\theta= a/f_S$ with $c_3=1$ (or replacing $f_a$ by $f_S$ with $c_3=N_{\rm DW }$ -- for more see the last paragraph of this subsection), is defined up to the domain wall number of the PQ singlet $S$, $f_S=N_{\rm DW } \fa$.
The derivative interaction term
proportional to $c_1 $ preserves the PQ symmetry.
The $c_2$-term is
related to the phase of the quark mass matrix, and the $c_3$-term
represents the anomalous coupling.  The axion-lepton interaction term
$\mathcal{L}_{\rm leptons}$ is analogous to the axion-quark interaction term.

Two prototype field theory models for {\it very light} axions (so-called invisible axions) have
been considered in the literature.  At the SM level, one considers the
six SM quarks, $u, d, s, \ldots$, as strongly interacting matter fermions.
Above the electroweak scale $\vew\simeq 246\,\gev$, one
additionally introduces beyond-the-Standard Model (BSM) heavy vector-like quarks
($Q_i, \overline{Q}_i$), which in the interaction
Lagrangian in \eq{eq:efflagr} are already integrated out.

At the field theory level, the axion is present if  quarks
carrying the net PQ charge $\Gamma$ of the global U(1)$_{\rm PQ}$
symmetry exist.  In the Kim-Shifman-Vainstein-Zakharov~(KSVZ)
model~\cite{Kim79,Shifman80}, one introduces only heavy quarks as PQ charge-carrying quarks.
This results in $c_1=c_2=0$, and $c_3=1$ below the $\vew$, or below the QCD scale $\Lambda_{\rm QCD}$.
The gluon anomaly term (the $c_3$ term), induced by an effective heavy
quark loop, then solves the strong CP problem.  The axion field is a
component of the SM singlet scalar field $S$.
String axions from $B_{MN}$~\cite{Witten84,Witten85,ChoiKimFa,ChoiKimStAxs,Svrcek06} behave like the
KSVZ axion since they are defined by the QCD-anomaly coupling at low energy.

In the Dine-Fischler-Srednicki-Zhitnitskii (DFSZ) model~\cite{DFSZ81,Zhitnitsky80},
one instead does not assume any net PQ charge in the sector beyond the SM (BSM),
 but the SM quarks
are assigned the net PQ charge, \ie, $c_1=c_3=0$ and $c_2\ne 0$ below the electroweak scale $\vew$.
Here also, the axion is predominantly a part of the SM singlet scalar field $S$.
String axions from matter~\cite{Kim88,KimPRL13,KimDW14,KimAgamma14} have a component mimicking the DFSZ axion in addition to a KSVZ axion component. These field theory models can be realized in terms of a fundamental pseudoscalar field or a composite pseudoscalar field \cite{KimComp85,ChoiKimComp85,KimNilles91,ChunKN92NP,BabuChoi94}.

Several specific implementations of the KSVZ and the DFSZ frameworks can be found
in Refs.~\cite{Kim98} and~\cite{KimRMP10}; these, however,
require a whole host of additional BSM fields.
Realistically, any references to the properties of the KSVZ and the DFSZ models can serve,
at best, as guidelines.
In this respect,  unfortunately, there exists only two references clarifying the axion-photon-photon
coupling at the Lagrangian level from a string-derived BSM framework,
one with the approximate PQ symmetry~\cite{ChoiKimIW07}
and the other with the exact PQ symmetry~\cite{KimAgamma14}.
We can view the axion models by classifying by the terms defining the PQ symmetry.
The PQ symmetry in the Peccei-Quinn-Weinberg-Wilczek (PQWW) axion  is defined by the renormalizable couplings in terms of the SM fields only. For `invisible axions', it is sometimes succinct to present the case in terms of SUSY models.
A SM singlet $X$ must be included for an `invisible axion'.
The PQ symmetry of $X$ is defined by the QCD anomaly term.
For the invisible axion, we consider an effective theory of the SM fields plus $X$ below the GUT scale.
Suppose the axion coupling arises only in the $aG\tilde{G}$ term in the effective theory.
Then, its realization at the field theory level by renormalizable interactions is the KSVZ model
by the heavy quark coupling $X Q\OVER{Q}$ at the mass scale of $X$.
Its realization by nonrenormalizable interactions in string theory is the model-independent axion~\cite{Witten84}.
If the PQ charge is defined by the $d=4$ superpotential term $H_uH_d X^2/M_{\rm GUT}$~\cite{KimNilles84},
its realization by the renormalizable couplings is a SUSY DFSZ model.\footnote{
A similar comment applies to the neutrino mass for which the effective operator, defining the lepton number of $H_u$, is given by the Weinberg operator $\ell\ell H_uH_u/M$~\cite{Weinberg79} which may be generated by the seesaw model with renormalizable couplings.
}
Without SUSY, one can present a similar argument for the DFSZ.

Before one considers spontaneous symmetry breaking of U(1)$_{\rm PQ}$,
the axion Lagrangian can be said to have the axion shift symmetry
(which is just a phase rotation) $a\rightarrow a + {\rm constant}$,
and the physical observables are invariant under the PQ phase rotation.
Below $\fa$, the PQ rotational symmetry is broken, which is
explicitly reflected as a breaking of the axion shift symmetry through
the appearance of the $c_2$ and the $c_3$ terms in
Eq.~(\ref{eq:efflagr}). However, the $c_2$ term enters into the phase, and a
discrete shift of the axion field can bring it back to the original phase.
The $c_3$ term is the QCD vacuum angle term, and if the vacuum
angle is shifted by $2\pi$, then it comes back to the original
value. Thus, even though the U(1)$_{\rm PQ}$ is broken, one of its
discrete subgroups, \ie, the one corresponding to the common
intersection of the subgroups corresponding to the $c_2$ and the $c_3$
terms, can never be broken. As a result, the combination $c_2+c_3$ is
invariant under the axion shift symmetry, and $c_2+c_3$ is defined to
be an integer signifying the unbroken discrete subgroup of U(1)$_{\rm PQ}$~\cite{KimRMP10}.
It is called the domain wall number $N_{\rm DW}=|c_2+c_3|$~\cite{Sikivie:1982qv}.

\subsection{Weakly interacting massive particle (WIMP)}
\label{Subsec:ThWIMP}

The first cosmological study of a heavy stable particle interacting weakly with the
visible-sector particles was performed by Lee and Weinberg~\cite{Lee:1977ua} based on $2\to 2$ interactions.
This was followed by studies of the SUSY photino by
Goldberg~\cite{GoldbergH83} and general neutralino by
Ellis {\it et al.}\,~\cite{Ellis:1983ew} and has been reviewed
extensively in the case of SUSY models in, {\it e.g.} \cite{Jungman96}.

A particle $X$ of mass $m_X$ is absolutely stable if there are no particles lighter
than the sum of masses of those particles whose total quantum number is equal to that of $X$:
\ie $m_X<\sum m_i$.
Both the proton and the electron are  examples of stable particles
whose stability arises from a symmetry principle: in the former case, from baryon number conservation
whilst in the latter case, from electric charge conservation.
For the proton, a global symmetry is frequently used.
The proton is the lightest particle carrying a global U(1)$_B$ quantum number, the baryon number $B$.
Since there is no lighter  baryon number carrying color-singlet particle below the proton mass,
the proton is absolutely stable if $B$ is exact.
However, in theories where $B$ is broken --  {\it e.g.} in GUTs --
the proton can decay to lighter particles such as $p\to e^+\pi^0$.
If one uses a discrete symmetry, a similar argument can be used:
if there is no combination of lighter particles with the same discrete quantum number of
$X$ with mass below $m_X$, then $X$ is absolutely stable if the discrete symmetry is exact.
If the discrete symmetry is broken, then $X$ is not absolutely stable.

In the literature~\cite{Ishimori10,Ishimori12}, nonabelian discrete symmetries have been considered,
mainly for the lepton mass matrix texture.
However, these may not be so useful for the case of the WIMP because `nonabelian' by definition
includes many non-singlet representations~\cite{Lisanti:2007ec,Adulpravitchai:2011ei}
while in the WIMP context we discuss only one absolutely stable particle.
If one considers a  nonabelian discrete symmetry spontaneously broken down to $\Z_2$,
the DM stability is not due to the nonabelian nature but to the group $\Z_2$~\cite{Meloni:2011cc}.

The simplest example of a discrete symmetry is $\Z_2$ or parity $P$ because then all the
visible-sector particles are simply assigned  $0$ (or $+$) modulo 2 quantum number of $\Z_2$
(or parity $P$).
Because most of the visible-sector particles are assumed to be lighter than the WIMP,
the WIMP is assigned $1$  modulo 2 quantum number of $\Z_2$ (or $-$ of parity $P$).
The WIMP which is responsible for CDM is the lightest $\Z_2=1$ (modulo 2) particle,
or the lightest $P=-1$ particle.
This case is very elementary because then one may classify particles into two sectors:
the visible sector with $\Z_2={\rm even}$ and the other sector with $\Z_2={\rm odd}$.
For a SUSY WIMP, an exact $\Z_{2R}$ has been used such that the lightest $\Z_{2R}$-odd particle
can be the WIMP~\cite{HallSuzuki,HallLykken}.
With a bigger discrete symmetry, classification of particles according to the quantum numbers
of the discrete symmetry is more complex, but may also result in a stable WIMP.

\subsection{Discrete and global symmetries}
\label{Subsec:DiscGlob}

Discrete symmetries are useful in two respects.  First, they limit
possible interaction terms in the Lagrangian, which can simplify the
study of cosmic evolution.  Second, not all the possible discrete
symmetries are ruled out from string theory and gravitational
interactions. Since there does not exist a universally accepted
quantum gravity theory at present, our discussion may only proceed via
classical aspects of the gravity sector by topology change, \ie in
connection with wormholes and black holes which can take information
out from the visible Universe.  In this case, some discrete symmetries
are allowed while others are not.  In Fig.~\ref{fig:Worm}, an
illustration is shown where a wormhole connects the visible Universe
with a shadow world, where the flux lines of a U(1) gauge boson are
shown.  If the neck of the wormhole is cut to separate out the shadow
world, the visible Universe recovers the charges and an observer $O$
in the visible Universe does not consider that gauge charges are taken
out from his Universe. So, gauge symmetries are considered to be
unbroken by the gravitational effects~\cite{Giddings88,Gilbert89}.  On
the other hand, global symmetries do not accompany flux lines and the
observer $O$ notices that global charges are lost if the wormhole neck
is cut.  Thus, global symmetries are not considered to be respected by
gravity and indeed the PQ global symmetry has been considered in this
context~\cite{Barr92,Kamionkowski92,Holman92,Ghinga92,Dobrescu97}.

\begin{figure}[!t]
\begin{center}
\includegraphics[width=0.25\linewidth]{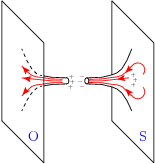}
\end{center}
\caption{A cartoon for a wormhole connecting the visible Universe and a shadow world.  } \label{fig:Worm}
\end{figure}

Using a bottom-up approach, discrete gauge symmetries have been considered widely in SUSY models for
obtaining the stability of the proton~\cite{Gilbert89,Ibanez92,BanksDine92}, {\it e.g.} $R$-parity.
Also, recently, the SUSY $\mu$-problem has been solved (under the unification assumption)
with the discrete symmetry $\Z_{4R}$ and more generally $\Z_{nR}$ where $n>2$ is a divisor of
24~\cite{ChoiKW97_2,BabuZ4,LeePLB11,LeeNPB11,Tamvakis12}.

On the other hand, the top-down approach relies on all the spectrum in the model and the
conclusions of the above bottom-up study do not apply~\cite{KimPLB13}.
String compactification allows a consistent massless spectrum with singlets included if needed~\cite{BanksDine92}.
In addition, it is known that certain discrete symmetries are allowed in string compactification~\cite{Hamidi87,Dixon87,Kobayashi07}.
This can be used to rule out some compactification models from the observed discrete symmetry.
The most widely worked-out top-down studies from $\EE8'$ heterotic string with suitable
MSSM spectra are $\Z_{12-I}$~\cite{ChoiKimBk,KimKyae07,KimJH07,HuhKK09}, $\Z_{6-II}$~\cite{Lebedev08},
and $\Z_2\times \Z_2$~\cite{Foerste04,Kappl09}.
Recently, discrete symmetries have been studied from the $\Z_{12-I}$~\cite{KimPRL13,KimPLB13}
and $\Z_{6-II}$~\cite{Nilles13} compactification of the heterotic $\EE8'$.
Even the tiny dark energy can be obtained from this kind of approximate global symmetry starting from an exact discrete symmetry by making a QCD anomaly-free global U(1)~\cite{KimNilles14,KimJKPS14}.

There is an easy way to construct $\Z_N$ and $\Z_{NR}$ discrete
symmetries from string compactification~\cite{KimPLB13}.  First, find
out a gauge U(1) which is a subgroup of $\EE8'$.  Second, assign VEVs
to some SM singlets with an even integer U(1) quantum number; take
{\it e.g.} the quantum number to be $N$. In SUSY, if the
superpotential terms carry the quantum number 0 modulo $N$, then the
discrete group is $\Z_N$.  If the superpotential terms carry the
quantum number 2 modulo $N$, then the discrete group is $\Z_{NR}$.

For proton stability, the $R$-parity is used to forbid dimension-3
superpotential terms such as $u^cd^cd^c$.  But the dimension--4
superpotential term $qqq\ell$ is not forbidden.  The coefficient of
the term $qqq\ell$ must be less than $10^{-7}$ from the bound on the
proton lifetime~\cite{PDG14}.  However, the dimension-4 Weinberg
superpotential term $\ell\ell H_uH_u$~\cite{Weinberg79} is needed for
neutrino masses via the seesaw mechanism~\cite{Minkowski77}.
Forbidding $qqq\ell$ but allowing $\ell\ell H_uH_u$ has been studied
under `proton hexality'~\cite{Dreiner06,BabuAppB03} and
$\Z_{4R}$~\cite{LeeNPB11}. In general, $Z_{NR}$ can achieve this goal
in the top-down approach.

A simple way to understand how $R$-parity might arise comes from
SO(10) SUSY GUTs.  In the case where matter superfields fill out a
complete 16-dimensional spinor of SO(10) and MSSM Higgs fields live in
the 10, then the superpotential only admits $\rm matter-matter-Higgs$
couplings, while $R$-parity violating terms are all of the form $\rm
matter-matter-matter$ or $\rm matter-Higgs$. Depending on how SO(10) is
broken, the exact $R$-parity conservation may or may not survive down
to low energies~\cite{Lee:1994je,Aulakh:1999cd,Aulakh:2000sn,Martin:1996kn}.
This simplicity of SO(10) results from ${\rm (matter)}\to {\rm spinor} ~
{\mathcal S}(={\bf 16})$ and ${\rm (Higgs)}\to {\rm vector} ~ {\ V}
(={\it e.g.}~ {\bf 10})$ assignments of SO(10) because one needs an even
number of SO(10) spinors to construct an SO(10) vector.  In any SM or
GUT theory arising from the compactification of heterotic
E$_8\times$E$_8'$, similar spinor and vector assignments of the gauge
group E$_8\times$E$_8'$ were used in~\cite{KimJH07,Lebedev08}.
Recently, the (unstable) axino CDM scenario with R-parity violation
(to generate the baryon asymmetry via the Affleck-Dine mechanism) has been discussed in
detail \cite{Ishiwata14}. For the bulk of this review-- if not specified explicitly--
R-parity conservation has been implicitly assumed.

Most DM candidates suggested so far, {\it e.g.} Wimpzilla DM, minimal
DM, Kaluza-Klein DM, Chaplygin DM (see Sec.~\ref{Sec:NonSUSY}), rely
on exact or almost exact discrete symmetries.

\subsection{BCM supersymmetrized}\label{Subsec:SUSYColl}

\begin{figure}[!t]
\begin{center}
\includegraphics[width=0.25\linewidth]{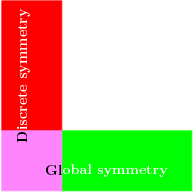}
\end{center}
\caption{A cartoon for the terms satisfying the discrete (red) and global (green) symmetries.
The lavender part satisfies both the discrete and global symmetries.  }
\label{fig:DiscrGlob}
\end{figure}

As an important application of discrete symmetries, a discrete subgroup of gauge  U(1)s can be used to obtain some approximate global symmetries.
The approximate PQ  symmetry U(1)$_{\rm PQ}$~\cite{ChoiKimIW07,ChoiKS09} and the approximate U(1)$_R$
symmetry~\cite{Kappl09} from string compactification have been considered before.
Fig.~\ref{fig:DiscrGlob} shows how an approximate global symmetry is obtained from a top-down approach. Gauge symmetries are not spoiled by gravitational interactions.
If a discrete symmetry results from a subgroup of gauge symmetries of string compactification,
there is no spoiling of the discrete symmetry by gravity~\cite{KimPLB13}.
One can consider a series of interaction terms allowed by the discrete symmetry.
The vertical red band of Fig.~\ref{fig:DiscrGlob} represents the infinite tower of terms not spoiled by gravity.
If one considers the few lowest order terms of the red column, there can be an accidental global symmetry. With this global symmetry, one can consider an infinite series of terms shown as the horizontal green band of Fig.~\ref{fig:DiscrGlob}.
The terms shown in lavender satisfy both of the discrete and global symmetries.
But, the horizontal green terms outside the lavender are neglected because they are spoiled by gravity.
However, the vertical red band terms  not spoiled by gravity break the global symmetry.
Then, there results an approximate global symmetry~\cite{KimPLB13,KimNilles14,KimJKPS14,BabuAppA03}.

SUSY models containing a very light axion can provide clues to the magnitude of the axion decay constant $\fa$.
In Ref.~\cite{Kim83} it was speculated that $\fa$ is related to the MSSM Higgs/higgsino mass parameter $\mu$ as $f_a\sim \sqrt{\mu M_P}$. Along the line of the preceding paragraph: to avoid the wormhole breaking of the PQ symmetry~\cite{Barr92},
$f_a$ has been given from the $S_2\times S_2$ symmetry~\cite{KimPRL13}.

As stated in the Introduction,
the supersymmetric extension of the axion field includes the
fermionic partner $\axino$ called the {\it axino} \cite{Tamv82,NillesRaby82,Frere83},
as well as a real scalar field $s$ called the {\it saxion}.
 The axion supermultiplet $A$ is represented in two-component notation as
\dis{ A=\frac{1}{\sqrt2}(s+ia)+\sqrt2 \axino
  \vartheta + F_A \vartheta\vartheta,
}
where $F_A$ stands for an auxiliary field of $A$
and $\vartheta$ are the Grassmann superspace coordinates.
The interaction of the axion supermultiplet is obtained by
supersymmetrizing the axion interaction in \eq{eq:efflagr}.  In
particular, the interaction of the axion supermultiplet $A$
with the vector multiplet $V_a$, which is a SUSY version of the $c_3$ term in
\eq{eq:efflagr}, is given by
\dis{ {\mathcal L}^{\rm eff}
  =-\sum_{V}\frac{\alpha_V\, C_{aVV}}{2\sqrt2\pi \fa}\int A\,{\rm
    Tr}\,[V_a V^a] + \textrm{h.c.},
\label{Leff3}
}
where $\alpha_V$ denotes a gauge coupling, $C_{aVV}$ is a model-dependent
constant and the sum is over the SM gauge groups.
From this, the relevant axino--gaugino--gauge-boson and axino--gaugino--sfermion--sfermion
interaction terms can be derived and are given by~\cite{Choi:2011yf} (in four component spinor notation)
\dis{
{\mathcal L}^{\rm eff}_\axino&=i\frac{\alpha_s}{16\pi \fa}\overline{\axino}
\gamma_5[\gamma^\mu,\gamma^\nu]\tilde{G}^b
G^b_{\mu\nu} +\frac{\alpha_s}{4\pi \fa}\overline{\axino}\tilde{g}^a
\sum_{\tilde{q}}g_s \tilde{q}^*T^a\tilde{q}\\
&+i\frac{\alpha_2C_{aWW}}{16\pi \fa}\overline{\axino}\gamma_5[\gamma^\mu,\gamma^\nu]\tilde{W}^b
W^b_{\mu\nu} +\frac{\alpha_2}{4\pi \fa}\overline{\axino}\tilde{W}^a
\sum_{\tilde{f}_D}g_2 \tilde{f}_D^*T^a\tilde{f}_D\\
 &+ i\frac{\alpha_YC_{aYY}}{16\pi \fa}\overline{\axino}
 \gamma_5[\gamma^\mu,\gamma^\nu] \tilde{Y}Y_{\mu\nu} +\frac{\alpha_Y}{4\pi \fa}\overline{\axino}\tilde{Y}
\sum_{\tilde{f}}g_Y \tilde{f}^*Q_Y\tilde{f},
\label{eq:Laxino}
}
where the terms proportional to $\alpha_2$ correspond to the \SUW~ and
the ones proportional to $\alpha_Y$ to the \UY~ gauge groups,
respectively.  $C_{aWW}$ and $C_{aYY}$ are model-dependent couplings
for the \SUW~ and the \UY~ gauge group axino--gaugino--gauge-boson anomaly
interactions, respectively, which are defined after the standard
normalization of $\fa$, as in  \eq{aGG} for the \SUC\ term. Here,
$\alpha_2$, $\widetilde{W}$, $W_{\mu\nu}$ and $\alpha_Y$,
$\widetilde{Y}$, $Y_{\mu\nu}$ are, respectively, the gauge coupling,
the gaugino field and the field strength of the \SUW~ and the \UY~ gauge
groups. $\tilde{f}_D$ represents the sfermions of the \SUW-doublet, and
$\tilde{f}$ denotes the sfermions carrying the \UY\ charge.

Similarly, one can derive supersymmetrized interactions of the axion
supermultiplet with a matter multiplet as a generalization of the $c_1$
and the $c_2$ terms in \eq{eq:efflagr}. Ref.~\cite{Bae11} considered a
generic form of the effective interactions and clarified the issue of
the energy scale dependence of axino interactions.  At some energy scale
$p$, which is larger than the mass of the PQ-charged and gauge-charged
multiplet $M_\Phi$, the axino-gaugino-gauge boson interaction is
suppressed by $M_\Phi^2/p^2$. This suppression is manifest in the DFSZ
axion model, and even in the KSVZ model, if the heavy quark mass is relatively
low compared to the PQ scale, in which case of
course the heavy quark is not integrated out.

However, SUSY must be broken at low energy.
Then, the SUSY relation between the axino and the axion is modified.
In fact, the most important axino parameter in cosmological considerations --  the axino
mass $\maxino$ --  does not even appear in Eq.~(\ref{eq:Laxino}).
SUSY breaking generates the masses for the axino and the saxion and
modifies their definitions. The saxion mass is set by the SUSY soft
breaking mass scale, $M_{\rm SUSY}$~\cite{Tamv82,Nieves:1985fq}. The
axino mass, on the other hand, is strongly model dependent.  An
explicit axino mass model with SUSY breaking was first constructed long
 ago~\cite{Kim83} with the superpotential $W$ with the PQ symmetry
transformation $S\to e^{i\alpha} S$ and $\OVER{S}\to
e^{-i\alpha}\,\OVER{S}$,
\dis{ W=\sum_{i=1}^{n_I} Z_i(S
  \OVER{S}-f_i^2),~n_I\ge 2.\label{eq:SUSYglobBr}
}
With $n_I=1$, the U(1)$_{\rm PQ}$ symmetry is spontaneously broken, but SUSY remains
unbroken.
The case $n_I=2$ breaks SUSY, which however gives $\maxino=0$~\cite{Kim83}.

As first pointed out by Tamvakis and Wyler~\cite{Tamv82}, the axino
mass is expected to receive at least a contribution on the order of
$\maxino \sim \mathcal{O}(M_{\rm SUSY}^2/\fa)$ at the tree level in the
spontaneously broken global SUSY. In the literature, a whole range of
axino mass was considered;  in fact it can be even much
smaller~\cite{Frere83,KimMasiero84,Moxhay:1984am,ChunKN92,Goto:1991gq},
or much larger, than the magnitude of $M_{\rm SUSY}$~\cite{ChunLukas95}.
Because of this strong model dependence, in cosmological studies one often
assumes  axino interactions as given by the U(1)$_{\rm PQ}$ symmetry and
treats the axino mass as a free parameter.

Recently, the issue of a proper definition of the axion and the axino
was studied in the most general framework, including the non-minimal
K$\ddot{\rm a}$hler potential~\cite{KimSeo12}. In that study, the axino
mass is given by $\maxino=\mgravitino$ for $G_A=0$,
where $G=K+\ln |W|^2$ and $G_A\equiv \partial G / \partial A$. For $G_A\neq 0$,
the axino mass depends on the details of the K$\ddot{\rm a}$hler potential, and
the case given by Eq.~(\ref{eq:SUSYglobBr}) was shown to belong to one of these
examples. In the gauge mediation scenario, the gaugino mass is the
dominant axino mass parameter. In the case of gravity mediation, the axino mass
is likely to be greater than the gravitino mass $m_{3/2}$, but one cannot rule out
lighter axinos~\cite{KimSeo12}.

One important, but often overlooked, aspect of the axino is that its
definition must be given at a mass eigenstate level.  The coupling to
the QCD sector given in the first line of Eq.~(\ref{eq:Laxino}) can
plausibly be that of the axino, but it does not give the axino
mass. This is because the axino is connected to two kinds of symmetry
breaking: the PQ global symmetry breaking and the SUSY breaking.
These, in general, are not orthogonal to each other.
The PQ symmetry breaking produces an almost massless pseudo-Goldstone boson (the axion), while
SUSY breaking produces a massless goldstino. The massless goldstino is
then absorbed into the gravitino to make it heavy via the super-Higgs
mechanism. This raises the question of what the axino really is. This
issue is shown in Fig.~\ref{fig:AxinoDef} taken from Ref.~\cite{KimSeo12}.
The axino must be orthogonal to the massless goldstino component.
Therefore, for the axino to be present in a
spontaneously broken supergravity theory, one has to introduce at
least two chiral fields~\cite{Kim83}. Even though its name refers to
the axion-related QCD anomaly, one must select the component that is
orthogonal to the goldstino. If there are two SM singlet chiral
fields, this is simple because there is only one component left beyond the
goldstino. However, if more than two chiral fields are involved in
SUSY breaking, more care is needed to identify the orthogonal mass
eigenstate.  Among the remaining mass eigenstates beyond the
goldstino, a plausible choice for the axino field is the component
whose coupling to the QCD anomaly term is the biggest. For two initial
chiral fields in Fig.~\ref{fig:AxinoDef}, $\axino'$ has the anomaly
coupling of Eq.~(\ref{eq:Laxino}); hence, the $\axino$ coupling to the
QCD sector is equal to or smaller than those given in
Eq.~(\ref{eq:Laxino}). The remaining coupling is the one to the $s=\pm\frac12$
components of a massive gravitino. Therefore, for the two initial chiral
fields,  axino cosmology must include the gravitino as well, if
$\axino'$ is not identical to $\axino$. The ``leakage'' is parametrized by
the $F$-term of the initial axion multiplet $A$. With more than two
initial chiral fields, the situation involves more mass parameters.
One notable corollary of Ref.~\cite{KimSeo12} is that the axino CDM
relic abundance for $\maxino<\mgravitino$ is an over-estimate if $A$ obtains
the $F$-term.

\begin{figure}[t]
  \begin{center}
  \begin{tabular}{c}
   \includegraphics[width=0.70\textwidth]{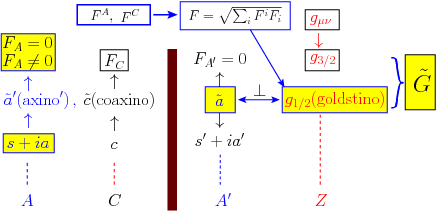}
  \end{tabular}
  \end{center}
\caption{  (Color online) Axion (blue) and goldstino (red) multiplets. The axion direction $a$ is defined
 by the PQ symmetry, and the goldstino ($ {g}_{1/2}$) and axino ($\tilde a$)
    directions are defined by the fermion mass eigenvalues.  The
    primed fields are not mass eigenstates.}
\label{fig:AxinoDef}
\end{figure}
\vspace*{2\baselineskip}

\section{Thermal production}
\label{Sec:ThPD}

As stated in Sec.~\ref{Sec:Introduction}, we define {\em thermal
  production} (TP) of dark matter relics as the
mechanism in which they are produced from particles in thermal
equilibrium in such a way that their resultant energy spectrum
is the same as that of the particles in the thermal equilibrium (up to some
normalization).  DM relics can subsequently freeze out from the
thermal equilibrium or they can be already decoupled, {\it e.g.}
when produced from scatterings or decays of thermal particles.

After decoupling from the thermal plasma, the number density of DM
relics $\nx$ is redshifted and the comoving abundance, or yield $Y$ --
defined as the ratio of number density to entropy density --
\dis{
Y\equiv \frac{n}{s},
}
is conserved as long as the comoving entropy is conserved.
The comoving abundance $Y$ of some species can be converted to
their present relative relic
density using the expression
\dis{
\Omega h^2 \equiv \frac{\rho}{\rho_c/h^2}\simeq 0.27\, \bfrac{Y}{10^{-11}}\bfrac{\VEV{E}}{100\gev},
}
where $\VEV{E}\equiv \rho/n$ is the average energy of the species and
is approximately equal to their mass when they become non-relativistic.

\subsection{Hot Relics}
\label{Subsec:HotRelic}

In this case DM relic particles were in thermal equilibrium
during an early epoch, and then decoupled at a temperature $\Tdec$ which is larger
than their mass $\mx$; see Fig.~\ref{fig:WIMP}.  Since they were still relativistic when they
were produced, they are called ``hot relics''. The energy spectrum
froze out when they decoupled so that the distribution is the same as
when they were in the thermal equilibrium; the number density is
only redshifted after freeze-out. In this case, the comoving abundance
only depends on the effective degrees of freedom of entropy $g_{*S}$
at the time of decoupling. Using Eqs. (\ref{eq:nR}) and (\ref{eq:s}),
one obtains
\dis{ Y_X=\left.\frac{n_X}{s}\right|_{\Tdec} = \frac{45
    \zeta(3)}{2\pi^4}\frac{g_{\rm eff}}{g_{*S}(T_{\rm dec})} , }
where $\zeta(3)\simeq 1.202$ and $g_{\rm eff}=g$ (boson) and $g_{\rm
  eff}=3g/4$ (fermion), with $g$ denoting the degrees of freedom of
the field $X$.

A well-known example of hot relics are light active (SM) neutrinos which decouple at
$\Tdec \simeq 1 \mev$ when $g_{*S}=10.75$.
Their relic density at present is given by
\dis{
\Omega_\nu h^2 =  \frac{\sum m_\nu}{91.5\ev},
}
assuming that they are now almost non-relativistic
({\it i.e.} the bulk of their
energy density is tied up in their rest mass but their velocity distribution still
typically exceeds their escape velocity so that they are not gravitationally bound).

\subsection{Cold Relics: case of WIMPs}
\label{Subsec:WIMP}

When $\Tdec<\mx$, WIMPs decouple when their typical
velocities are still semi-relativistic, $v\simeq c/3$.
The relic abundance and freeze-out temperature can be calculated from
the Boltzmann equation,
\dis{
  \frac{d n_X}{d t} + 3 H n_X &= g_X\int C[f_X] \frac{d^3 p}{(2\pi)^3},
\label{Boltzmann2}
}
where $n_X$ and $g_X$ are respectively the number density and spin degrees of freedom of $X$
while $C[f]$ is the collision operator.
In a homogeneous and isotropic Universe, $n_X$ is defined from the phase space density $f_X$ by
\dis{
n_X = g_X\int  \frac{d^3 p}{(2\pi)^3}f_X(E,t).
}
For definiteness-- in the process of self-annihilation with the type $X+X \rightarrow 3+4$
where we assumes that the species 3 and 4 are in the thermal equilibrium--
the Boltzmann equation can be written as
\be
\frac{d\nx}{dt}=-3H\nx-\langle\sigma_{\rm ann} v\rangle (\nx^2-n_{\rm eq}^2).
\ee
\begin{figure}[t]
  \begin{center}
  \begin{tabular}{c}
   \includegraphics[width=0.65\textwidth]{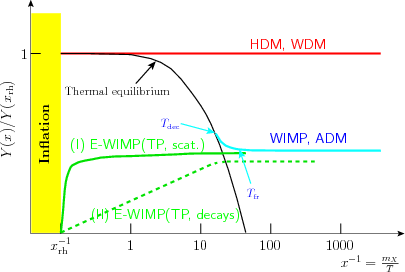}
  \end{tabular}
  \end{center}
  \caption{Relative yield for thermal production of HDM, WDM, CDM WIMPs and ADM, as
    well as E-WIMPs.  In case (I)
    E-WIMPs are thermally created in scatterings involving heavier
    particles in thermal equilibrium starting from $x^{-1}_{\rm reh}=
    \mx/\treh$, while in case (II) in decays of heavier particles in
    thermal equilibrium. In the case of WIMPs a small difference
    between $\Tfr$ and $\Tdec$ is also marked; see text. }
\label{fig:WIMP}
\end{figure}

Initially, the WIMPs are in thermal equilibrium and their number
density follows a Maxwell-Boltzmann distribution Eq.~(\ref{eq:nX}),
decreasing exponentially as the temperature decreases.
WIMPs freeze out (see Fig.~\ref{fig:WIMP}) when the scattering term in Eq. (\ref{eq:Boltzmann}) becomes comparable to the Hubble term,
\dis{
\langle\sigma_{\rm ann} v\rangle n_{X,eq} \simeq H(\Tfr).
\label{eq:freeze}
}

Using the Friedmann relation (\ref{eq:Friedmann}) with $\rho_{\rm rad}=\pi^2 g_*T^4/30$,
this expression may be solved to find the freeze-out temperature of WIMPs
in a radiation-dominated Universe,
\be
T_{\rm fr}^{\rm RD}\simeq \mx\, {\left[\ln\left(\frac{3\sqrt{5}\langle\sigma_{\rm ann}  v\rangle
      M_P \mx^{3/2}}{\pi^{5/2}(\Tfr^{\rm RD})^{1/2} g_*^{1/2}(\Tfr^{\rm RD})}\right)\right]^{-1} }\,.
      \label{eq:Tfr}
\ee
When solved iteratively, Eq. (\ref{eq:Tfr}) usually yields
$x_{\rm fr}=\Tfr^{\rm RD}/\mx\sim1/25$,
except near resonances and thresholds; see Fig.~\ref{fig:WIMP}.
An approximate solution of the Boltzmann equation giving the
present-day CDM relic density is given in Eq.~(\ref{eq:oh2approx}).

In the case where WIMP annihilations proceed dominantly via $s$--wave,
then $\langle\sigma_{\rm ann} v\rangle$ is approximately constant and
the comoving abundance may also be obtained from
Eq.~(\ref{eq:freeze}),\footnote{The freezeout temperature $\Tfr$
  (below which the yield stays constant) is almost the same as the
  decoupling temperature $\Tdec$ (which is when out-of-equilibrium
  commences).  In Fig.~\ref{fig:WIMP} a small difference between
  $\Tfr$ and $\Tdec$ for WIMPs is shown.  }
\be
Y_X=\frac{(90/\pi^2g_*(\Tfr^{\rm RD}))^{1/2}}{4\VEV{\sigma_{\rm ann}v } M_P \Tfr^{\rm RD}  }.
\label{eq:Y_X}
\ee

In the case when $\langle\sigma_{\rm ann} v\rangle$ is $p$--wave dominated, then it is temperature dependent.
Gondolo and Gelmini~\cite{Gondolo:1990dk} showed that
\be
\langle\sigma_{\rm ann} v\rangle=
\frac{1}{4xK_2^2(\frac{1}{x})}\int_2^\infty da \sigma (a)
a^2(a^2-4) K_1\left(\frac{a}{x}\right),
\ee
where $x=T/m_X$, $a=\sqrt{s}/m_X$ and the $K_i$ are modified Bessel
functions of order $i$.  More involved expressions containing the case
of co-annihilations can be found in Ref.~\cite{Edsjo:1997bg}.  Once
the temperature-dependent thermally averaged WIMP annihilation cross
section times velocity is found, then the Boltzmann equation can be
integrated to find the present-day WIMP number density
\be
\nx(T_0)=\frac{1}{\mx}\left(\frac{T_0}{T_\gamma}\right)^3(T_\gamma )^3
\sqrt{\frac{4\pi^3g_*G_N}{45}}\left[ \int_0^{T_f}\langle \sigma_{\rm
    ann} v_{\rm rel}\rangle dT/m \right]^{-1} ,
\ee
where $T_\gamma =2.725$ K is today's cosmic microwave background temperature.

In the case where freeze-out occurs during a matter-- or decay--dominated epoch,
then simple expressions such as Eq. (\ref{eq:Y_X}) are modified.
For instance, when WIMPs decouple during a matter-dominated phase,
then the comoving abundance --
after the matter to radiation transition due to the decay of
dominating matter -- is~\cite{CKLS08,Baer_mixed}
\be
Y_X =\frac32  \frac{(90/\pi^2g_*(\Tfr^{\rm MD}))^{1/2}}{4\VEV{\sigma_{\rm ann}v } M_P \Tfr^{\rm MD} }
\bfrac{T_D}{\sqrt{T_e \Tfr^{\rm MD}}},
\ee
where
\be
T_{\rm fr}^{\rm MD}\simeq
\mx \left[\ln\left(\frac{3\sqrt{5}\langle\sigma_{\rm ann} v\rangle
    M_P m_X^{3/2}}{\pi^{5/2}T_e^{1/2}g_*^{1/2}(\Tfr^{\rm MD})}\right)\right]^{-1}
    \ee
is the freeze-out temperature in the matter dominated Universe,
$T_D$ is the temperature of radiation when the dominating matter decays
and $T_e$ is the matter-radiation equality temperature $T_e=\frac{4}{3}m_MY_M$,
with $m_M$ and $Y_M$ denoting the unstable matter mass and yield, respectively.


\subsection{Cold relics: case of E-WIMPs}
\label{Subsec:eWIMP}

If the relic particles' interactions are {\it extremely} weak, then
they may never be in the thermal equilibrium in the early Universe.
This is the case of particles labeled in the literature as E-WIMPs~\cite{Choi:2005vq} (or
alternatively super-WIMPs~\cite{feng03} or FIMPs~\cite{Hall:2009bx}).
The E-WIMP freeze-out temperature
is larger than the reheating temperature after inflation,
$\Tfr>\treh$.  Therefore, the primordial population of E-WIMPs is
inflated away, and the particles regenerated after reheating are
already decoupled from thermal plasma.  However, this does not mean
that they are completely decoupled from the plasma.  Even though their
interaction rates are tiny, E-WIMPs can still be produced at
significant rates in scatterings (case~I in Fig.~\ref{fig:WIMP}) or
decays (case~II) involving heavier thermal particles, so that they
still may give rise to substantial, or even dominant, contributions to
E-WIMP abundance.

Due to their suppressed interactions with ordinary matter, the E-WIMP
number density is small enough and one may usually neglect
the back reaction of E-WIMP annihilation. In this case the comoving
abundance can easily be
obtained by integrating the production rate over the temperature from
the reheating temperature to the present one,
\dis{
Y(T_0) = \int_{T_0}^{\Treh} \frac{\Gint n_{eq}}{s(T) H(T) T} dT,
}
where $\Gint =n\langle\sigma v\rangle$ in general depends on the energy of the participating
particles and thus the temperature of the background.
Also, $T_0$ here and below is any low temperature
below which entropy is assumed to be conserved.

A specific example of  E-WIMPs is the axino of SUSY models augmented with
the PQ symmetry.  Axino interactions with SM particles and their
superpartners are  strongly suppressed by the axion decay constant $\fa$.

The quantity that is very important in axino astroparticle physics
and cosmology, and at the same time most poorly known, is its mass
$\maxino$. In the literature there exist several theoretical
calculations of the axino mass~\cite{ChunKN92,ChunLukas95,KimSeo12}. A
method for calculating the axino mass applies to any goldstino (the
superpartner of a Goldstone boson). A goldstino related to the
Goldstone boson has a root in a global U(1) symmetry and receives its
mass below the SUSY breaking scale. SUSY breaking triggers the
super-Higgs mechanism and is related to the gravitino mass
$\mgravitinoalt$; this issue was recently clarified in
Ref.~\cite{KimSeo12}. Even though a typical expectation for the axino
mass is to be of order $\mgravitinoalt$, the theoretically allowed mass
range encompasses a much wider range from sub-\ev\ to multi-\tev,
allowing axino LSPs to be hot, warm or cold DM.
We will discuss these cases in more detail below.

In an early paper~\cite{KimMasiero84}, a very light HDM-like axino from the decay of a
photino was shown to constrain the photino mass dependence on the axion decay constant $\fa$.
In~\cite{RTW91}, Rajagopal, Turner and Wilczek considered axinos with $\maxino$ in the keV range.
Axinos in this mass range can give the right amount of DM if produced
from freeze-out in thermal equilibrium and can constitute WDM in the standard Big Bang cosmology.
However, this kind of thermal axino is cosmologically irrelevant if the
reheating temperature $\treh$ after inflation is much lower than the
Peccei-Quinn (PQ) symmetry breaking scale $\fa$ (See, however, Refs. \cite{MarshJE14,Gondolo14}). In this case, the
population of primordial axinos is strongly diluted by cosmic
inflation.

Axinos can, however, be subsequently re-generated after reheating
in spite of their exceedingly small interaction strength.
In axion models such as KSVZ,  the relic abundance
of thermal axinos changes linearly with the reheating temperature and
depends on a SUSY
axion model. This special feature might allow for a
glimpse of the earliest time after inflation through the reheating
temperature inferred from the relic density of axino DM.\footnote{It
  is also worth mentioning that, due to the strongly suppressed
  interaction strength, it is not necessary to assume $R$-parity
  conservation for very light axinos to constitute DM.  In connection
  with the recent $3.5\kev$ X-ray line from the Andromeda
  galaxy and Persus  galaxy cluster \cite{Bulbul:2014sua,Boyarsky:2014jta,Boyarsky:2014ska},
  a possible  solution in terms of warm decaying axino DM has been pointed out in
  the presence of $R$-parity violation~\cite{Kong:2014gea,ChoiSeto14}.
See also~\cite{Babu:2014pxa}.
However there is a different claim with no statistically significant line emission near 3.5 keV~\cite{Jeltema:2014qfa}.
}
Alternatively, in the DFSZ SUSY axion model, the direct coupling of axinos
to Higgs and gauge bosons leads to maximal production rates at temperatures
$T\sim m_{\ta}$ so that axinos are thermally produced via the ``freeze-in''
process~\cite{Hall:2009bx}.

The remaining physical state of the axion supermultiplet is the
$R$-parity even scalar saxion $s$. Its mass $m_s$ comes from soft SUSY
breaking and hence is expected to be of order $\mgravitinoalt$.
In the early Universe, it is expected to eventually decay to SM particles:
$s\to gg$ in the KSVZ model and $s\to$ gauge/Higgs bosons in DFSZ.
An important cosmological implication of saxion production and decay to SM particles is the
possibility of late-time (post freeze-out) entropy production and a
dilution of frozen-out cosmic particles and the cosmic energy density.
In axion cosmology, the effect leads to an increase in the
cosmological upper bound on
$\fa$~\cite{Kim91,KimKim95,Kawasaki:2007mk,Kawasaki:2011ym,Baer:2011eca}
for a given assumption on $\theta_i$.

Depending on their couplings, saxions may also decay into axion or axino pairs.
In the first case, the axions can affect the cosmic microwave
background temperature anisotropy by contributing an additional relativistic
component~\cite{ChangKim96,Hasenkamp:2011em,Choi:2012zna,Graf:2012hb,Bae:2013qr,Graf:2013xpe},
often parameterized by the allowed number of additional species of
neutrinos $\Delta N_{eff}$.  There is some weak evidence for a
non-zero value of $\Delta N_{eff}$ beyond the SM value, but a
conservative limit gives $\Delta N_{eff}\alt 1.6$~\cite{Ade:2013zuv,Bae:2013qr}.
In the case where saxions decay to axinos or other SUSY particles,
their late decays may augment the abundance of the LSP dark matter, be
it axinos themselves, or via cascade decays to neutralinos, gravitinos or something else.

\subsubsection{\bf $\mathbf{\Gint \sim T^3}$}

For the case of a constant E-WIMPs scattering cross section $\VEV{\sigma v}=\sigma_0$ --
which happens when the interaction is induced by non-renormalizable terms
as in the case of axions, KSVZ axinos or gravitinos --  particles are
predominantly produced at the highest temperature (case~I in
Fig.~\ref{fig:WIMP}). Then, the comoving abundance is proportional to
the highest temperature in the integration range~\cite{Choi:1999xm}:
\dis{
Y(T_0) = \int_{T_0}^{\Treh} \frac{\VEV{\sigma v}n_{eq}^2}{s(T) H(T) T} dT
\sim  \frac{135\sqrt{10} M_P}{2\pi^7 g_*^{3/2}} \sigma_0\Treh.
}

\subsubsection{\bf $\mathbf{\Gint \sim T^n}$ with $n>3$}

If $\Gint\sim T^n$ with $n>3$, then the abundance will depend on
higher powers of the reheating temperature as ${\treh}^{n-2}$.
Dark matter produced this way will have thermal spectrum but will not
be in thermal equilibrium, just like in the case of $n=3$. A specific example
of $n=5$ in a model with heavy $Z'$ gauge boson was given in Ref.~\cite{Mambrini:2013iaa}.

\subsubsection{\bf $\mathbf{\Gint \sim T^n, n<3}$ and FIMPs}

In this case (case~II in Fig.~\ref{fig:WIMP}), the production of
relics takes place predominantly at low temperatures $T\sim m$, just
before the Boltzmann suppression kicks in.  When the scatterings
proceed through renormalizable interactions, the corresponding cross
section depends on the temperature as $\sigma = \kappa /T^2$, where
$\kappa$ is some constant. In this case, the thermal production of
decoupled particles is independent of the reheating temperature and is
given by
\dis{
Y(T_0) \simeq \frac{135\sqrt{10} \zeta(3)^2 \sqrt{g}M_P}{2\pi^7}  \frac{\kappa}{T_1},
}
where $T_1$ is the temperature of the order of the mass of the
particles participating in the scatterings. These sort of DM particles have been dubbed
FIMPs by Hall \etal\,~\cite{Hall:2009bx} for feebly interacting massive particles, but perhaps
a more appropriate
name would be ``frozen-in massive particles''.

\subsubsection{\bf E-WIMPs from decays}

A further case occurs  when E-WIMPs are produced from decays of thermal particles with mass $M$
and a decay rate $\Gamma$ (case~II in Fig.~\ref{fig:WIMP}).
In this situation, the comoving abundance is given by~\cite{Choi:1999xm}
\dis{
Y(T_0)\simeq  \frac{405\sqrt{10} \zeta(5)M_P}{8\pi^4 g_*^{3/2}} \frac{ \Gamma}{M^2}.
}
These processes have been used for the production of axinos, right-handed scalar
neutrinos, etc.

\subsection{Asymmetric dark matter (ADM)}
\label{Subsec:AsymDM}

The idea behind asymmetric dark matter (ADM)~\cite{Boucenna13,Petraki:2013wwa,Zurek:2013wia}
is based on an asymmetry between DM particles and their antiparticles (``anti-DM'').
In the early Universe, only the number density difference between the two
(asymmetric component) remains after the annihilation of the symmetric
components of DM and anti-DM.  In this case, the relic density of ADM
is set by the asymmetry in their initial populations, and not by the
thermal freeze-out.  This is similar to the mechanism of generating
the baryon number density which relies on an initial baryon asymmetry.

Some examples of ADM include
technibaryons~\cite{Nussinov:1985xr,Barr:1990ca},
mirror dark matter~\cite{Blinnikov:1982eh,Hodges:1993yb,Foot:2001hc,An:2009vq,Blinnikov:2014hra},
scalar neutrinos~\cite{Hooper:2004dc}, pure higgsinos~\cite{Blum:2012nf}
and others.

The origins of the asymmetry of ADM and of the baryon asymmetry can be
related~\cite{Kaplan:1991ah}. In this case, the mass of ADM is
calculable in a specific model as
\dis{ m_{\rm ADM} = \frac{\eta_{B}}{\eta_{\rm ADM}}
  \frac{\Omega_{\rm ADM}}{\Omega_{B}}\, m_p ={\mathcal O}(1-10)m_p,
}
where the baryon and ADM asymmetries are parametrized by
\dis{
\eta_i \equiv  \frac{n_i - n_{\bar{i}} }{s},~{\rm for~} i=B~{\rm or~ADM}
}
with entropy density $s$.  If the asymmetry of both ADM and baryons is
the same, $\eta_{\rm ADM}=\eta_{B}\simeq 10^{-10}$, the energy density
of DM and baryons are related by $\Omega_{\rm DM}/\Omega_{B}\simeq 5$
~\cite{Ade:2013zuv}, which implies the existence of a light DM with mass
around $5\gev$. In this regard, it is interesting to note that the
light DM has been supported by several claims of signals in direct
detection experiments: DAMA/LIBRA~\cite{Bernabei:2010mq},
CoGeNT~\cite{Aalseth:2010vx,Aalseth:2012if},
CRESST~\cite{Angloher:2011uu} and CDMS-II (Si)~\cite{Agnese:2013rvf},
which however have been
contradicted~\cite{HuhDel13,DelNobile1,DelNobile2,DelNobile3} by other
experiments: CDMS-II~\cite{CDMSII,CDMSIIGe}, XENON10~\cite{XENON10},
XENON100~\cite{XENON100:2011,XENON100:2012}, and the recent null
result from CDMSLite~\cite{Agnese:2013jaa} and
LUX~\cite{Akerib:2013tjd}.

To generate the asymmetry from an initially symmetric Universe in the
sector of either DM or baryons, the mechanism should satisfy the well
known Sakharov conditions~\cite{Sakharov:1967dj}.  First, the
asymmetry can be created in one of the sectors at high temperatures
and then subsequently transferred to the other sector, or both
asymmetries can be created together at the same moment. At low
temperatures, the interactions for generating and transferring
asymmetries are frozen.

\begin{figure}[!t]
\begin{center}
\includegraphics[width=0.55\linewidth]{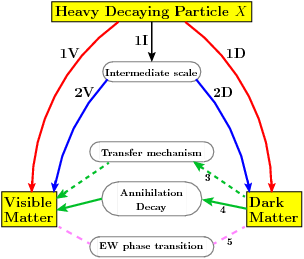}
\end{center}
\caption{Schemes for ADM production summarized in~\cite{Boucenna13}.
} \label{fig:ADMsummary}
\end{figure}

In Fig. \ref{fig:ADMsummary}, various mechanisms for DM production in the ADM models
(Table 1 of Ref.  \cite{Boucenna13})  are illustrated.
One obvious mechanism is to relate the visible (V) sector baryon asymmetry to the
dark (D) sector asymmetry via a heavy particle decay both to V and D sectors.
These are denoted as {\bf 1} and {\bf 2} categories, where {\bf 2} stops
over at the intermediate sector {\bf I}.
A well-known mechanism is the Kitano-Low mechanism \cite{Kitano05}.
Categories  {\bf 3}, {\bf 4} and {\bf 5} use the  mechanisms without the decaying mother particle.
Category {\bf 3} uses one conserved quantum number which is carried by D and V sectors.
So, `transfer' means some DM quantum numbers are transferred to some visible sector quantum numbers,
as we know the lepton number is transferable to the baryon number during the
electroweak phase transition era with $B-L$ conservation.
Category {\bf 4} uses DM+DM annihilation processes or DM decay to produce the SM particles.
Category {\bf 5} is the case where some D sector particles carry SU(2) electroweak charge.
Then the SU(2) phase transition relates the two asymmetries.
Namely, sphalerons were used to transfer the asymmetries between the two
sectors~\cite{Barr:1990ca,Kaplan:1991gb,Gudnason:2006yj}. These sphalerons
violate the baryon and DM symmetry while preserving some linear
combination of them, like the well known electroweak sphalerons which
violate $B$ and $L$ number while preserving $B-L$.
At some level, {\bf 3} also has the feature described in {\bf 5}.
In this case, the sphalerons transform the DM asymmetry to baryon
asymmetry, or in the opposite way to share the asymmetry between two
sectors. After sphalerons decouple, both asymmetries are frozen in
each sector.

The other way is that the transfer proceeds through higher-dimensional
operators which are gauge singlets of the $B-L$
operator~\cite{Kaplan:2009ag}. When the operator decouples as the
Universe expands, the asymmetries freeze in. This mechanism can be
extended to non-standard cosmologies~\cite{Iminniyaz:2013cla,Gelmini:2013awa}.

In ADM models, the symmetric component must annihilate efficiently.
Those interaction may enhance the scattering cross section with nuclei
in direct detection experiments.  The large interactions could potentially
explain the claimed light DM signals with a spin-independent
scattering cross section of order $\sim 10^{-40} \cm^2$.  Some studies
attempted to reconcile these signals with other null experiments in
the ADM
scenario~\cite{Chang:2010yk,Farina:2011pw,Hooper:2011hd,DelNobile:2011je,Okada:2013cba},
while ADM limits have been derived from direct detection in \cite{Cui:2012mq,MarchRussell:2012hi}.
The new physics element in the ADM models -- light scalar particles which mediate the
self-annihilation or the non-renormalizable interactions -- can also
affect collider
phenomenology~\cite{Kaplan:2009ag,Blennow:2012de,Bhattacherjee:2013jca,Kim:2013ivd}.

ADM when accumulated in a star can scatter off nuclei and transport
energy and modify the density and temperature of the star.  Such
processes can be used to constrain ADM
models~\cite{Iocco:2012wk,Lopes:2012af}.  The accumulation of ADM in
astrophysical objects, such as neutron stars, the Sun, and brown and
white dwarves can give strong constraints on the ADM scattering cross
section with baryons~\cite{Kouvaris:2012dz,Bell:2013xk,Jamison:2013yya,Zurek:2013wia},
especially for the case of scalar DM.

If the ADM is self-interacting~\cite{Ellwanger:2012yg,Pearce:2013ola} or decaying,
then various astrophysical signatures may arise.
For example, decaying ADM might explain the cosmic positron excess~\cite{Feng:2013vva}
or produce signatures in the gamma-ray sky~\cite{Masina:2012hg}.

\vspace*{2\baselineskip}

\section{Non-thermal production of dark matter}
\label{Sec:NTProduction}

When the energy spectrum of the produced DM is different from a
thermal distribution, then we call the process a {\it non-thermal
  production} (NTP) mechanism.  The BCM of oscillating (pseudo)scalar
fields discussed in Subsections~\ref{Subsec:ThCollective}
and~\ref{Subsec:SUSYColl} is an example of non-thermal dark matter
production.  Another important type of NTP mechanism is production of
DM via decays of heavy particles which are already out-of-equilibrium.
More non-thermally produced DM candidates such as primordial black
holes (PBH) and super-heavy particles such as Wimpzillas may be
produced by gravitational effects; these are discussed in
Sec.~\ref{Sec:NonSUSY}.

\subsection{Dark matter from bosonic coherent motion}

\subsubsection{Scalar fields in the early Universe}

The equation of motion for a scalar field in the expanding Universe can be written as
\be
\ddot{\phi}+3H(T)\dot{\phi}+\frac{\partial V(\phi )}{\partial\phi}=0.
\label{eq:axionEoM}
\ee
For small values of $\phi$, the potential energy term is approximated by
\be
V(\theta )\simeq \frac{1}{2}m ^2(T)\phi^2
\label{eq:axion_potl}
\ee
where we have introduced in Eq. (\ref{eq:axion_potl}) a possible {\it
  temperature-dependence} of the boson mass $m(T)$ which is $\sim 0$
at a sufficiently high temperature.  In the limit where the potential
term is negligible, the solution is $\phi\sim {\rm constant}$.  But
once the potential term becomes comparable to the Hubble term,
Eq. (\ref{eq:axionEoM}) becomes the equation of a damped harmonic
oscillator, with the Hubble term providing the friction. In this case,
the solutions become oscillatory.

By introducing the scalar field energy density $\rho
=\dot{\phi}^2/2+V(\phi )$ and by averaging terms over one oscillation
cycle, $\langle\dot{\phi}^2\rangle =m^2\langle\phi^2\rangle =2\langle
V\rangle =\langle \rho\rangle$, the equation of motion can be recast
in terms of time-averaged quantities as
\be
\dot{\rho}+3\frac{\rho}{R}\dot{R}-\frac{\rho}{m}\dot{m}=0 \ \ \ {\rm or}\ \ \ \frac{d}{dt}\left(\frac{\rho R^3}{m}\right)=0.
\ee
One immediate consequence of this result is that the oscillating field
$\rho\sim R^3$ so that it behaves like {\it non-relativistic
  matter}. Another is that the oscillatory phase begins at a
temperature $T_{\rm osc}$ defined by $3H(T_{\rm osc})\simeq m(T_{\rm
  osc})$. The solution for the scalar field energy density is
\be
\rho (T) =\rho (T_{\rm osc})\left(\frac{R_{\rm osc}}{R}\right)^3\frac{m(T)}{m(T_{\rm osc})} .
\ee
The solution $\phi (t)$ of Eq. (\ref{eq:axionEoM}) we refer to is a
bosonic coherent motion.  It is ubiquitous in early Universe cosmology \cite{BCMreview14}
and can be applied not only to the motion of the axion field, but also
to the case of inflatons, moduli, saxions, Affleck-Dine fields, etc.

\subsubsection{Axion production via BCM}

For the case of the axion field, we take $a=\theta f_a$ where $\theta$
is the axion misalignment angle.  Also, for the axion field, $m_a$ is
temperature dependent and is negligible for $T>T_{\rm QCD}\sim 1$ GeV;
later, it ``turns on'' at lower temperatures after the QCD phase
transition when the chiral anomaly becomes
relevant~\cite{BaeHuhKim09}.  Axion field oscillations begin at a
temperature $T_{\rm osc}\sim 0.9$ GeV when the axion mass and Hubble
constant are comparable: $3H(T_{\rm osc})\simeq m(T_{\rm osc})$.
Also, at $T_{\rm osc}$ one finds $\rho_{\rm osc}=\dot{a}^2/2+V\simeq
m^2(T_{\rm osc})a_i^2/2$ where $a_i$ is the initial axion field
strength $a_i=\theta_i f_a$.

A slightly more refined calculation which allows for anharmonic corrections (not just small oscillations) and which is valid for $f_a\alt \MG $ gives the present day axion energy density as~\cite{BaeHuhKim09}
\be
\rho_a(T_\gamma)\simeq \frac12 m_a(T_\gamma) m_a(T_{\rm osc})\left(\frac{R(T_{\rm osc})}{R(T_\gamma)} \right)^3
\theta_i^2f_{\rm anh}(\theta_i) .
\ee
Here, $f_{\rm anh}$ is the anharmonic correction term parametrized by
$f_{\rm anh}(\theta_i)\approx \left[\ln (\frac{e}{1-\theta_i^2/\pi^2})\right]^{7/6}$ and
the initial misalignment angle $\theta_i$  takes on a range from $-\pi \rightarrow \pi$
and $T_{\gamma}$ is the present temperature of radiation.

In~\cite{BaeHuhKim09}, $\Omega_a$ has been calculated  for $f_a\alt \MG $:
\dis{ \Omega_a\simeq \left\{ \begin{array}{l}
      0.503\left(\frac{\theta_i^2 F(\theta_i)}{\gamma} \right) \left(\frac{0.678}{h} \right)^2 \left(\frac{f_{a}}{10^{12}\gev} \right)^{1.182},~{\rm for}~\Lambda_{\rm QCD}=320\,\mev,\\[0.5em]
      0.444 \left(\frac{\theta_i^2 F(\theta_i)}{\gamma}
      \right)\left(\frac{0.678}{h}
      \right)^2\left(\frac{f_{a}}{10^{12}\gev} \right)^{1.184},~{\rm
        for}~\Lambda_{\rm QCD}=380\,\mev,
      \\[0.5em]
      0.399 \left(\frac{\theta_i^2 F(\theta_i)}{\gamma}
      \right)\left(\frac{0.678}{h}
      \right)^2\left(\frac{f_{a}}{10^{12}\gev} \right)^{1.185},~{\rm
        for}~\Lambda_{\rm QCD}=440\,\mev\,,
\end{array}\right.
\label{eq:Oh2axion}
}
where $\gamma$ is the entropy increase ratio from $t_i$ to present.

\begin{figure}[t]
  \begin{center}
  \begin{tabular}{c}
    \includegraphics[width=0.5\textwidth]{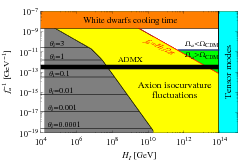}
   \end{tabular}
  \end{center}
  \caption{The axion decay constant versus $H_I$ for the QCD axion corresponding to the pink part of Fig.~\ref{fig:ALPs}.
Here, we used the updated result of~\cite{BaeHuhKim09} --  Fig.~\ref{fig:BaeHuh14} --
to denote the lines of $\theta_i$. The shaded parts are not allowed~\cite{Gondolo10}. }
\label{fig:Gondolo}
\end{figure}

\begin{figure}[t]
  \begin{center}
  \begin{tabular}{c}
    \includegraphics[width=0.5\textwidth]{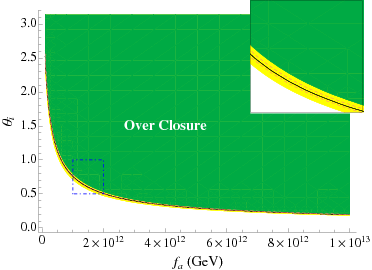}
   \end{tabular}
  \end{center}
  \caption{$\theta_i$ vs. $\fa$ plot modified from the figure of~\cite{BaeHuhKim09} with new data from PLANCK~\cite{Ade:2013zuv} and updated quark masses~\cite{PDG14}.   }
\label{fig:BaeHuh14}
\end{figure}

Visinelli and Gondolo introduced also the cosmic inflation constraint
on the allowed parameter space of the axion-only CDM as shown in
Fig.~\ref{fig:Gondolo} for the case where the BCM is assumed to
account for all CDM~\cite{Gondolo10}.  The misalignment angle of
Fig.~\ref{fig:Gondolo} can be read from the closure boundary of
Fig.~\ref{fig:BaeHuh14}. A common assumption is that $\theta_i$ takes
a specific value, such as $\pi/\sqrt{3}$ (from averaging over many
distinct subdomains in the case where the axion field forms {\it
  after} inflation~\cite{Turner86}); this leads to an upper bound on
the PQ scale of $f_a\alt 10^{11}$ GeV to avoid overproduction of
axions. However, in the case where the axion field forms {\it before}
the end of inflation, the entire Universe would have a single value of
$|\theta_i|\sim 0\to \pi$ in which case any value of $f_a$ is
permitted ranging from the astrophysical bound $f_a\agt 10^9$ GeV all
the way up to $M_P$ or beyond. The region with $\fa>10^{12}$ GeV is
called the {\it ``anthropic region''} since in this case unusually
tiny values of $\theta_i$ would be
required~\cite{Pi84,Tegmark06,Tegmark08}.  The horizontal bars of
$\theta_i$ values in Fig.~\ref{fig:Gondolo} are determined by
requiring that $\Omega_a$ of
Eq. (\ref{eq:Oh2axion})~\cite{BaeHuhKim09} saturates the measured DM
abundance.

\subsubsection{Impact of recent BICEP2 measurement of tensor-to-scalar ratio}

The BICEP2 collaboration has recently claimed a measurement of tensor
$B$-modes in the CMBR.  Their measurement corresponds to a tensor to
scalar ratio $r=0.2^{+0.07}_{-0.05}$~\cite{BICEP2I} at the end of
inflation. This corresponds to the value of the Hubble constant at the
end of inflation as $H_I=1.1\times 10^{14}$ GeV~\cite{MarshJE14}.
Even more recently, the Planck collaboration presented the dust polarisation
in the BICEP2 experiment field~\cite{Adam:2014bub}.
The extrapolation from Planck 353 GHz data to 150 GHz gives a dust power
with the same magnitude as reported by BICEP2.
It is expected that the BICEP2 result will be corrected during an ongoing joint
analysis between Planck and BICEP2.

Taking the BICEP2 result at face value, then reading off from
Fig.~\ref{fig:Gondolo}, this would appear to exclude the case where
the axion field forms before the end of inflation (where $\theta_i$ is
a free parameter), and instead one must average over many disparate
domains of $\theta_i$ such that the average $\langle\theta_i\rangle
=\pi/\sqrt{3}$.  Then the axion-only CDM possibility excludes the
small axion mass range $m_a<10$\,nano-eV.  More importantly, for the
QCD axion only a narrow range of cosmic axion is allowed, $m_a\gtrsim
14\mu \ev$~\cite{MarshJE14,Gondolo14}, where we included a factor
of five uncertainty from the numerical analysis of
Ref.~\cite{KawasakiWall12}.  In these studies, axion creation via the
axionic string-wall annihilation was also taken into account,
following the numerical analysis of Ref.~\cite{KawasakiWall12}.  At
present, the amount of axionic DM creation via numerical studies is
not a settled issue yet, see, e.g.,
Refs.~\cite{Hagmann91,Shellard94,Chang99,Hagmann01}.  One of the main
reasons for the uncertainty in estimating this amount is the
difficulty of simulating the Vilenkin-Everett annihilation mechanism
of the horizon scale string-wall system~\cite{Vilenkin82,BarrChoiKim}.
Or there may be no domain wall problem for the QCD axion via BSM
physics \cite{BarrKim14}.  However, their conclusion that ``the anthropic region is
closed'' may be valid in so far as the reheating temperature after
inflation goes up to the region $\treh>10^{12}$ GeV, which depends on
the existence of any massless boson (because the invisible axion is
massless at this high temperature) at the scale where BICEP2 data was
relevant.\footnote{Even if the BICEP2 value of $r$ is settled to
  $\frac13$ of the original value, $H_I$ reduces by a factor of
  $(\frac13)^{1/2}$, and hence most probably the domain wall problem
  we commented here may be still valid.}

\begin{figure}[t]
  \begin{center}
  \begin{tabular}{c}
   \includegraphics[width=0.35\textwidth]{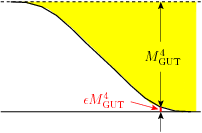}\hskip 1cm
   \end{tabular}
  \end{center}
  \caption{Two scalars contributing to the CC. }
\label{fig:IsoCurv}
\end{figure}

However, the assumption that axions are the only CDM component is unlikely to be realized at
the GUT scale.  There are numerous SM singlet fields and the
isocurvature perturbation may involve several scalar components: {\it
  e.g.}  Fig.~\ref{fig:IsoCurv} is shown for the case of two scalars
where the dominant component is provided by $\Phi$ and a negligible
component is provided by $\chi$.  If $\langle\chi\rangle$ breaks the
PQ symmetry and $\fa\lesssim 10^{11}\gev$, the contribution to the CC
from the $\chi$ field is negligible and practically there is no
isocurvature constraint if $10^9\gev<\fa<10^{11}\gev$.

\subsubsection{Detection of axions and ALPs}

There have been several experimental approaches suggested to detect axion CDM assuming they provide
100\,\% of the CDM density of the Universe~\cite{KimRMP10}.

The standard mode for axion (and also ALP) detection makes use of the low energy effective Primakoff
interaction parametrized by $c_{a\gamma\gamma}/\fa$:
\dis{
{\mathcal L}_{a\gamma\gamma}=-c_{a\gamma\gamma}\frac{\alpha_{\rm em}\,a}{8\pi\fa}\,\frac{\epsilon_{\mu\nu\rho\sigma}}{2} F_{\rm em}^{\mu\nu}F_{\rm em}^{\rho\sigma}
\equiv -c_{a\gamma\gamma}\frac{\alpha_{\rm em}\,a}{8\pi\fa}\,F_{\rm em}\tilde{F}_{\rm em}\propto {\bf E}\cdot{\bf B}.
}
For axions, the low energy value $c_{a\gamma\gamma}$ includes the QCD chiral symmetry breaking effect, $-1.98$, which is the value for $m_u/m_d\simeq 0.48$~\cite{Manohar12},
\dis{
c_{a\gamma\gamma}=\OVER{c}_{a\gamma\gamma}-1.98,
}
where $\OVER{c}_{a\gamma\gamma}$ is determined above the electroweak scale.
The values of ${c}_{a\gamma\gamma}$s are presented in Ref.~\cite{Kim98,KimRMP10} for several different KSVZ and DFSZ models. There exists one string calculation of  ${c}_{a\gamma\gamma}$ in Ref. \cite{KimAgamma14}:  $\OVER{c}_{a\gamma\gamma}=1123/388\to c_{a\gamma\gamma}\simeq 0.91$.

With the BICEP2 reported value  of a GUT scale energy density, it may
be worthwhile to consider some  models with ultraviolet
completion to gain perspective on possible values of the ${c}_{a\gamma\gamma}$.
There are two such estimations, one with an approximate PQ symmetry~\cite{ChoiKimIW07} and the other with an
exact PQ symmetry resulting from the anomalous U(1) gauge symmetry in string models~\cite{KimAgamma14}.
Since $m_a=(0.59\times 10^7/f_{a,{\rm GeV}})$\,[eV], the functional relationship between the interaction strength versus $m_a$ is
\dis{
y = \left|\frac{\alpha_{\rm em}c_{a\gamma\gamma}}{8\pi }\cdot \frac{1}{f_{a\,\gev}}\right|
&=  \left|4.92\times 10^{-11} c_{a\gamma\gamma} \right| m_{a,\rm eV},~{\rm  coeff.~ of}~ {F_{\rm em}\tilde{F}_{\rm em}} \\
&\to \left|1.57\times 10^{-10}  c_{a\gamma\gamma} \right| m_{a,\rm eV},~{\rm in~case~ of}~ {\bf E\cdot B} \label{eq:cagvsmass}
}
where $\fa$ and $m_a$ are given in units of GeV and eV, respectively. In the log-log plot, this is just a line.

The axion haloscope makes use of the Primakoff effect to convert cosmic axions into photons
which build up as transverse magnetic (TM) modes
within a supercooled cavity placed within a magnetic field~\cite{Sikivie83}.
The TM mode builds following the oscillation of the axion vacuum, which is depicted in Fig.~\ref{fig:AxDetCartoon}. This introduces the key difference in the detection methods between the QCD axion and ALPs.
It takes a certain time to build up the maximum TM mode amplitude.
After this maximum TM mode is established, the coherent motion of the axion field and electromagnetic fields are described by the axion-electrodynamics.
Recently, a complete solution of this axio-electrodynamics has been given in Ref.~\cite{HongJ14}.

\begin{figure}[t]
  \begin{center}
  \begin{tabular}{c}
   \includegraphics[width=0.30\textwidth]{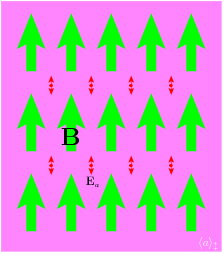}
   \end{tabular}
  \end{center}
  \caption{The scheme depicting {\bf E} field oscillation, following the axion vacuum. }
\label{fig:AxDetCartoon}
\end{figure}

The collected TM modes in the cavity can loose energy by the conversion (TM mode)$\to$(axions) in the cavity, and by the Joule heating through the cavity wall.
These define $Q$-factors denoted by $Q_a$ and $Q_c$, respectively.
$Q_c$ is the quality factor for the cavity and for the $l$-th resonance it is equal to $Q_l=\sqrt{m_a R \sigma R/2}$. $Q_J$ at resonance could be similar and $Q_l$.
Since we must consider the maximum loss of energy to define the quality of the cavity,
the built-up TM mode cannot exceed the minimum of $Q_J$ and $Q_a$.
In these types of cosmic axion search experiments, the feasibility depends crucially on the quality of the cavity wall, which has been discussed in Refs.~\cite{Sikivie83,Sikivie85,Krauss85,HongJ14}.
The so-called `invisible' axion might then be detected in these Sikivie-type cavity detectors
immersed in a strong magnetic field~\cite{Sikivie83}.
Experimental bounds on axions and also on axion-like particles (ALPs) are shown in Fig.~\ref{fig:ALPs}
in the coupling vs. mass plane.

The ALPs are hypothetical particles defined to have $(a_{\rm ALP}/f_{\rm ALP})F^{\rm em}\tilde{F}^{\rm em}$ couplings which need not relate $f$ and $m$ (they are related in the axion case, Eq. (\ref{eq:cagvsmass})).
In connection with the recent 3.5\,keV line from galaxy clusters,
an ALP possibility with $m_{\rm ALP}\sim 7$\,keV and $f_{\rm ALP}\sim 4\times 10^{14}\,\gev$
has been proposed in Ref.~\cite{LeeHM14,Higaki14} (which is barely consistent with Fig.~\ref{fig:ALPs}) but an ALP with $g_{a\gamma\gamma}\sim (3-10)\times 10^{-18}\,\gev^{-1}$ is consistent~\cite{Jaeckel14}.

At present, the cavity experiments are closing in on the theory-preferred region~\cite{SnowALPs}.
Here, we see that the ADMX experiment is just beginning to probe a small portion of the
theoretically-favored region of the QCD axion.
The allowed parameter space of Fig.~\ref{fig:Gondolo} also shows the region currently explored by the
ADMX experiment~\cite{Duffy:2006aa}.

\begin{figure}[t]
  \begin{center}
  \begin{tabular}{c}
   \includegraphics[width=0.75\textwidth]{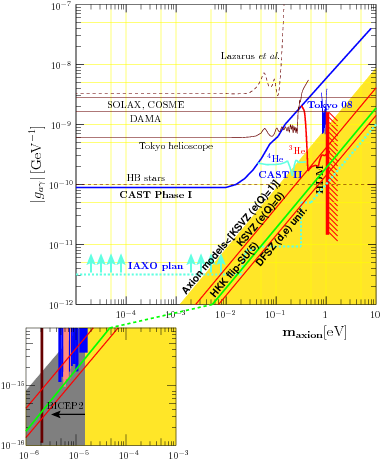}\hskip 1cm
   \end{tabular}
  \end{center}
  \caption{Experimental constraints on axions and ALPs~\cite{SnowALPs}. }
\label{fig:ALPs}
\end{figure}

In addition, it is important to confirm the oscillating nature of the axion field as a BCM,
which was envisioned some years ago for the case of the oscillating electric dipole moments of
electron and nucleon~\cite{HongJ90,HongJ91}.
Recently, the nuclear magnetic resonance (NMR)
 technique is used to detect the oscillating nucleon EDM from the GUT scale $\fa$~\cite{Graham13,Budker13}.
Also, LC circuits have been suggested to detect the feeble oscillating magnetic field using a SQUID detector~\cite{Sikivie14}.
Even in atomic physics, bounds on oscillating EDMs of the electron and proton were
given \cite{AtominPV14}, $|d_0^e|<0.7\times 10^{-14}\,\gev$ from Dy and $|d_0^p|<3.1\times 10^{-8}\,\gev$ from Cs, where $|d_0|$ is a function of $c_1$'s of Eq. (\ref{eq:efflagr}), the misalignment angle $\theta_1$ and the energy difference of two levels in the second order perturbation. At the moment, the limit is not stringent enough to exceed the present limit on $f_a$.

\subsection{DM production via decay of heavy unstable particles}
\label{Subsec:DecayofX}

Heavy unstable particles produced in the early Universe can decay
and produce dark matter, as shown in Fig.~\ref{fig:DefNTDM}, in the
form of a non-thermal population $X_{\rm NTH}$ if they decay
out-of-equilibrium, or a thermal one $X_{\rm TH}$ if they decay while
remaining in thermal equilibrium.  The heavy particles themselves
can be produced either thermally, like WIMPs, or non-thermally -- like
the inflaton, moduli or curvaton, etc.  Non-thermal production from
particle decay has been considered in many models of dark matter
including: axinos from NLSP binos~\cite{CKR00} or staus;
neutralinos from
axinos~\cite{CKLS08,Baer_mixed}, saxions~\cite{Baer:2011uz},
gravitinos~\cite{Baer:2010kd}
and inflatons~\cite{Dev:2013yza};
gravitinos from bino, stau, sneutrino
decays~\cite{Borgani:1996ag,feng03-prl} or Q-ball
decays~\cite{Shoemaker:2009kg}; and KK-gravitons from the decay of KK
U(1) hypercharge gauge bosons~\cite{feng03-prl} or in the string
compactification~\cite{Allahverdi:2013noa}.

\begin{figure}[!t]
  \begin{center}
  \begin{tabular}{c}
   \includegraphics[width=0.4\textwidth]{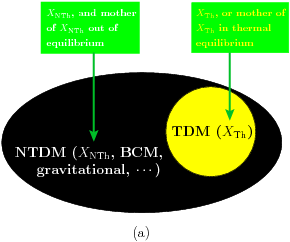}\\[1em]
   \includegraphics[width=0.65\textwidth]{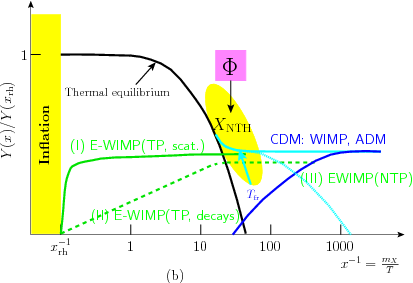}
   \end{tabular}
  \end{center}
  \caption{Thermal and non-thermal dark matter. (a) A cartoon is shown
    for defining thermal and non-thermal DM, and (b) non-thermal DM
    population $X_{\rm NTH}$ is distinguishable from thermal
    population if it is produced more abundantly than thermally produced DM.  }
\label{fig:DefNTDM}
\end{figure}

In the standard inflationary scenario, the decay of the inflaton fills
the Universe with relativistic particles, commonly denoted as
``radiation''.  With the subsequent evolution of the Universe, some of
these particles (some after becoming non-relativistic) decouple from
the thermal plasma and the ratio of the number density of the
decoupled particles to the entropy density is preserved. If the
frozen-out particle is unstable, it finally decays to lighter
particles which may include the DM.

In this way of producing non-thermal particles, the
abundance of DM is directly related to the abundance of the decaying
mother particle $\Phi$ via
\dis{
\left.Y^{\rm NTP}_{\dm}\right|_{T=\TD} = \frac{\alpha_X}{ r_S} {\rm BR}(\Phi\rightarrow \dm)\,  \left.Y_{\Phi}\right|_{T=\TD},
\label{n_NTP}
}
where $\alpha_X$ is the number of DM produced per one $\Phi$,
$r_S\equiv S_f/S_0$ is the ratio of the entropy before and after the $\Phi$ decay and
${\rm BR}(\Phi\rightarrow \dm)$ is the branching ratio of the DM production from the decay of $\Phi$.
Here, $\TD$ denotes the temperature of radiation at the time of $\Phi$ decay.
Assuming an instant decay, it is given by
\dis{
T_D= \bfrac{\pi^2 g_*}{90}^{-1/4}\sqrt{\Gamma_\Phi M_P},
}
where $\Gamma_\Phi$ is the total decay rate of $\Phi$ and $g_*$ is the relativistic degrees of freedom.
Here, the number density of the decaying particles, $n_\Phi$,
is determined from the earlier dynamics of the heavy particle $\Phi$  after inflation.
The DM particles produced from the decay can have a different evolution depending on their properties
at the time of production-via-decay, and we can classify them in the following three cases:

{\bf (1)} $T_D > \Tfr$ (Decay before freeze-out): \\

DM particles are produced before  the freeze-out temperature of DM, $\Tfr$.
They are thermalized with the plasma and  the decay  has no effect on the DM abundance. Therefore,
\dis{
Y_{\dm} = Y^{\TP}.
}

{\bf (2)} $T_D < \Tfr$, $\VEV{\sigma_{\rm ann}  v} n_\dm > H_D  $ (Reannihilation of DM):\\

The heavy particle decays after DM particles are frozen out.
However, the DM produced from the decay are abundant  enough to make them reannihilate. In other words, for
\dis{
\left.Y_{\dm}\right|_{T=\TD} > \bfrac{90}{\pi^2 g_*}^{1/2} \frac{1}{4\VEV{\sigma_{\rm ann} v}} \frac{1}{M_P \TD},
}
some just produced DM would annihilate. Here, $Y_\dm= n_\dm/s$ with
entropy density $s$ given by Eq.~\ref{eq:s}
and $\VEV{\sigma_{\rm ann} v}$ denoting the thermally-averaged
annihilation cross section times velocity, as usual.  In this case,
the final abundance is obtained by solving the Boltzmann equation in a
form re-cast in terms of the yield $Y$,
\dis{
\frac{d Y_\dm}{d t} = -\VEV{\sigma_{\rm ann} v} Y_\dm^2 s.
}
Integrating from time $t=t_D$ to $t$ gives
 \dis{
 Y_\dm^{-1}(T)=Y_\dm^{-1}(T_D)-\langle\sigma_{\rm ann}v\rangle\left(\frac{s}{H}-\frac{s(T_D)}{H(T_D)}\right)\simeq
Y_\dm^{-1}(\TD) +\frac{\VEV{\sigma_{\rm ann} v}s(\TD)}{H(\TD)},
\label{eq:DMdensityNonTher}
 }
 where $Y_\dm(\TD)$ is the sum of the DM densities from thermal
 production and non-thermal production.  The non-thermal production is
 given by \eq{n_NTP} and thermal production depends on the specific
 production mechanism.  When the second term of the right-hand-side of
 Eq. (\ref{eq:DMdensityNonTher}) dominates, then the relic density of
 reannihilated DM is approximated by~\cite{Fujii:2002kr}
 \dis{
 \Omega_\dm^{\rm reann} h^2 \simeq 0.14 \bfrac{90}{\pi^2 g_*(\TD)}^{1/2} \bfrac{m_\dm}{100\gev} \bfrac{10^{-8}\gev^{-2}}{\VEV{\sigma_{\rm ann} v} } \bfrac{2\gev}{\TD}.
}

{\bf (3)} $\TD < \Tfr$, $\VEV{\sigma_{\rm ann}  v} n_\dm < H_D  $ (No reannihilation): \\

In this case, the abundance of DM is just the simple sum of the thermal production and the non-thermal production,
\dis{
Y_\dm = Y^{\TP} +Y^{\NTP}.
}

\subsection{Cosmological constraints}\label{Subsec:ConstraintsCos}

Here we summarize the constraints from cosmology on the dark matter produced via decay of heavy particles.

\subsubsection{Big Bang Nucleosynthesis}

One of the great successes of the Big Bang cosmology is that the calculated abundances of light element
production in the early Universe agrees well with the bulk of measured values.
However, the presence of late-decaying unstable particles in the early Universe can severely upset
the current theory/experiment match in BBN. The late-time decay of heavy particles in the early Universe gives rise to highly energetic particle showers which can
disrupt the abundance of the light elements such as D, ${}^4$He, ${}^7$Li, and ${}^6$Li.
The constraint on the electromagnetic showers~\cite{Kawasaki:1994sc,cefo02} has been extended to include
hadronic showers~~\cite{Jedamzik:2004er,kkm04,Jedamzik:2006xz}.
Since the produced particles redistribute their energy quickly, the constraint is only given to the total energy released.
The energy released into electromagnetic showers is simplified as
\dis{
\xi_{em}\simeq \sum_i \epsilon^{em}_i B^{em}_i Y_X,
}
where the sum is over all the decay modes and $\epsilon^{em}$ and $B^{em}$ is the electromagnetic energy
released from $X$ and the branching ratio for each mode. Similarly for the hadronic energy
\dis{
\xi_{had}\simeq \sum_i \epsilon^{had}_i B^{had}_i Y_X,
}
is constrained. For example for the decay via $X\rightarrow DM + (Z,h,H,A)$,
\dis{
\sum_i \epsilon^{em}_i B^{em}_i  \simeq \frac{\sum_k \epsilon_k \Gamma(X\rightarrow DM+k)B^k_{had}   }{\Gamma_X},
}
where
\dis{
\epsilon_k \simeq \frac{m_X^2-m_{DM}^2 +m_k^2}{2m_X},\qquad {\rm for} \, k= Z,h,H,A.
}

The constraints from BBN on late-decaying massive particles (as calculated by Jedamzik~\cite{Jedamzik:2004er})
are shown in Fig.~\ref{fig:bbn} in the $\Omega_Xh^2$ vs. $\tau$ plane for a particle with mass $m_X=1$ TeV.
Here, $\Omega_Xh^2$ is the would-be relic abundance of particle $X$ had it not suffered a late decay.
The various red contours correspond to different values of the $X$ particle's hadronic branching fraction: $\log_{10}(B_h)$.
The region to the lower-left is allowed since in these cases the abundance of the $X$ particles
is reduced or $\tau_X$ is sufficiently short-lived so that it decays before BBN is fully underway.
The blue curves are the same as red but allow for a more liberal constraint on the $^6{\rm Li}/^7{\rm Li}$ abundance.
For a charged NLSP, the constraints are even stronger~\cite{Pospelov:2006sc}.
\begin{figure}[!t]
  \begin{center}
  \begin{tabular}{c}
   \includegraphics[width=0.65\textwidth]{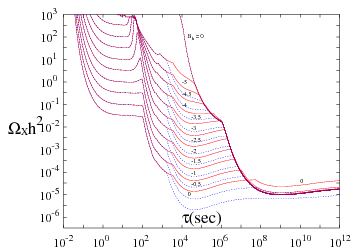}
   \end{tabular}
  \end{center}
  \caption{Constraints on late decaying relics $X$ with mass $m_X=1$ TeV due to their disruption of
light element abundances as calculated in standard Big Bang Nucleosynthesis~\cite{Jedamzik:2006xz}.
Results are plotted in the $\Omega_X h^2$ vs. lifetime $\tau_X$ plane, with
the region to lower-left being allowed.
The various solid red contours for $^6$Li/$^7$Li  $\lesssim 0.66$, with the numbers ($X$ particle's hadronic branching ratio
$\log_{10}(B_h)$) shown just below, correspond to different values of the logarithmic BR.
The blue curves are the same as red but allow for a more liberal constraint on the $^6$Li/$^7$Li abundance, $\lesssim 0.1$.
}
\label{fig:bbn}
\end{figure}
%

\subsubsection{CMB distortion}

The late-time injection of electromagnetic energy from decaying particles may distort the Planck distribution of the CMB spectrum.
After the cosmic age $\tau_X > 10^6\sec$, the photon number changing processes such as double Compton scattering and
thermal bremsstrahlung become inefficient so that energy is exchanged only through elastic scattering.
As a result, the CMB spectrum is distorted from a Planckian distribution with non-vanishing chemical potential $\mu$:
\dis{
f_\gamma (E) = \frac{1}{e^{E/{kT}+\mu }-1}.
}
The current bound is~\cite{Fixsen:1996nj,Khatri:2012tv}
 \dis{
 |\mu| < 9\times 10^{-5}.
 }
 This can be translated to the constraint on the released energy by~\cite{Hu:1993gc}
 \dis{
 \xi_{\rm em} < 1.59 \times 10^{-8} e^{(\tau_{dC}/\tau_X)^{5/4}}\bfrac{1\sec}{\tau_X}^{1/2}.
 }
At late times $\tau_X \gtrsim 4\times 10^{11} \Omega_b h^2\simeq 8.8\times 10^9 \sec$,
even elastic Compton scatterings are not sufficient  enough and only the  injected energy is constrained via the $y$ parameter given by
\dis{
y =\frac{\delta \epsilon}{4\epsilon } = 7.04\times \frac{c^2}{kT(t_{\rm eff}) \xi_{\rm em}},
}
where $T(t_{\rm eff})$ is the CMB temperature at time $t_{\rm eff}=  [\Gamma(1-\beta)]^{1/\beta}\tau_X$, with $T\propto t ^{-\beta}$.
The bound on $y$ is
\dis{
|y|<1.2\times 10^{-5},
}
from which the following constraint is obtained,
\dis{
\xi_{\rm em} < 7.84 \times 10^{-9} \bfrac{\pi \tau_X}{1\sec}^{-1/2}\simeq 4.42 \times 10^{-9} \frac{1}{\sqrt{\tau_X}}.
}

\subsubsection{Large Scale Structure formation}

The large kinetic energy of the non-thermally produced DM components can have large free streaming length, below which the growth of the perturbation is suppressed compared to the pure CDM case.
The particles are classified as warm or even hot depending on the free streaming length $\lambda_{\rm FS}$. For two body-decay  $X' \rightarrow X+(\rm massless~particle)$, it is given by~\cite{Cembranos:2005us}
\dis{
\lambda_{\rm FS} \simeq & 1.0 \mpc \bfrac{\tau}{10^6 \sec}^{1/2}\left( \frac{m_{X'}}{2m_X}  - \frac{m_X}{2m_{X'}} \right)\\[0.5em]
 &\cdot\left\{1+0.14 \log \left[ \bfrac{10^6\sec}{\tau}^{1/2}\left( \frac{2m_{X'}m_X}{m^2_{X'} - m^2_X}   \right) \right]  \right\} .
}

This warm DM property of the decay products
can help to solve the small-scale structure problems
such as cuspy halos, dense cores, and large numbers of
subhalos~\cite{Lin:2000qq,Hisano:2000dz}.\footnote{However, the effect of gas and stars may significantly affect the DM distribution~\cite{Pontzen:2014lma}.
The gas outflows from a galaxy powered by energy released from stars and black hole accretion
change the distribution of the gas and stars. The change of the gravitational potential lead to the change of the DM distribution. This process makes the prediction of the standard $\Lambda$CDM consistent with observation.
}
The late-time decay of heavy particles are naturally obtained in the super WIMP dark matter scenarios
involving gravitinos or KK gravitons~\cite{Cembranos:2005us,Kaplinghat:2005sy}
or right-handed sneutrinos or KK right-handed neutrinos~\cite{Borzumati:2008zz} or vice versa~\cite{Ibe:2012hr}.

The large velocity of warm- or mixed-dark matter is however constrained by cosmology such as the limits from the reionization  of the Universe~\cite{Jedamzik:2005sx,Barkana:2001gr} and the Lyman-$\alpha$ forest~\cite{Narayanan:2000tp}.
The strongest come from the WDM constraint using the Lyman-$\alpha$ data
where the recently improved constraint gives a lower bound to the mass of WDM  of around $2\kev$~\cite{Viel:2005qj,Seljak:2006qw,Viel:2006kd,Boyarsky:2008mt}, which corresponds to a characteristic comoving scale of
\dis{
k_{\rm FS} \sim 15.6 	h \mpc^{-1}\bfrac{m_{\rm WDM}}{1\kev}^{4/3}\bfrac{0.12}{\Omega_{\rm DM} h^2}^{1/3}.
}
The updated  constraints on WDM from Lyman-$\alpha$ forest measured from high-resolution spectra  of 25 quasars ($z>4$)
puts the lower bound at $4\kev$~\cite{Viel:2007mv,Viel:2013fqw}.
In the mixed cold/warm DM scenario, the WDM fraction is constrained to be $\Omega_{\rm WDM}/\Omega_{\rm DM} < 0.35$
in the larger velocity region~\cite{Boyarsky:2008xj}.

\subsubsection{Dark Radiation}

The large kinetic energy of DM produced by late-time heavy particle decays can contribute
additional relativistic degrees of freedom to the particle content of the Universe.
Such contributions to the energy density are called {\it dark radiation}, and are usually
parametrized as the additional neutrino species $\Delta N_{eff}$ beyond the SM value
which is calculated to be $N_{eff}^{SM}=3.046$~\cite{Mangano:2005cc}.
Indeed, up until recently, cosmological data seemed to favor the existence of dark radiation beyond that which is predicted by the Standard Model.
Previous data from WMAP7, the South Pole Telescope (SPT) and the Atacama Cosmology Telescope (ACT) suggested
$N_{eff} \simeq 3.5-4.5$~\cite{wmap,Dunkley11,Keisler11}, indicating a source of dark radiation beyond the SM.
A variety of papers have recently explored this possibility~\cite{vb,ichikawa,Hasenkamp:2012ii,Cicoli:2012aq,Hooper:2011aj,Graf:2012hb}.
More recently, the ACT~\cite{act} has released additional data which, when combined
with the measurement of baryon acoustic oscillations (BAO) and the Hubble constant, reduced the $N_{eff}$ value to:
\begin{equation}
N_{eff}=3.50\pm 0.42\qquad \mbox{(WMAP7+ACT+BAO+$H_0$)}.
\end{equation}
On the other hand, recent SPT~\cite{spt} and WMAP9~\cite{wmap9} analyses reported rather higher values,
\begin{eqnarray}
N_{eff}&=3.71\pm0.35\qquad &\mbox{(WMAP7+SPT+BAO+$H_0$)},\\
N_{eff}&=3.84\pm0.40\qquad &\mbox{(WMAP9+eCMB+BAO+$H_0$)}.
\end{eqnarray}
From the above numbers, it is clear there exists tension between the latest ACT and SPT/WMAP9 values for $N_{eff}$.
While the ACT result has only $1.1\sigma$-level deviation from the standard value, $N_{eff}=3.046$,
the SPT and WMAP9 results show almost a $2\sigma$-level deviation.\footnote{It is worth pointing out that in the WMAP9 analysis `eCMB' denotes the extended CMB, which uses the old data sets of SPT (2011) and ACT (2011).
Also, each $N_{eff}$ value is obtained from different data sets for BAO and $H_0$.
Hence it is hard to determine the most updated result.
Meanwhile, in Ref.~\cite{divalentino}, an independent analysis was made, which consistently combines the most recent data sets from ACT and SPT with WMAP9 data.
The results obtained in this case for ACT+WMAP9+BAO+$H_0$ and SPT+WMAP9+BAO+$H_0$ are consistent with the latest values reported by ACT~\cite{act} and SPT~\cite{spt}.
}
In addition, the Planck~\cite{Ade:2013zuv} collaboration has also reported their first results.
While from  CMB+BAO data the collaboration obtains $N_{eff} = 3.30 \pm 0.27$, once $H_0$ data is included,
slightly larger values are preferred:
\begin{equation}
N_{eff}=3.52^{+0.48}_{-0.45}\qquad\mbox{(95\%; {\rm Planck}+WP+highL+$H_0$+BAO)}.\footnote{Here `WP' denotes a WMAP polarization low multipole
likelihood at $l\leq 23$ and `highL' denotes high-resolution CMB experiments which include ACT and SPT.
The detailed explanations for the dataset are provided in Ref.~\cite{Ade:2013zuv} and references therein.}
\end{equation}
Thus the Planck result shows at most a $2\sigma$ deviation from the standard value.
Due to the tension between the different analyses and the fact that all current results are compatible with the SM model
value within $2\sigma$, it is hard to consider these results as a strong evidence for dark radiation.
Even so, a conservative constraint of
\be
\Delta N_{eff} \equiv N_{eff} - N_{eff}^{SM} < 1.6 \;.
\ee
may at least be invoked which is sufficient to highly constrain many cosmological models which
give rise to dark radiation~\cite{Hooper:2011aj,Kelso:2013paa,Kelso:2013nwa}.
In many cases, new light particle(s) may accompany the non-thermal production of  DM so that
NT production of DM and dark radiation are generated at the same time.
The first discussion of axion dark radiation produced by saxion decay was presented in Ref.~\cite{ChoiRad97};
some recent analyses can be found in~\cite{Choi:2012zna,Graf:2012hb,Hasenkamp:2012ii,DiBari:2013dna,Graf:2013xpe,Bae:2013qr}.

\vspace*{2\baselineskip}

\section{Non-thermal SUSY dark matter}
\label{Sec:SUSY}

\subsection{Gravitino dark matter}\label{gravitinodm:sec}

The gravitino, $\gravitino$, is the massive spin-$3/2$ particle
predicted in theories of local supersymmetry, or supergravity.  Its
mass $\mgravitinoalt$ depends on the SUSY breaking mechanism.  In
gravity-mediation models, $\mgravitinoalt$ is expected at the TeV scale and in AMSB scenarios at the 100 TeV scale, while in
GMSB models it can be much lighter, ranging from  the keV to the GeV scale.  If the
gravitino is the LSP in $R$-parity conserving models, then it is
stable and is a good candidate for dark matter.  Without inflation, a
cosmological constraint on stable gravitinos requires that its mass
must be less than $1\kev$~\cite{Primck82} to avoid an over--abundance.

Any population of primordial gravitinos is expected to be diluted by inflation~\cite{Ellis:1982yb}.
However, gravitinos may be regenerated via a variety of thermal and
non-thermal processes, as discussed above.
The thermal production depends linearly on the reheating
temperature~\cite{nos83,khlopov+linde84,Ellis:1984eq,Ellis:1984er,jss85,Kawasaki:1994af,kakhori99}
and thus gives upper bound on the reheating temperature to avoid overproduction of gravitino DM.
A recent calculation estimated the thermally-produced gravitino abundance
as~\cite{bbb00,Pradler:2006qh,Rychkov:2007uq}:
\be
Y_{\tG}^{\rm TP} = \sum_{i=1}^{3}y_i g_i^2(\treh)
\left(1+\frac{M_i^2(\treh)}{3m_{\tG}^2}\right)\ln\left(\frac{k_i}{g_i(\treh)}\right)
\left(\frac{\treh}{10^{10}\ {\rm GeV}}
\right) \label{yieldG},
\ee
where $y_i=(0.653,1.604,4.276)\times 10^{-12}$, $k_i=(1.266,1.312,1.271)$, $g_i$ are the
gauge couplings evaluated at $Q=\treh$ and $M_i$ are the gaugino masses also evaluated
at $Q=\treh$.

The thermal population of gravitinos may be augmented by heavier sparticle production
followed by their decays in the early Universe.
In this case, the decay-produced abundance depends on the number density  and lifetime of the NLSP.
Since gravitinos interact hardly at all, their production from decays would not undergo re-annihilation.
Then the thermal plus decay-produced gravitino relic abundance is expected to be
\be
\Omega_{\tG}h^2=\Omega_{\tG}^{\rm TP}h^2+\frac{m_{\tG}}{m_{\rm NLSP}}\Omega_{\rm NLSP}h^2
\ee
where $\Omega_{\rm NLSP}h^2$ is the would-be NLSP relic abundance had it not decayed.
Thus the gravitinos produced in NLSP decays inherit the NLSP number density.
Both thermal and non-thermal gravitino production in the CMSSM was considered
in~\cite{rrc04,Cerdeno:2005eu,Bailly:2009pe,Heisig:2013sva}
and the maximum allowed reheating temperature and the viable regions of parameters of the model were found.

There are other non-thermal gravitino production mechanisms.
Gravitinos can be produced during preheating after inflation due to the oscillating
inflaton field~\cite{kklp99,grt99}:
however, explicit calculation shows that production is not very efficient~\cite{nps01}.
Gravitinos can be produced via late decay of inflaton~\cite{Endo:2006xg,Endo:2007sz},
moduli~\cite{kyy04,Nakamura:2006uc}, Q-balls~\cite{Seto:2005pj,Doddato:2012ja,Kasuya:2012mh}
or by a heavy scalar field in gauge-mediated scenarios~\cite{Asaka:2006bv,Hamaguchi:2009hy,Ferrantelli:2009zv,Dalianis:2013pya}.

Gravitinos which are produced non-thermally from NLSP decays are
accompanied by very energetic electromagnetic and/or hadronic
particles. The extremely weak interaction of gravitinos with ordinary
particles usually causes the decay of the NLSP to occur with lifetime
longer than seconds-minutes, \ie during or after the BBN
epoch. For models where gravitinos are produced at large rates from
NLSP decays, there exist strong constraints on late-decaying NLSPs
which might disrupt the successful predictions of standard BBN. The
released energy can affect the light element abundances during and
after BBN~\cite{Jedamzik:2004er,kkm04,Jedamzik:2006xz} and destroy the
standard prediction of the light element abundances.  These
constraints were shown previously in Sec.~\ref{Sec:NTProduction},
Fig.~\ref{fig:bbn}.  While this gravitino problem has been considered
since the early 1980s
~\cite{weinberg-grav82,nos83,khlopov+linde84,Ellis:1984eq,Ellis:1984er,jss85,Kawasaki:1994af}
it has recently been updated in~\cite{cefo02,kkm04,rrc04,Kohri:2005wn}
and
~\cite{Cerdeno:2005eu,Jedamzik:2006xz,Kawasaki:2008qe,Steffen:2006hw,Pradler:2006qh,Covi:2009bk,Hasenkamp:2010if,Boubekeur:2010nt,Roszkowski:2012nq,Bobrovskyi:2012dc,Hasenkamp:2013opa}.
In the MSSM with gravitino DM, the neutralino is disfavored as NLSP
~\cite{feng03-prl,feng03,eoss03-grav,fst04,fst04-2} unless
$\mgravitinoalt\lsim1\gev$~\cite{Cerdeno:2005eu,Kawasaki:2008qe}.  The
stau NLSP with mass around $\tev$ region was argued to be most
consistent with  cosmological and collider
constraints~\cite{ChoiKY07}.  The rather severe BBN constraint may be
avoided if there was late time entropy production which diluted the
abundance of quasi-stable NLSPs~\cite{Ibe06}.

A quasi-stable negatively charged NLSP (such as a stau) can enter a bound state with nuclei
and severely affect early BBN era nuclear reactions. These considerations seem to constrain the charged
NLSP lifetime to values shorter than around $10^3$ seconds~\cite{Pospelov:2006sc}.
More detailed studies followed in~\cite{Kaplinghat:2006qr,Kohri:2006cn,Hamaguchi:2007mp,Jedamzik:2007qk} and
then~\cite{Cyburt:2006uv,Pradler:2006hh,Pradler:2007is,Kersten:2007ab} for the case of gravitino DM with a charged NLSP.
Possible impact on resolutions of the cosmic lithium problems were also
investigated in~\cite{Jedamzik:2005dh,Bailly:2008yy,Bailly:2010hh}.
A coloured NLSP such as a stop or gluino could also enter bound states and lead to even
stronger constraints~\cite{Kusakabe:2009jt}.
These authors require that the lifetime of NLSP be shorter than  200 sec for a heavy NLSP.

Due to their highly suppressed interactions, gravitinos can be stable enough to be DM even when
$R$-parity is broken~\cite{Buchmuller:2007ui,Takayama00}.
For bilinear $R$-parity violation,  the occasional gravitino DM decays can produce photons and massive gauge bosons
and the final decay products may constitute a source of  gamma rays~\cite{Ibarra:2007wg,Choi:2009ng},
high energy cosmic rays~\cite{Ishiwata:2008cu,Ibarra:2008qg,Ibarra:2008jk,Takayama09,Ibe:2013nka,Delahaye:2013yqa}
and neutrinos~\cite{Covi:2008jy}.  When the gravitino mass is below the $W$-boson mass,
gravitinos decay dominantly to photon plus neutrino; however, in some cases three-body decays
may also be comparable~\cite{Choi:2010xn,Choi:2010jt}.

At the LHC one might in principle be able to probe and distinguish between gravitino DM and axino DM by
scrutinizing the quasi-stable NLSP tracks and decays~\cite{Brandenbeug05}.
For the broken $R$-parity case, the neutralino NLSP~\cite{Bobrovskyi10,Bobrovskyi11} and the stop NLSP~\cite{Covi14}
cases were studied in the search for a gravitino DM signal at LHC.

\subsection{Non-thermal WIMPs}

The thermal relic density of higgsino LSP agrees with the present DM
relic density when its mass is close to $1\tev$~\cite{Profumo:2004at}
and in the case of the wino it is around 3\tev~\cite{Hisano:2006nn}.
Shortly after the introduction of anomaly-mediated SUSY-breaking
models, with their concomitant wino-like LSP which generated a
thermal under--abundance of WIMP DM, Moroi and
Randall~\cite{Moroi:1999zb} proposed non-thermal wino-like WIMP
production via the decay of relatively light, multi-TeV scale moduli
fields which should always occur in string
theory~\cite{Acharya:2009zt}.

The issue was analyzed under rather general conditions by Gelmini and
Gondolo~\cite{Gelmini:2006pw} who conclude that, given a SUSY
theory with {\it any} standard thermal WIMP abundance,
non-thermal effects from an arbitrary late-decaying scalar field have
the potential to bring the predicted WIMP relic density into accord
with measurement by adjusting just two parameters: the scalar field
$\phi$ decay temperature $T_D^\phi=\sqrt{\Gamma_\phi
  M_P}/(\pi^2g_*(T_D^\phi)/90)^{1/4}$ and the ratio $b/m_\phi$ where
$m_\phi$ is the scalar mass and $b$ is its decay branching fraction into
WIMPs.  The static and subsequent oscillatory motion of the scalar
field was discussed previously in Subsec.~\ref{Subsec:ThCollective}.
At early times the scalar field is static, but at the oscillation
temperature $T_{\rm osc}^\phi\simeq $ defined by $3H(T_{\rm osc}^\phi
) \simeq m$, the field $\phi$ begins to oscillate and in fact
behaves as a matter fluid~\cite{Turner:1983he} with $\rho_\phi\propto
R^{-3}$.  If $T_D^\phi$ is long enough,  $\rho_\phi$ can be large
enough for $\phi$ to dominate the Universe.  Its decay into DM particles will obviously {\it
  increase} the DM yield beyond thermal expectations, while its decay
to SM particles will lead to entropy dilution, thus {\it decreasing}
the yield below the thermal expectation.  Non-thermal generation
of DM with a full numerical calculation was presented
in~\cite{Arcadi:2011ev}.

Since the decay must happen after WIMP freeze-out, the decay
temperature of the non-thermal production of WIMP should be below
around $m_{\chi}/25$.  Thus the non-thermal WIMP scenario
naturally occurs for the very weakly interacting heavy particles such
as moduli~\cite{Giudice,Moroi:1999zb}, Polonyi fields~\cite{Nakamura:2007wr},
gravitinos~\cite{Kohri:2005ru,Schmitz1212}, or axinos~\cite{CKLS08}.
When the number density of the produced WIMPs is so large that
$n\VEV{\sigma_{\rm ann} v } > H$, then WIMPs can reannihilate and the
relic density is determined by the temperature after decay,
$T_D$,~\cite{CKLS08,Baer_mixed}
\dis{
\Omega_{\chi} h^2 \simeq 0.14 \bfrac{90}{\pi^2 g_*(T_D)}^{1/2}\bfrac{m_{\chi}}{100\gev}
\bfrac{10^{-8}\gev^{-2}}{\VEV{\sigma_{\rm ann} v}}\bfrac{2\gev}{\TD}.
}

The mechanism was in particular applied to the case of wino-like
DM~\cite{Moroi:2013sla,Baer:2010kd} which is expected from AMSB
models~\cite{Moroi:1999zb}, string models~\cite{Acharya:2009zt}, pure
gravity-mediation models and spread supersymmetry~\cite{Hall:2012zp}.
These latter cases are expected to have very heavy squarks and
sleptons ($\sim 25-100\tev$), but with a sub-TeV AMSB-like spectrum of
gauginos and with a wino as LSP.  Higgsino-like WIMP production
  from moduli decay in mirage mediation
  models~\cite{Allahverdi:2012wb} and in more general string
  compactifications~\cite{Allahverdi:2014ppa} has also been examined.
  WIMP production from visible sector scalar field decay has been
  considered in~\cite{Allahverdi:2012gk}.

The non-thermal WIMP scenario often invokes annihilation cross
sections that are larger than standard values  for thermal WIMP DM.  The
large cross section has been used to explain the cosmic ray anomalies
such as positron excess in the
PAMELA~\cite{Nagai:2008se,Dutta:2009uf,Kane:2009if}, AMS-02, and
gamma-ray line signal in the Fermi-LAT.  Since wino-like LSP has
rather large direct detection cross sections and annihilation rates,
it has been claimed to be excluded by indirect WIMP detection
experiments~\cite{Cohen:2013ama,Fan:2013faa,Hryczuk:2014hpa}, although
it becomes again viable when, e.g., the reheating temperature is
comparable with, or lower than, the freezeout temperature~\cite{Roszkowski:2014lga}.

\begin{figure}[!t]
  \begin{center}
  \begin{tabular}{c}
   \includegraphics[width=0.65\textwidth]{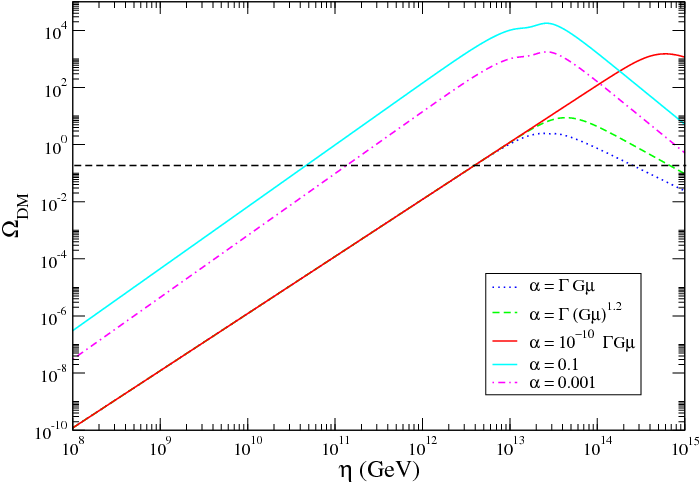}
   \end{tabular}
  \end{center}
  \caption{Dark matter relic density produced from cosmic string decay via loop cusping as a function of the symmetry breaking VEV $\eta$, figure taken from~\cite{Cui:2008bd}. The  curves with different colors correspond to different values of the initial loop size parameter $\alpha$. The black dashed horizontal line denotes the observed dark matter relic density.}
\label{fig:String}
\end{figure}

WIMP production from cosmic string has been considered with/without
friction terms in~\cite{Cui:2008bd}.  The dominant production
mechanism is concluded to be `cusping': \ie near the apex of a cusp (a
portion of string overlapping itself).  It is known that, for a
relatively small size of string, DM production is most significant
right after the freezeout time, $t_{\rm fr}$.  In
Fig.~\ref{fig:String} $\Omega_{\rm DM}$ produced by cusping is
presented as a function of the VEV $\eta$ at which scale gauge strings
are formed by the Kibble mechanism.  If $\eta\lesssim 10^{10}\,\gev$,
non-thermal DM production by cusping is shown to be below the observed
CDM density. Including frictional interactions, string production of
non-thermal DM exceed the observed CDM abundance for
$10^{11}\,\gev\lesssim \eta\lesssim 10^{13-15}\,\gev$.  However, DM
candidates such as WIMPs rely on an almost exact discrete symmetry
$\Z_2$ as discussed in Subsection~\ref{Subsec:ThWIMP}, which is
realized by spontaneous breaking of a gauge U(1)$'$.

\subsection{Axino dark matter}
\label{Subsec:AxinoDM}

The axino $\ta$, stable or almost stable on cosmological time scales,
is a well-motivated DM candidate because it occurs from the axion
solution of the strong CP problem. The axino as the superpartner of
the axion was first considered right after the recognition that softly
broken supersymmetry (SUSY) was relevant for particle
physics~\cite{NillesRaby82,Tamv82,Frere83}.  Axinos belong to the
category of E-WIMPs, and this radically changes their cosmological
properties compared to thermal WIMPs.  Relic axinos can be produced in
a hot plasma or in decays of heavy particles in the early Universe.  A
very interesting case is when axinos constitute CDM, the possibility
first considered in Refs.~\cite{CKR00,CKKR01}.  Alternatively, axinos may
serve as a CDM-generating mother particles, a possibility first
considered in Ref.~\cite{CKLS08} and used in GUTs~\cite{HuhDecay09}.
In both cases, the non-thermal production of axinos or WIMPs is a key
mechanism for CDM production.  The axino as hot DM (HDM), warm DM
(WDM), and CDM in cosmology, astrophysics and collider experiments has
been studied in many
papers~\cite{CKKR01,CRS02,CRRS04,Steffen04,Strumia10,flaxino,Baer,Baer:2010wm,Baer_mixed,Wyler09,Wyler09-2,Wyler09-3,Baer:2011uz,Kang:2008jq,ChoiKYJKPS}.

A specific example of an E-WIMP is the axino from the PQ-augmented
MSSM (PQMSSM).  Axino interactions with SM and MSSM particles are
suppressed by the axion decay constant $\fa$.  The quantity that is
most relevant for the axino in astroparticle physics, and at the same
time most poorly known, is its mass $\maxino$. There exist several
theoretical calculations of the axino
mass~\cite{ChunKN92,ChunLukas95,KimSeo12}. A method for calculating
the axino mass applies to any goldstino (the superpartner of a
Goldstone boson). A goldstino related to the Goldstone boson has a
root in a global U(1) symmetry and receives its mass below the SUSY
breaking scale. SUSY breaking triggers the super-Higgs mechanism and
is related to the gravitino mass $\mgravitinoalt$: this issue was
recently clarified in Ref.~\cite{KimSeo12}.  The typical expectation
for the axino mass is of order $\mgravitinoalt$.  More generally, the
phenomenologically allowed mass range encompasses a much wider set of
values which may be relevant for hot, warm or cold axino DM.  We will
discuss this in more detail below.

The saxion mass $m_s$ enters in the PQMSSM as a soft SUSY breaking term: thus, its mass is
also expected to be of order $\mgravitinoalt$.
Saxions can be produced both thermally and as a BCM.
Depending on the saxion mass and couplings and model, the saxion can decay into a variety of modes.
A notable cosmological implication of saxion production and decay to SM particles
is the possibility of late-time (post freeze-out) entropy production and
consequently a dilution of frozen-out cosmic particles and the cosmic energy density.
When applied to axion cosmology, the effect leads to an increase in the cosmological upper bound on
$\fa$~\cite{Kim91,ChangKim96,Kawasaki:2007mk,Kawasaki:2011ym,Baer:2011eca} for a given assumption on $\theta_i$.
If kinematically allowed, saxions can also decay into SUSY particles, thus potentially increasing the abundance of LSPs.
Depending on couplings, saxions may also decay to axion pairs or axino pairs.
In the first case, $s\to aa$ decay may affect the cosmic microwave background temperature
anisotropy by contributing an additional relativistic
component~\cite{Hasenkamp:2011em,Choi:2012zna,Graf:2012hb,Bae:2013qr,Graf:2013xpe}
(parameterized by the allowed number of additional species of neutrinos $\Delta N_{eff}$ as discussed previously).
There is some possible evidence for an elevated value of $\Delta N_{eff}$ beyond SM values,
but a conservative limit gives $\Delta N_{eff}\alt 1.6$.
In the case where saxions decay to axinos or other SUSY particles, then their late decays
may augment the abundance of LSP dark matter, be it axinos, neutralinos, gravitinos or something else.

In this Subsection, we will adopt a phenomenological point of view and
treat the axino mass $m_{\ta}$ as a free parameter ranging from eV to
multi-TeV values.  A schematic representation of several DM candidates
including the axino was shown previously in the strength of
interaction vs. mass plane in Fig.~\ref{fig:DMtype}.
The various candidates were labelled by color (red, pink and blue) depending on whether they
would comprise HDM, WDM or CDM.
Depending on the axino mass and the production mechanism,
cosmic axinos may fall into any of these different categories as discussed below.

\subsection{Axino production in the early Universe}
\label{sec:AxinoProd}

As mentioned earlier in Fig.~\ref{fig:DefNTDM},
there are two generic ways of producing relic axinos in the early Universe \cite{CoviKim09}:
\bi
\item thermal production from scatterings and decays of particles in thermal equilibrium, and
\item non-thermal production from out-of-equilibrium decays of heavier
  particles wherein the parent particles themselves may be produced
  either thermally or non-thermally.  \ei

\subsubsection{Thermal production of relic axinos}
\label{sssec:ThProd}

Primordial axinos decouple from thermal equilibrium at the temperature~\cite{RTW91}
\be
\Tdec= 10^{11}\,\gev \bfrac{\fa}{10^{12}\gev}^2\bfrac{0.1}{\alpha_s}^3.
\label{eq:DecTemp}
\ee
For $\treh >\Tdec$ the axino relic density from thermal production is estimated to be
\be
\Omega_{\ta}h^2\simeq \frac{m_{\ta}}{2\ {\rm keV}}
\ee
so that one would {\it overproduce} dark matter unless the axino mass is bounded to be smaller
than $\sim 0.2 \kev$~\cite{RTW91}.
Such high values of $\treh \agt 10^{11}$ GeV tend also to overproduce gravitinos
which can lead to violations of BBN predictions and/or overproduction of DM.
Also, in inflationary cosmology, any population of primordial axinos may be strongly diluted by
the exponential expansion in which case the $\kev$ mass upper bound of
Ref.~\cite{RTW91} no longer holds.
However, axinos can be re-generated during reheating.

When the reheating temperature $\treh$ is below the decoupling temperature, axinos do not
reach thermal equilibrium.
However, axinos can still be produced from scatterings in a thermal plasma.
The calculation follows a similar procedure to that used to estimate
gravitino regeneration and decay~\cite{Ellis:1984eq,Kawasaki:1994af}.
If the axino mass ranges from an MeV to several GeV, then the correct axino CDM density is
obtained (in the case of the KSVZ axion/axino model) with $\treh$ less than about $5\times
10^4\gev$~\cite{CKKR01}.

Thermal production of axinos is described by the Boltzmann equation,
\dis{
  \frac{d n_\axino}{d t} + 3 H n_\axino &= \sum_{i,j} \VEV{\sigma (i+j
    \rightarrow \axino+\ldots) v_{\rm rel}} n_i n_j+ \sum_{i} \VEV{\Gamma (i\rightarrow \axino +\ldots) } n_i,
\label{Boltzmann}
}
where the first term on the right-hand side corresponds to scatterings and the
second one to decays, $\sigma (i+j \rightarrow
\axino+\ldots) $ is the scattering cross section for particles $i,j$
into final states involving axinos, and $n_i$ stands for the number
density of the $i$th particle species. $\Gamma (i\rightarrow
\axino +\ldots) $ is the corresponding decay width into final states including axinos.

To solve the Boltzmann equation, a large number of axino production diagrams must be calculated which include various radiation and decay processes.
In the KSVZ model, where axinos dominantly couple via a derivative coupling
to strongly interacting SUSY/SM particles,
then the axino abundance is proportional to $\treh$~\cite{CKKR01,Steffen04,Strumia10,Gomez:2008js}.
For instance, Ref.~\cite{Steffen04} gives
\be
Y_{\ta}^{\rm TP}({\rm KSVZ})\simeq 2\times 10^{-7} g_s^6 \ln\left(\frac{1.108}{g_s}\right)\frac{\treh}{10^4\ {\rm GeV}}
\left(\frac{10^{11}\ {\rm GeV}}{f_a}\right)^2 \label{eq:Yksvz}
\ee
where $g_s$ is the strong coupling renormalized around $\treh$.

Alternatively, in the SUSY DFSZ axion model, where the axion
superfield directly couples to PQ charged MSSM Higgs doublets via the
superpotential Higgs/higgsino parameter $\mu$~\cite{KimNilles84},
the axino abundance is largely independent of $\treh$~\cite{Chun:2011zd,Bae11,Bae:2011iw}
\be
Y_{\ta}^{\rm TP}({\rm DFSZ})\simeq
10^{-5}\zeta_{\ta}\left(\frac{\mu}{{\rm
      TeV}}\right)^2\left(\frac{10^{11}\ {\rm GeV}}{f_a}\right)^2,
\label{eq:Ydfsz}
\ee
where $\zeta_{\ta}$ is a model-dependent constant of order 1.

In Fig.~\ref{fig:YTR} (taken from Ref.~\cite{Choi:2011yf} and updated using
Refs.~\cite{CKKR01} and~\cite{CRS02}), we show the
thermally-produced axino yield $Y$ resulting from scatterings and decays
involving strong interactions in the KSVZ model.
For different values of $\fa$, the curves move up or down proportional to $1/\fa^2$. The contributions
from \SUW~ and \UY~ interactions are suppressed by the gauge coupling
because the cross section $\sigma \propto \alpha^3$.  (For
comparison, in Fig.~\ref{fig:YTR}, we also show the yield from bino-like neutralino decay
after freezeout, which is subdominant at larger $\treh$, but becomes
important at low $\treh$.)

In the case of scatterings, we compare three different prescriptions
for treating the infrared (IR) divergence. They have been used in the
literature. In Ref~\cite{CKKR01}, an effective thermal mass
approximation was used to regulate the infrared divergence from
massless gluons.  A more consistent method using a hard thermal loop (HTL)
approximation was used in Ref.~\cite{Steffen04}. The technique is,
however, valid only in the region of a small gauge coupling, $g_s\ll
1$, which corresponds to the reheating temperature $\treh \gg
10^6\gev$ where, as we shall see, axinos as DM are too warm.
In Ref.~\cite{Strumia10} fully re-summed finite-temperature propagators for
gluons and gluinos were used which extended the validity of the
procedure down to $\treh\gsim10^4\gev$. However, the gauge invariance
at the next-to-leading order is not maintained. We conclude that there
currently remains a factor of a few uncertainty in the thermal yield
of axinos at high $\treh$.

As noted in Ref.~\cite{Bae11}, when the temperature is higher than the
mass $M_Q$ of the PQ-charged and gauge-charged matter in the model
which induces the axino-gaugino-gauge boson interaction, the
interaction amplitude is suppressed by $M_Q^2/T^2$, in addition to the
suppression by the PQ scale $\fa$.  This is most notable in the DFSZ
model where the higgsino mass $\mu$ is around the weak scale, and the
temperature is higher than this scale.

Axinos can also be efficiently produced through decays of thermal
particles via the second term in \eq{Boltzmann} when $\treh$ is
comparable to the mass of the decaying particles. At larger $\treh$,
the contribution from decays becomes independent of temperature and, in
any case, is strongly subdominant relative to that from scatterings. At
lower $\treh$, the production becomes exponentially reduced due to the
Boltzmann suppression factor for the population of  decaying
particles in a thermal plasma.
\begin{figure}[!t]
  \begin{center}
  \begin{tabular}{c}
   \includegraphics[width=0.60\textwidth]{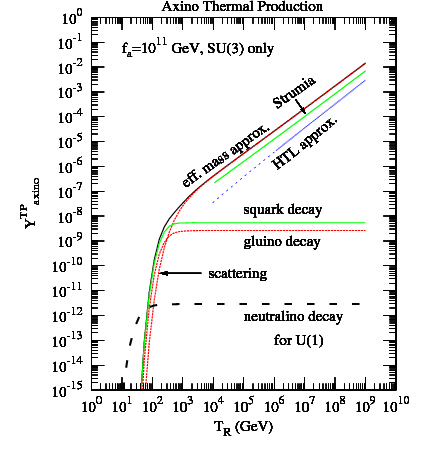}
   \end{tabular}
  \end{center}
  \caption{ Thermal yield of axinos, $Y_\axino^{\rm TP}\equiv
    n_\axino/s$, versus $\treh$ in the SUSY KSVZ model.
    For strong interactions, the
    effective thermal mass approximation (black) is used. We use
    the representative values of $\fa=10^{11}\gev$ and
    $m_{\squark}=\mgluino=1\tev$. For comparison, we also show the HTL
    approximation (dotted blue/dark grey) and that of Strumia
    (green/light grey).  We also denote the yield from squark (solid
    green/light grey), gluino decay (dotted red), and
    out-of-equilibrium bino-like neutralino decay (dashed black) with
    $C_{aYY}=8/3$.  Figure taken from~\cite{Choi:2011yf}. }
\label{fig:YTR}
\end{figure}

One of the relevant decay channels is a two-body decay of a gaugino into an
axino and a gauge boson~\cite{CKKR01}: {\it e.g.} $\chi\rightarrow
\gamma \ta$ or $Z\ta$.
The gaugino-axino-sfermion-sfermion interaction in \eq{eq:Laxino}
generates three-body decays of a gaugino into an axino and two
sfermions, which is subdominant to the two-body decay.  In the KSVZ
model, an effective dimension-4 axino--quark--squark coupling is
generated at one-loop levels and the squark decay can produce significant numbers
of axinos~\cite{CRS02}. Axino production from thermal gluons, neutralinos and squarks
is also shown in Fig.~\ref{fig:YTR}.

In the DFSZ framework, the dominant thermal axino production mechanism comes from scatterings
involving \SUW~ interactions and  from the decays of higgsinos into axino plus
a Higgs boson due to a tree-level axino--Higgs--higgsino interaction
term~\cite{KimNilles84,Chun:2011zd,Bae11} that is proportional to the higgsino
mass $\mu$: {\it e.g.}
\be
{\mathcal L}^{\rm DFSZ}\ni c_H\frac{\mu}{f_a}\ta [\tilde{H}_dH_u+\tilde{H}_uH_d]+h.c.
\ee
Axino production from higgsino decays in thermal equilibrium is comparable to --
or for large $\mu$ can even be larger than --
that from squark decays for which a coupling   already exists
at a tree level due to the $c_2$ interaction term, which is
proportional to the mass of the quark. Generally, in the DFSZ
framework, axino production from thermal decays dominates~\cite{Chun:2011zd} over production from
scatterings~\cite{Chun:2011zd,Bae11,Bae:2011iw,CKKR01}, which is
suppressed by the quark mass at higher temperature~\cite{Bae11}.
Therefore, the axino abundance is independent of the reheating
temperature if the reheating temperature is high enough compared to the higgsino mass,
as shown in Eq. (\ref{eq:Ydfsz}).

\vskip 0.5cm
\subsection{Non-thermal axino production via sparticle decays}
\label{Subsec:AXinoNTP}

When axinos are the lightest SUSY particle, they are likely to
comprise at least a portion of the dark matter.  Axinos can be
produced thermally as discussed above, but also non-thermally via
decay of heavier SUSY particles -- especially the NLSP -- which would
be present in the thermal plasma after NLSP freeze-out.  The NLSP,
considered here to be the lightest MSSM particle, would couple to the
axino via interactions suppressed by $1/f_a$.

If axinos are produced via NLSP decays very late --  at times later than one
second after the Big Bang --  then the injection of
high energy hadronic and electromagnetic particles can affect the
abundance of light elements produced during Big Bang Nucleosynthesis (BBN)~\cite{Jedamzik:2004er,kkm04}.

In the case of NLSP decays to axino, the BBN constraint can be severe
especially for large values of $\fa$ as discussed in
Refs~\cite{CKKR01,CRRS04}
and~\cite{Wyler09,Wyler09-2,Wyler09-3}. However, as long as
$\fa\lsim10^{12}\gev$, the lifetime of bino-like NLSP in the mass
range of a few hundred GeV is less than 1 second (and for the stau, it
is similar) which liberates axino DM from the BBN problem.
Constraints from BBN may also be applicable when the axino is heavy
and unstable, in which case it decays into lighter MSSM particles and
SM particles~\cite{CKLS08,Bae:2013qr}.

Non-thermal production of axinos follows the scheme presented in
Fig.~\ref{fig:DefNTDM} where $\Phi$ stands for the NLSP.  After the
NLSP freezes out from the thermal plasma, then the final step is the
decay of the NLSP to the axino LSP.  This decay occurs at a much later
time scale compared to that for producing NLSPs owing to the small
coupling ($\sim 1/\fa$) for NLSP decay into axinos.  In this case the
number density of axinos is the same as one of the NLSP, and the NTP
of axinos from NLSP decay is simply given by
\dis{
\omegaantp h^2= \frac{\maxino}{m_{\rm NLSP}} \Omega_{\rm NLSP} h^2\simeq
2.7\times 10^{10} \bfrac{\maxino}{100\gev} Y_{\rm NLSP} .
\label{eq:NTP}
}

\begin{figure}[!t]
  \begin{center}
  \begin{tabular}{c}
\includegraphics[width=10cm]{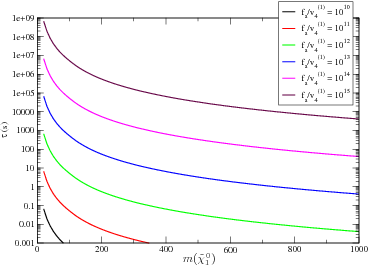}
   \end{tabular}
  \end{center}
\caption{ Lifetime (in seconds) of a bino-like $\tz_1$ with a $\ta$
as LSP versus $m_{\tz_1}$, for various choices of $(f_a)/v_4^{(1)}$ (in GeV units)~\cite{Baer:2010kw}. Here, $v_4^{(1)}$ is the bino component of the lightest
neutralino in the notation of~\cite{Baer:2010kw} and also $C_{aYY}=8/3$ is used.
}
\label{fig:tau}
\end{figure}

For the case of the lightest neutralino as NLSP, we show in Fig.~\ref{fig:tau}
the neutralino lifetime versus $m_{\tz_1}$ for various values
of $f_a$ divided by the bino content of the neutralino~\cite{Baer:2010kw}.
As can be seen, the neutralino tends to decay before about 1 second (onset of BBN)
for $f_a\alt 10^{12}$ GeV and heavier $m_{\tz_1}$ values~\cite{CKR00,CKKR01,Baer:2010kw}.

In the case where $\Phi$ of Fig.~\ref{fig:DefNTDM} stands for other
non-thermal relics ({\it e.g.} an inflaton, moduli, saxion,
Q-balls~\cite{Roszkowski:2006kw} {\it etc}.), then their decays can
also produce axinos; this implies that our estimation of axino
production from NLSP decay is a conservative one for the axino relic
density.

In Refs.~\cite{CKR00,CKKR01} axinos were considered primarily as CDM
but in the latter reference cases when they could be HDM and WDM were also discussed.
Squark decays in a thermal plasma were also considered
in~\cite{CRS02}, and studies in the CMSSM with a
neutralino and a stau as NLSP~\cite{CRRS04} followed by detailed
calculations in Refs.~\cite{Wyler09,Wyler09-2,Wyler09-3}.  It was shown that
tau--stau--axino couplings are not important for thermal production,
but important for non-thermal production when the stau is the NLSP.
Cases for a colored NLSP were considered in
Refs.~\cite{Berger:2008ti,Covi:2009bk}; however, their contribution
is negligible due to their late freezeout.  For CDM axinos ($10\kev$
for TP or $10\mev$ for NTP), relatively low reheating temperatures are
preferred as shown in Fig.~\ref{fig:TR_ma}.  At the time of their
production, axinos are relativistic but their velocity is subsequently
red-shifted and they have a small free streaming length by the time of
structure formation.


\begin{figure}[t]
  \begin{center}
  \begin{tabular}{c}
   \includegraphics[width=0.50\textwidth]{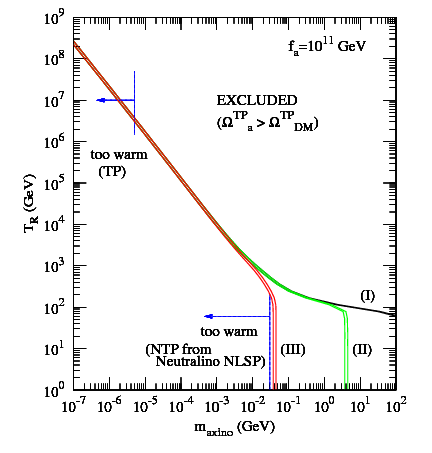}
   \end{tabular}
  \end{center}
  \caption{$\treh$ versus $\maxino$ for $\fa=10^{11}\gev$ in the KSVZ
    model. The bands inside the adjacent curves correspond to a correct
    relic density of the DM axino with both TP and NTP included. The relic
    density of CDM is derived from Planck data~\cite{Ade:2013zuv}
    $\Omega_{\rm CDM} h^2 = 0.1199\pm0.0027\, (68\%\, \textrm{CL})$.
    To parametrize the non-thermal production of axinos, we used
    $Y_{\rm NLSP}=0$ (I), $10^{-10}$ (II), and $10^{-8}$ (III) in
    \eq{eq:NTP}. The upper right-hand area of the plot is excluded
    because of the over--abundance of axinos. The regions disallowed by
    structure formation are marked with vertical blue dashed lines and
    arrows, respectively, for TP ($\maxino\lsim5\kev$) and NTP
    ($\maxino\lsim30\mev$, for a neutralino NLSP).  }
\label{fig:TR_ma}
\end{figure}

In Fig.~\ref{fig:TR_ma}~\cite{Choi:2011yf}, an upper limit on the reheating temperature vs. axino mass --  considering both TP and NTP --
is shown in the KSVZ model for $\fa=10^{11}\,\gev$ and for three different values of  $Y_{\rm NLSP}=0, 10^{-10}$, and $10^{-8}$. For a small axino mass, less than some $10\mev$, thermal production is dominant and depends on the reheating temperature.
However, for a larger mass, NTP provides the dominant contribution.
The regions above/to the right of the curves are excluded due to the over--abundance of axino DM.

\begin{figure}[t]
  \begin{center}
  \begin{tabular}{c}
   \includegraphics[width=0.5\textwidth]{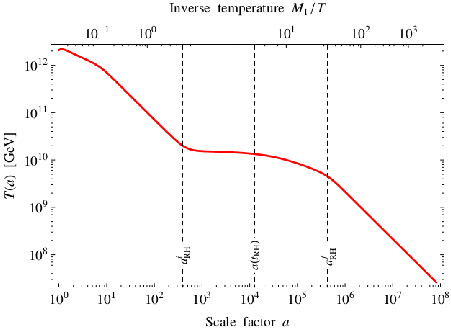}
   \end{tabular}
  \end{center}
 \caption{
 For leptogenesis,  $\treh$ is estimated with some parameters to reach $\treh\approx 5\times 10^9\,\gev$ after the horizon expands by 0.4\,million times, starting from $H_I\sim 2\times 10^{12}\,\gev$~\cite{Buchmuller12,Domcke13}.
 }
\label{fig:leptogenesis}
\end{figure}

The axino in the eV or even sub-eV mass range -- belonging to the HDM
category -- have been considered early on in Ref.~\cite{KimMasiero84}.
An interesting HDM axino case is axino production via photino decay,
which may contribute to $\Delta N_{eff}$ and be relevant both in the
standard Big Bang and in the inflationary cosmology.  This is because
the photino abundance is calculated from the photino decoupling
temperature, which is below the reheating temperature after inflation;
hence, the photino abundance is independent of the cosmological
scenarios.  A related sub-eV mass fermion useful for DM is the
gravitino for $\mgravitinoalt\lsim 1\kev$~\cite{Primack82}.  Because the
decoupling temperature of the gravitino is close to the Planck mass,
primordial gravitinos were diluted out in the inflationary
Universe.\footnote{ If the recent BICEP2 determination of $H_I\simeq
  1.1\times 10^{14}\gev$~\cite{MarshJE14} is confirmed, then the
  sub-keV gravitino hot DM possibility may be ruled out.}  However,
axinos can decay to sub-eV gravitinos~\cite{ChunKim94} which are
possible in the gauge-mediated SUSY breaking scenario.  In the
unstable axino case, sub-eV gravitinos can become HDM in the Universe,
which is called the `axino-gravitino cosmology'~\cite{KimKim95}.  The
axino-gravitino cosmology seems a natural scenario for gauge mediated
SUSY breaking models~\cite{DineNelson93,Giudice99}.

A further possibility for axino CDM arises in the Asaka-Yanagida
scenario~\cite{Asaka:2000ew}.  In this case, it is assumed $m({\rm
  sparticle})>m({\rm gravitino})>m({\rm axino})$. Then, any
possibility of overproduction of gravitinos by sparticle decays is
avoided since axinos inherit the gravitino number density and the
energy density is reduced by the ratio $m_{\ta}/m_{3/2}$. Also, BBN
constraints on sparticle$\rightarrow$ gravitino decays can be avoided
since now the quicker sparticle$\rightarrow$ axino decays can bypass
the gravitino. This scenario allows $\treh$ values above the $10^9$
GeV level which seems required for simple
leptogenesis~\cite{Baer:2010gr} and the recent estimate of high value
of $H_I$~\cite{MarshJE14,Gondolo14}, but which is otherwise
constrained by the gravitino BBN and overproduction problems.

Axinos in the keV range constitute WDM (for $\maxino < 2\kev$) in the
standard Big Bang cosmology~\cite{RTW91}.  But in the currently
standard inflationary cosmology, the primordial population of WDM
axinos is diluted away if the reheating temperature $\treh$ after
inflation is much lower than $\fa$.  In this case, keV-mass axinos
cannot become WDM in the inflationary cosmology for $\treh<\fa$.  If
the BICEP2 value for $H_I\simeq 1.1\times 10^{14}\gev$ (with $r\simeq
0.2$)~\cite{BICEP2I} is accepted, $\treh$ after inflation cannot be
much lower than $\fa$.  For leptogenesis, roughly $\treh\approx
0.5\times 10^{10}\,\gev$ was obtained starting from $H_I\sim 2\times
10^{12}\,\gev$ in Ref.~\cite{Buchmuller12,Domcke13} as shown in
Fig.~\ref{fig:leptogenesis}.  The keV axino as  WDM
may be still feasible.   Warm DM was recently advocated in a review for a solution of the
cuspy-core problem in the $\LCDM$ cosmology~\cite{Pontzen:2014lma}.

Axinos from NTP via the decay of heavier particles can have a large
free-streaming length~\cite{Seto:2007ym,Choi:2012zna}. In this case
the $\mev$ mass axinos has to be warm enough to suppress the
small scale structures that can be probed by using
Lyman-$\alpha$~\cite{Boyarsky:2008xj} and
reionization~\cite{Jedamzik:2005sx} data.  The blue arrow line in
Fig.~\ref{fig:TR_ma} shows this region for axinos produced from
neutralino decay.  This constraint however is relaxed if the
decay-produced axino population is subdominant to the population of
cold axions.

\begin{figure}[t]
  \begin{center}
  \begin{tabular}{c}
   \includegraphics[width=0.5\textwidth]{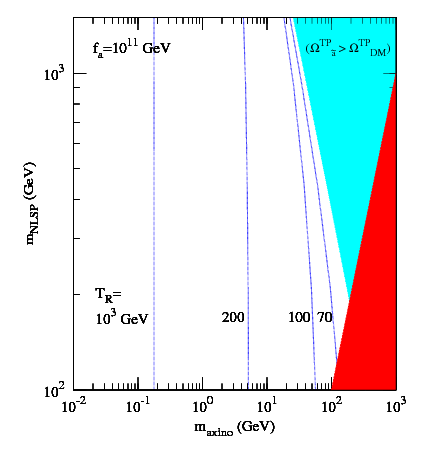}
   \end{tabular}
  \end{center}
 \caption{Contours of the reheating temperature that gives the correct relic density of axino dark matter
in the SUSY KSVZ model in the NLSP--axino mass plane~\cite{Choi:2011yf}. Here, we have assumed $Y_{\rm
      NLSP}=10^{-12}\left({\mnlsp}/{100\gev}\right)$, typical of
    neutralino NLSP, and have taken $\fa=10^{11}\gev$. The cyan wedge in the
    upper right-hand corner is excluded by the overdensity of DM,
    while in the red wedge below it, the axino is not the LSP.     }
\label{fig:ms_ma_A}
\end{figure}

In Fig.~\ref{fig:ms_ma_A} we show the contours of the reheating
temperature needed to explain the correct DM relic density for
$\fa=10^{11}\gev$ and $Y_{\rm NLSP}$ with a typical value of bino-like
neutralino NLSP in the SUSY KSVZ model.  The wedges on the right hand
side are disallowed by overdensity of DM (blue) and axino is not LSP
(red). For small axino mass the TP dominates and the contour is
vertical.  However, when NTP of axinos begins to contribute then
$\Treh$ depends on the mass of the NLSP as well as on the axino mass.

In $R$-parity violating models axinos LSPs can decay with a lifetime
much longer than the age of the Universe.  Then, photons from axino
decay may be a DM signature and may explain some astrophysical
anomalies~\cite{Hooper:2004qf,Chun:2006ss,Endo:2013si,Dey:2011zd,Hasenkamp:2011xh}.

\subsection{Mixed axion-axino CDM}

Models with axinos as DM necessarily contain axions also so that one expects mixed axion-axino CDM,
{\ie \it two dark matter particles}.
In such a case, the {\it total} dark matter abundance is constrained
to equal its measured value $\Omega_{a\ta}h^2\simeq 0.12$.
For the range of $f_a\alt 10^{12}$ GeV, then one would expect
\be
\Omega_{a\ta}h^2=\Omega_a h^2+\Omega_{\ta}^{\rm TP}h^2+\Omega_{\ta}^{\rm NTP}h^2
+\Omega_{\ta}^{\tG}h^2
\ee
where $\Omega_{\ta}^{\rm NTP}=\frac{m_{\ta}}{m_{\tz_1}}\Omega_{\tz_1} h^2$ and
$\Omega_{\ta}^{\tG}=\frac{m_{\ta}}{m_{\tG}}\Omega_{\tG}^{\rm TP} h^2$: {\it i.e.}
we expect dark matter to be comprised of axions via vacuum misalignment,
thermally produced axinos and axinos produced non-thermally by both neutralino and gravitino decays.

In Fig.~\ref{fig:Oh2}, we show the axino abundance as calculated within the KSVZ model.
In the upper frame,  the relative importance of the four individual contributions is shown
as a function of $f_a$
for an mSUGRA/CMSSM scenario with $m_{\tG}=1$ TeV, $m_{\tz_1}=122$ GeV and a would-be value of $\Omega_{\tz_1}^{\rm TP}h^2=9.6$.
For the axion/axino sector we take $\theta_i=0.05$ and $m_{\ta}=100$ keV.
The value of $\Treh$ is adjusted such that $\Omega_{a\ta}h^2=0.12$, which is shown in the lower frame of Fig.~\ref{fig:Oh2}.
For low $f_a$ values, the TP axino contribution is dominant since the axino-gluino-gluon
coupling is large and the corresponding $\Treh$ can be read in the lower figure.
But as $f_a$ increases, the axion component grows and  it becomes dominant at $f_a\sim 4\times 10^{13}$ GeV.
In this case, $\Treh$ can reach as high as $\sim 10^{11}$ GeV with mainly axion CDM.
Such large values of $\Treh$ are consistent with the values needed to sustain
thermal leptogenesis (which requires $\Treh\agt 10^9$ GeV).
However, for such high $f_a$, then the $\tz_1$ becomes so long-lived that it violates
the bounds from BBN on late decaying neutral particles (as indicated in the figure).
%
\begin{figure}[t]
  \begin{center}
  \begin{tabular}{c}
\includegraphics[width=10cm]{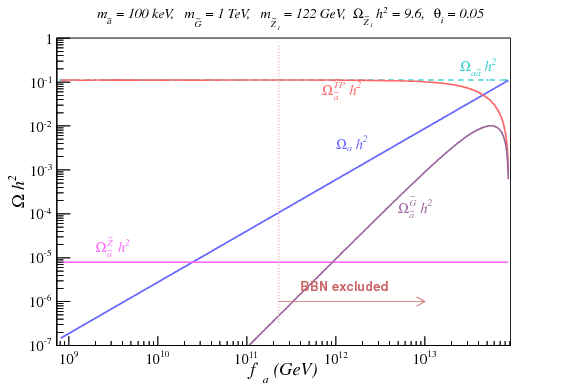}\\
\includegraphics[width=10cm]{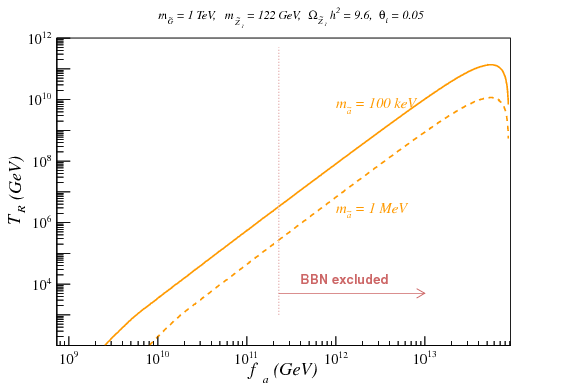}
   \end{tabular}
  \end{center}
\caption{Upper frame: Contribution of axions and TP and NTP
axinos as a function of $f_a$, for an mSUGRA point with $m_0=1000$ GeV, $m_{1/2}=300$ GeV, $A_0=0$, $\tan\beta =10$ and $\mu >0$,
and fixing $m_{\ta}=100$ keV and $\theta_i=0.05$; $\Treh$ is adjusted such that $\Omega_{a\ta}h^2=0.12$.
Lower frame: the value of $\Treh$ that is needed to achieve $\Omega_{a\ta} h^2 = 0.12$
for $m_{\ta} = 0.1$ and 1 MeV, for the same mSUGRA point and $\theta_i$.
Figure taken from~\cite{Baer:2010gr}. }
\label{fig:Oh2}
\end{figure}

The contribution $\Omega_{\ta}^{\rm NTP}h^2=\frac{m_{\ta}}{m_{\tz_1}}\Omega_{\tz_1} h^2$~\cite{CKR00} is of crucial importance since in SUSY models with a large value of $\Omega_{\tz_1} h^2\gg 0.12$ --
which one might naively expect are excluded --  the prefactor mass ratio can reduce
the apparent abundance by huge factors --  and the model becomes phenomenologically allowed.
In the era of ever tighter LHC constraints on sparticle masses, the remaining parameter
space of models such as the CMSSM tend to predict $\Omega_{\tz_1}^{\rm TP} h^2\sim 10-10^4$.
The possibility of $\tz_1\to \ta\gamma$ decays brings such models back into the
regime of being cosmologically-allowed.

While for this case the values of $f_a\agt 10^{11}$ GeV become disallowed by BBN constraints,
at even higher $f_a$ values $\sim \MG\simeq 10^{16}$ GeV the region can again become allowed
if we include the effects of saxion production and decay in the calculation.
Saxion production via BCM becomes large at large $f_a$
(assuming the initial saxion field strength $s_i\sim f_a$), and if they decay
mainly to SM particles then they give rise to late-time entropy injection
which dilutes all relics present at the time of decay.
While one might expect the saxions at large $f_a$ to violate BBN constraints,
if the saxion mass is sufficiently large --  around a few TeV --
then their decay rate is hastened and can be again BBN allowed~\cite{Baer:2011eca}.
\subsection{Mixed axion-neutralino CDM}

In supergravity models, it is generally expected that the axino mass~\cite{ChunLukas95,KimSeo12}
would also be of order $\mgravitinoalt$, which also sets the mass scale for the superpartner spectrum.
Then, it is expected that the lightest MSSM particle would be LSP
and we would have $m_{\rm LSP}< m_{\ta}$. For this scenario, one would expect DM
to be comprised of an axion-LSP admixture. In most models, the lightest MSSM particle turns out
to be the lightest neutralino, so one then expects mixed axion-neutralino
dark matter, with the obvious consequence that {\it both} the axion and a WIMP
might be detected in DM search experiments. We will see in this case
that three distinct populations of axions arise: from thermal production, from BCM
and from decays. While the second of these would give rise to CDM axions, the decay process
would lead to production of dark radiation. Likewise, the neutralinos can be produced
thermally, but also non-thermally via axino~\cite{CKLS08,Baer_mixed}, gravitino and saxion decays.

Calculation of mixed axion-neutralino CDM production breaks up into two distinct cases
depending upon whether one assumes 1. a SUSY KSVZ axion model or 2. a SUSY DFSZ axion model.
Further model dependence arises from the form of the
axion-axino-saxion kinetic terms and self-couplings.
In four component notation, these are of the form~\cite{ChunLukas95}
\be
{\mathcal L}=\left(1+\frac{\sqrt{2}\xi}{v_{\rm PQ}}s\right)
\left[\frac{1}{2}\partial^\mu a\partial_\mu a+\frac{1}{2}\partial^\mu s\partial_\mu s
+\frac{i}{2}\bar{\ta}\partial \llap/ \ta\right]
\ee
where $\xi =\sum_i q_i^3v_i^2/v_{\rm PQ}^2$. Here $q_i$ and $v_i$ denote PQ charges and
VEVs of PQ fields $S_i$, and the PQ scale $v_{\rm PQ} = f_a/\sqrt{2}$ is given by $v_{\rm PQ}=\sqrt{\sum_i q_i^2v_i^2}$.
In the above interaction, $\xi$ is typically $\sim 1$, but in some cases can be as small
as $\sim 0$~\cite{ChunLukas95}.

\subsubsection{SUSY KSVZ with $\xi =0$}

In the SUSY KSVZ case with $\xi =0$, then the axion supermultiplet couples
to PQ charged heavy quark superfields $Q$ and $Q^c$. Upon integrating out the heavy quark fields,
one is left with effective saxion-gluon-gluon and axino-gluino-gluon {\it derivative}
couplings which give rise to $\treh$-dependent thermal production rates~\cite{Steffen04,Graf:2010tv,Graf:2012hb}:
\bea
\frac{\rho_{\ta}^{\rm TP}}{s}&\simeq & 0.9\times 10^{-5}g_s^6\ln\left(\frac{3}{g_s}\right)\left(\frac{10^{12}\ {\rm GeV}}{f_a}\right)^2
\left(\frac{\treh}{10^8\ {\rm GeV}}\right)m_{\ta} \\
\frac{\rho_{s}^{\rm TP}}{s}&\simeq & 1.3\times 10^{-5}g_s^6\ln\left(\frac{1.01}{g_s}\right)\left(\frac{10^{12}\ {\rm GeV}}{f_a}\right)^2
\left(\frac{\treh}{10^8\ {\rm GeV}}\right)m_{s} \\
\frac{\rho_{a}^{\rm TP}}{s}&\simeq & 18.6 g_s^6\ln\left(\frac{1.501}{g_s}\right)\left(\frac{10^{12}\ {\rm GeV}}{f_a}\right)^2
\left(\frac{\treh}{10^{14}\ {\rm GeV}}\right)m_{a}.
\eea
In addition, axions and saxions can be produced via BCM with the saxion abundance
estimated as~\cite{Kawasaki:2007mk,Baer:2011eca}
\be
\frac{\rho_{s}^{\rm BCM}}{s}\simeq  1.9\times 10^{-5}\ {\rm GeV}\left(\frac{{\rm min}[\treh,T_s]}{10^8\ {\rm GeV}}\right)
\left(\frac{f_a}{10^{12}\ {\rm GeV}}\right)^2
\left(\frac{s_i}{f_a}\right)^2 .
\ee
Notice that saxions are thermally produced at large rates in the lower range of $f_a\alt 10^{11}$ GeV
while they are dominantly produced by BCM at high $f_a>10^{12}$ GeV.

In the SUSY KSVZ model, the axinos decay via $\ta\to g\tg$, $\gamma\tz_i$ and $Z\tz_i$ (for $i=1-4$ neutralinos),
with the first of these typically dominating if it is kinematically open.
Of course, these (cascade) decays will all augment the neutralino abundance provided that the
axinos decay  after neutralino freeze-out: $T_D^{\ta}<\Tfr$. The saxion decays are expected to be
$s\to gg$ or $s\to\tg\tg$ if the latter mode is open.
The first of these leads to possible entropy dilution reducing the abundance of any relics present at the time of decay,
while the second of these may augment neutralino production.

Simple approximate analytic formulae for neutralino and axion production in a radiation-dominated (RD), matter-dominated (MD) or decay-dominated (DD) Universe can be found in Ref's~\cite{CKLS08,Baer_mixed}.
A more detailed treatment which also allows for a temperature dependent $\langle\sigma v\rangle$
(as occurs for a bino-like LSP which annihilates via $p$--wave processes)
requires simultaneous solution of eight coupled Boltzmann equations
for radiation, saxions (TP or BCM), axions (TP or BCM), axinos, neutralinos and gravitinos.
The calculation tracks the various energy densities as a function of scale factor, $R/\Rreh$,
where $\Rreh$ is the scale factor at temperature $\Treh$ at the end of inflation
when the Universe is (re)heated.

\begin{figure}[t]
\begin{center}
\includegraphics[width=11cm]{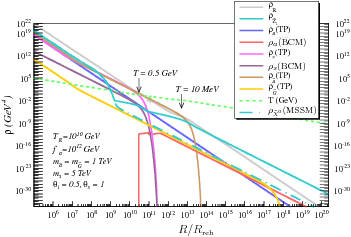}
\caption{Evolution of radiation, neutralino, axion, saxion, axino and gravitino
energy densities versus scale factor $R$.
We adopt an mSUGRA SUSY model with parameters
$(m_0,m_{1/2},A_0,\tan\beta,sign(\mu ))=(400\ {\rm GeV},400\ {\rm GeV}, 0,10,+)$.
We also take $m_{\tG}=1$ TeV and $\treh =10^{10}$ GeV and PQ parameters
$m_{\ta}=1$ TeV, $m_s=5$~TeV, $\theta_i=0.5$, $\theta_s=1$
with $\fa=10^{12}$ GeV.  Figure taken from~\cite{Baer:2011uz}.
}
\label{fig:bench}
\end{center}
\end{figure}

The evolution of various energy densities $\rho_i$ are shown in Fig.~\ref{fig:bench}.
As an example, we adopt a mSUGRA/CMSSM benchmark point
with parameters $(m_0,\ m_{1/2},\ A_0,\ \tan\beta,\ {\rm sign}(\mu ))=$ (400~GeV, 400~GeV, 0, 10, +).
The sparticle mass spectrum is generated by Isasugra
~\cite{isasugra,isagura1},
and has a bino-like neutralino with mass $m_{\tz_1}=162.9$ GeV and a standard relic abundance
from IsaReD~\cite{isared} of $\Omega_{\tz_1}^{std}h^2=1.9$
(it would thus be excluded by WMAP/Planck measurements assuming the standard neutralino freeze-out calculation).
We assume a gravitino mass $m_{\tG}=1$ TeV.
For the SUSY KSVZ model, the PQ parameters are $m_{\ta}=1$ TeV, $m_s=5$ TeV, $\theta_i=0.5$ and $\fa =10^{12}$ GeV
along with $\Treh=10^{10}$ GeV.
Also, $\theta_s = 1$, where $\theta_s f_a$ is the initial field amplitude $s_i$ for the BCM saxions.

From Fig.~\ref{fig:bench}, we see the Universe is initially radiation-dominated (gray curve).
At $R/\Rreh\sim 10^7$, the temperature drops to $T\sim 1$ TeV so that
the thermally-produced axinos, saxions and gravitinos become non-relativistic.
At $R/\Rreh\sim 10^9$, neutralinos begin non-relativistic and their abundance falls steeply; they soon freeze out
at around $T\sim m_{\tz_1}/25$. The dot-dashed curve shows the standard neutralino density in the MSSM
while the solid curves show the PQ augmented MSSM results.
In the PQMSSM, saxions --  and later still axinos --  begin decaying in earnest,
and augment the neutralino abundance.
At $T \sim 0.5$ GeV, the energy density of axinos surpass the radiation component and the
Universe becomes axino-dominated until the axino decays at $T \sim 10$ MeV.
In this case, the final neutralino abundance is enhanced far beyond its
standard value due to the augmentation by thermal axino and saxion production and cascade decay to neutralinos. Also near $T\sim 1$ GeV, we see that the axion oscillation temperature is reached, and an abundance of CDM axions arises (red curve).

In such calculations, it is possible to scan over PQMSSM parameters to find the ultimate combined
neutralino-plus-axion relic density.
Results are shown in Fig.~\ref{fig:BM3} of the final neutralino relic density $\Omega_{\tz_1}h^2$
in the PQMSSM, where each model is plotted versus $\fa$.
The blue points are labeled as BBN-allowed, while red points violate the BBN bounds as described previously.
Here, a mSUGRA/CMSSM point lying in the $A$-resonance annihilation region has been assumed with $2m_{\tz_1}\sim m_A$.
This requires mSUGRA parameters $(m_0, m_{1/2}) = (400\ {\rm GeV}, 400\,\gev)$ and $(A_0, \tan\beta, {\rm sign}(\mu )) =(0,55,+)$,
for which $\Omega_{\tz_1}^{std}h^2\sim 0.02$, \ie a standard thermal under-abundance of bino-like neutralinos.

\begin{figure}[t]
\begin{center}
\includegraphics[width=12cm]{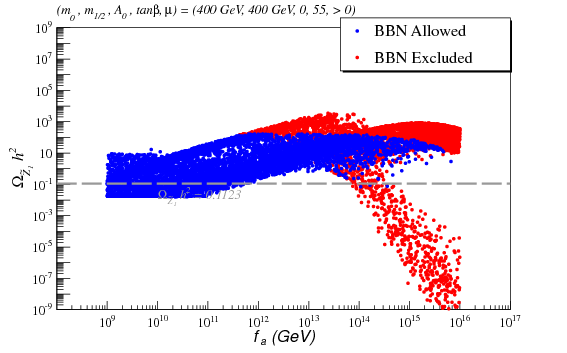}
\caption{Calculated neutralino relic abundance versus $\fa$ from
mSUGRA SUSY model BM3. The spread in dots is due to a scan over
PQ parameters $\fa$, $\treh$, $m_{\ta}$, $m_s$, $\theta_s$.  Figure taken from~\cite{Baer:2011uz}.
}
\label{fig:BM3}
\end{center}
\end{figure}
%

In this case, a scan over PQ parameters yields many points at low $\fa$ with
$\Omega_{\tz_1}h^2\sim 0.02-10$.
Thus, the (neutralino) standard under abundance (SUA) may be enhanced
up to the WMAP/Planck-allowed value, or even beyond.
As one pushes to higher $\fa$ values, the axino becomes so long-lived that it only decays after
neutralino freeze-out, and hence the neutralino abundance is always enhanced. Above $\fa\sim 10^{12}$ GeV, the neutralino abundance is enhanced into  the WMAP-forbidden region, with $\Omega_{\tz_1}h^2$
always larger than $0.12$. As we push even higher in $\fa$, then axino production is suppressed,
but BCM-production of saxions becomes large. Entropy dilution turns the range of $\Omega_{\tz_1}h^2$
back down again, and at $\fa\sim 10^{14}$ GeV, some BBN-allowed points again reach
$\Omega_{\tz_1}h^2\sim 0.12$. In this case, even rather large $\fa$ values approaching $\MG$ are allowed.

For the points with $\Omega_{\tz_1}h^2<0.12$, the
remaining DM abundance can be accommodated by axions via a suitable adjustment of
the initial axion misalignment angle $\theta_i$ such as to enforce
$\Omega_{\tz_1 a}h^2=\Omega_{\tz_1} h^2 + \Omega_a h^2 = 0.12$.
For lower $\fa<10^{12}$ GeV, then $\theta_i\sim 0.1-3$ is typically required while for
much higher $f_a\sim m_{GUT}$, then $\theta_i\sim 0.01-0.1$ although larger values
are allowed in cases of large entropy dilution.

\subsubsection{SUSY KSVZ with $\xi =1$}

This case is much the same as SUSY KSVZ with $\xi =0$ except that now $s\to aa$ and $s\to \ta\ta$
decays are also possible. The first of these is always open and tends to dominate the saxion branching fractions
leading to dark radiation. The second is only open if kinematically allowed, but can augment the neutralino abundance.

In Ref.~\cite{Bae:2013qr}, an eight coupled Boltzmann equation evaluation of mixed axion-neutralino
DM abundance was performed for two SUSY model benchmark point: one labeled SUA
was from radiatively-driven natural SUSY with a higgsino-like LSP and an expected standard thermal abundance $\Omega_{\tz_1}h^2\simeq 0.01$, \ie an order of magnitude below
the measured value. The other, an mSUGRA/CMSSM point with $m_h=125$ GeV and labelled
by standard over-abundance (SOA),
had a standard  over-abundance of bino-like neutralinos $\Omega_{\tz_1}h^2\simeq 7$, a factor of 70 too high. A scan over a large range of parameters $f_a$, $m_{\ta}$, $m_s$, $s_0$ (the initial saxion field strength) and $\treh$ found that the SOA point always generated too much neutralino density for $f_a\alt 10^{13}$ GeV.
For $f_a\agt 10^{13}$ GeV where entropy dilution could reduce $\Omega_{\tz_1}h^2$ to 0.12 or below,
then points were always excluded by generating too much dark radiation from BCM-produced saxions followed by $s\to aa$ decay. Thus, the SOA benchmark point remained always excluded in
moving from the MSSM model to the SUSY KSVZ model. While calculations were performed for a specific
CMSSM benchmark, similar behavior is expected for all SUSY models with a standard
over--abundance: moving from MSSM to SUSY KSVZ model with $\xi=1$
doesn't allow to save the model from exclusion.

In Fig.~\ref{fig:scan2}, the scan results for the neutralino relic density as a function of $f_a$
for the SUA benchmark point is shown.
Blue (red) points are allowed (excluded) by BBN constraints and have $\Delta N_{eff} < 1.6$,
while magenta points have $\Delta N_{eff} > 1.6$.
The green
points are both allowed by BBN and lie in the $1\sigma$ interval for $\Delta N_{eff}$ from the current WMAP9 results.
The standard thermal value for $\Omega_{\tz_1} h^2$ is shown by the dashed gray line and
we see that for $f_a \lesssim 10^{13}$ GeV the neutralino relic abundance is enhanced by
TP axino decays, $s\to \ta \ta$ and/or $s\to\tg\tg$ decays.
For larger values of $f_a$, there are several solutions with suppressed values of $\Omega_{\tz_1} h^2$ when compared to the MSSM value.
These points usually have suppressed axino and thermal saxion production
(due to the large $f_a$ value) and $s \to \tg \tg$ is forbidden ($m_s<2m_{\tg}$).
In this case, the injection of neutralinos from axino and saxion decays
is highly suppressed and easily compensated by the entropy injection from BCM-produced
saxions followed by decays to gluons.
However, as shown in Fig.~\ref{fig:scan2}, all these points have too large values of $\Delta N_{eff}$.
This is understood in that it is not possible to have entropy dilution
from coherent oscillating saxions without either violating the CMB constraint on dark radiation
or overproducing neutralinos ($\Omega_{\tz_1} h^2 \gg 0.12$).
The blue-shaded region in the Figure is excluded by applying the recent Xe-100 WIMP search
bounds~\cite{XENON100:2012}
to SUA with a re-scaled local abundance of WIMPs.
The upshot is that DM-allowed points for SUA-type models can be generated for $f_a\alt10^{12}$ GeV or
$f_a\agt 10^{15}$ GeV. The few DM-allowed points labelled as
green can in addition explain
any excess in $\Delta N_{eff}$ which might be found.

\begin{figure}[t]
\begin{center}
\includegraphics[width=13cm,clip]{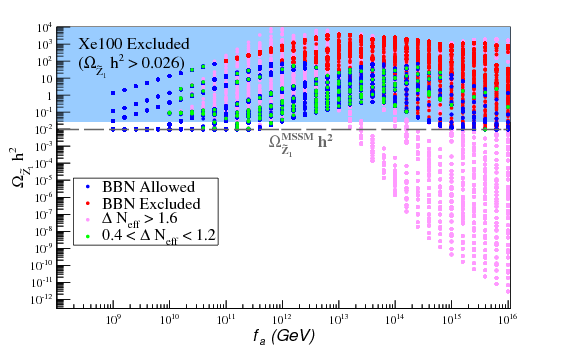}
\caption{$\Omega_{\tz_1} h^2$ as a function of the PQ breaking scale $f_a$
for the scan over the PQ parameter space defined in Ref.~\cite{Bae:2013qr},
assuming natural SUSY benchmark point with a standard thermal under-abundance of neutralinos (SUA).
Blue and red points have $\Delta N_{eff} < 1.6$, while green points have $0.4 < \Delta N_{eff} < 1.2$  and magenta points have
$\Delta N_{eff} > 0.1.6$.
Also, blue and green points are allowed by the BBN constraints on decaying saxions, axinos and
gravitinos, while red points are excluded.
The gray dashed line shows the standard thermal value $\Omega_{\tz_1}^{\rm TP} h^2$ in the MSSM.
The blue-shaded region is excluded by Xe-100 WIMP searches at $m_{\tz_1}=135.4$ GeV~\cite{XENON100:2012}
after applying a re-scaled local WIMP abundance. Figure taken from~\cite{Bae:2013qr}.
}
\label{fig:scan2}
\end{center}
\end{figure}

%
\subsubsection{SUSY DFSZ with $\xi =0$}

In the SUSY DFSZ model, the MSSM Higgs doublets carry PQ charges, and so couple directly to
the axion supermultiplet: the superpotential is of the form
\be
W_{\rm DFSZ}\ni\lambda\frac{S^2}{M_P}H_u H_d .
\label{eq:WDFSZ}
\ee
An advantage of this approach is that it provides a solution to the SUSY $\mu$ problem~\cite{KimNilles84}:
since the $\mu$ term is supersymmetric, one expects $\mu\sim M_P$ in contrast to
phenomenology  which requires $\mu\sim m_{\text{weak}}$.
In this Kim-Nilles solution, PQ charge assignments to the Higgs fields imply that the usual superpotential $\mu$ term
is forbidden.
Upon breaking of PQ symmetry, the field $S$ receives a vev $\langle S\rangle\sim f_a$, so that an
effective $\mu$ term is generated with $\mu\sim \lambda f_a^2/M_P$.
For $\lambda\sim 1$ and $f_a\sim 10^{10}$ GeV, then one may generate $\mu\sim 100$ GeV in accord with
requirements from naturalness~\cite{Baer:2013gva} whilst
$m_{\tq}\sim m_{3/2}\sim 10$ TeV in accord with LHC constraints and in accord with at least a partial
decoupling solution to the SUSY flavor, CP and gravitino problems~\cite{dine,Arkani-Hamed97,Agashe98,Cohen96}.

In SUSY DFSZ,  the direct coupling of the axion supermultiplet to the Higgs superfields
leads to thermal production rates which are independent of $\Treh$. The saxion and axino
thermal yields are then given by~\cite{Chun:2011zd,Bae11,Bae:2011iw}
\bea
Y_s^{\text{TP}} &\simeq & 10^{-7}\zeta_s\left(\frac{\mu}{\text{TeV}}\right)^2\left(\frac{10^{12}\ {\rm GeV}}{f_a}
\right)^2\\
Y_{\ta}^{\text{TP}} &\simeq & 10^{-7}\zeta_{\ta}\left(\frac{\mu}{\text{TeV}}\right)^2\left(\frac{10^{12}\ {\rm GeV}}
{f_a}\right)^2
\eea
where the $\zeta_i$ are model-dependent constants of order unity.
Saxions can also be produced as before via BCM.

Neutralinos can be produced (as before) thermally but also via axino, saxion and gravitino decays.
In SUSY DFSZ, the dominant axino decay modes include: $\ta\to \tz_i \phi$ (where $\phi=h,H,A$),
$\tz_i Z$ ($i=1-4$), $\tw_j^\pm H^\mp$ and $\tw_j^\pm W^\mp$ ($j=1-2$).
Summing over decay modes and neglecting phase space factors, the axino width is
\be
\Gamma_{\ta}\sim \frac{c_H^2}{4\pi}\left(\frac{\mu}{v_{\rm PQ}}\right)^2m_{\ta} ,
\ee
where $c_H$ is an order one parameter for the axino (saxion) coupling arising from Eq.~(\ref{eq:WDFSZ}) and
$v_{\rm PQ}\equiv\sqrt{\sum_iq_i^2v_i^2}\sim f_a$ in terms of PQ charges and VEVs.
This tends to greatly exceed the value obtained in SUSY KSVZ so that in SUSY DFSZ
axinos frequently decay before neutralino freezeout.

Saxions --  produced thermally and via BCM --  can decay via
$s\to hh$, $HH$, $hH$, $AA$, $H^+H^-$, $ZZ$, $W^+W^-$, $ZA$, $W^\pm H^\mp$, $\tz_i\tz_{i'}$,
$\tw_j^\pm\tw_{j'}^\mp$, and also to fermions and sfermions (complete decay formulae are given in~\cite{Bae:2013hma}).
For large $m_s$, the width is dominated by
\be
\Gamma (s\to\ {\rm Higgsinos})\simeq \frac{c_H^2}{32\pi}\left(\frac{\mu}{v_{\rm PQ}}\right)^2m_s .
\ee
In addition, for $\xi\sim 1$, the decay $s\to aa$ may be sizable, leading as before to dark radiation~\cite{Bae:2013qr},
or $s\to\ta\ta$ may occur as well, further augmenting the LSP abundance.
For $f_a\alt 10^{12}$ GeV, saxions also tend to decay before neutralino freeze-out.
A comparison of the saxion radiation  equality temperature $T_e^s$ against the decay temperature $T_D^s$ shows that saxions
dominate the energy density of the Universe only when $f_a\agt 10^{14}$ GeV for which $Y_s^{\text{BCM}}$ is large enough.


A coupled Boltzmann calculation of mixed axion-neutralino abundance has been performed in
Ref.~\cite{serce} for the SUA and SOA SUSY models in the SUSY DFSZ model with $\xi=0$
by scanning over $f_a$, $m_s$, $m_{\ta}$, $s_0$ and $\treh$. For models with $\Omega_{\tz_1}h^2\le 0.12$
(\ie allowed models), then the remaining CDM abundance was filled by axions via suitable adjustment
of $\theta_i$.
\begin{figure}
\begin{center}
\includegraphics[height=6cm]{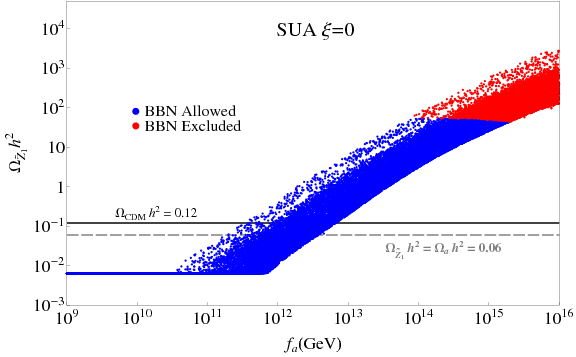}
\caption{The neutralino relic density from a scan over SUSY DFSZ parameter space for the SUA benchmark case with $\xi=0$~\cite{serce}.
The misalignment angle $\theta_i$ is adjusted so that the total CDM relic density $\Omega_{\tz_1 a}h^2=0.12$.
\label{fig:sua0}}
\end{center}
\end{figure}

From Fig. \ref{fig:sua0}, we see that at very low values of $f_a\sim 10^9-10^{10}$ GeV, then
axinos and saxions can be thermally produced at large rates, but since they both decay before
neutralino freezeout, the neutralino abundance adopts its standard value.
In this case, neutralinos comprise $\sim 5-10\%$ of dark matter whilst axions comprise
$90-95\%$ of the abundance. As $f_a$ increases, then axinos and saxions begin decaying after freezeout, and the
neutralino abundance becomes enhanced beyond its standard value.
For values of $f_a\agt 2\times 10^{12}$ GeV, then too much neutralino CDM is produced due to BCM-produced saxions
followed by their decays to SUSY particles (which always occur since $m_{\tz_1}$ is just 135 GeV while $m_s>2m_{\tz_1}$),
and for higher $f_a$ choices the model is always excluded.
BBN constraints kick in at higher $f_a$ values for DFSZ as compared to KSVZ: here, BBN excluded points only arise at $f_a\agt 10^{14}$ GeV.

A similar scan for the SOA benchmark cases was performed. This model has a bino-like LSP but with a huge value of
$\mu\sim 2.6$ TeV (leading to large EW finetuning).
The large $\mu$ parameter hastens saxion decays to Higgs and vector bosons.
In the context of the SUSY DFSZ model, the SOA benchmark is still excluded over all $f_a$ values until
$f_a\sim 10^{15}-10^{16}$ GeV, where huge entropy dilution from $s\to hh$ and other decays leads to
models that are both DM- and BBN-allowed.

\subsubsection{SUSY DFSZ with $\xi =1$}

A similar scan over SUSY DFSZ parameters has been performed for SUA and SOA models with $\xi =1$.
In the SUA case, the added $s\to aa$ decay mode dominates the saxion branching fraction and saxions then decay more
quickly than in the $\xi =0$ case.
In the SOA case with large $\mu$, then saxion decays to Higgs/higgsinos is enhanced and still
dominates the $s\to aa$ decay.
In both cases, at low $f_a$ the NTP-neutralinos arise from thermally produced axinos
followed by their decay to SUSY particles.
As $f_a$ increases, the longer-lived axinos enhance the neutralino abundance, but then
as their production rate becomes more and more suppressed, they contribute fewer and fewer neutralinos even though they
are longer lived. The neutralino abundance turns over, and the large range $f_a:10^9-10^{14}$ GeV can become both
DM-, BBN- and dark radiation-allowed. For yet higher $f_a$ values, then BCM-production of saxions dominates and the
NTP neutralino production again rises. For $f_a\agt 10^{14}$ GeV, many points become either BBN or dark radiation excluded.
For the SOA benchmark, as in the $\xi=0$ case, all values of $f_a\alt 10^{15}$ GeV remain excluded by overproduction of
neutralinos. For SOA and $f_a>10^{15}$ GeV, some points suffer huge entropy dilution and can become
DM-, BBN- and dark radiation-allowed.
\begin{figure}
\begin{center}
\includegraphics[height=6cm]{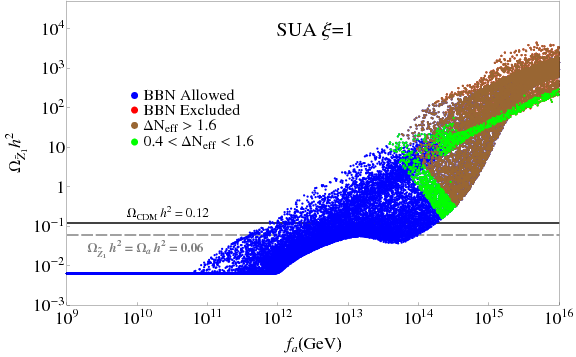}
\caption{The neutralino relic density from a scan over
SUSY DFSZ parameter space for the SUA benchmark case with $\xi=1$~\cite{serce}.
The misalignment angle $\theta_i$ is adjusted so that
the total CDM relic density $\Omega_{\tz_1 a}h^2=0.12$.
\label{fig:sua1}}
\end{center}
\end{figure}
%

\subsubsection{Impact on various cosmic anomalies}

The case where the axino mass is greater than the gravitino mass~\cite{HuhDecay09}
is particularly interesting as a possible explanation of recent data from
PAMELA~\cite{PAMELA09} and from AMS-02~\cite{AMS13} which could imply a TeV scale WIMP mass
if the WIMP is CDM.
In Ref.~\cite{HuhDecay09},  the superheavy axino was shown to
account for TeV-scale cosmic-ray positrons produced as decay products
of, for example, an NMSSM singlino $\tilde N$ to $\tilde e$ plus
$e^+$, but not to antiprotons. This is possible in a string-derived
flipped SU(5) grand unification model~\cite{KimKyae07}. In this case,
$\tilde e$ eventually decays to an LSP neutralino plus SM particles. Of
course, the final population of the LSP is not enough to account for
the present CDM density in the decaying DM scenario, but the mother
singlino density accounts for most of the CDM density, which is given
by the non-thermal production from superheavy axino decays.
The heavy axino decays to a singlino, and the singlino decays
to a positron and a selection, and  finally  the selection decays to the
LSP neutralino. This $\tilde N$--WIMP decay scenario was proposed to
explain the property that the PAMELA data does not contain any large excess
of antiprotons, but a significant flux of positrons~\cite{PAMELAprl09}.
On the other hand, if the decaying DM scenario is ruled out in favor of the
scattering production of positrons, the superheavy axino  may
predominantly decays  to the LSP plus SM particles, for example, in
the MSSM extended by the PQ symmetry with $S$ and $\OVER{S}$ of
Eq.~(\ref{eq:SUSYglobBr}), not introducing extra $N$-type singlets in
the NMSSM.

The decay of heavy axinos provides a non-thermal population of LSP DM,
such as gravitinos or neutralinos~\cite{CKLS08}. Therefore, the
abundance of axinos before decay is also constrained and gives a quite
strong limit on the reheating temperature in the SUSY KSVZ model~\cite{Cheung:2011mg}.

\subsubsection{Impact on cosmic structure}

The scenario of mixed axion-WIMP CDM --  where there is no relation between their abundances --
has been considered as a possibility to understand the rareness of observed dwarf galaxies~\cite{Marsh13,Medvedef13} in the $\LCDM$ cosmology.
Ref.~\cite{Marsh13} considered an ultra-light axion-like particle to reduce the number of dwarf galaxies (and in addition the cusp-core problem) with the mass range $10^{-24}$\,eV--$10^{-20}$\,eV.
They find that an axion of mass $m_a\approx 10^{-21}$\,eV contributing approximately 85\,\% of the total dark matter can introduce a significant kpc scale core in a typical Milky Way satellite galaxy.
Such an ultra-light axion mass, belonging to the anthropic region, would apparently be in conflict with the recent BICEP2 constraint~\cite{MarshJE14,Gondolo14}.
However, there exists discussion on reducing the BICEP2 constraint by diluting abundances via a large rate for late entropy production~\cite{KawasakiM14}.
Isocurvature and dark radiation constraints on SUSY/string models with multi-component DM arising  from
decays of heavy scalars have also been considered in~\cite{Iliesiu14}.
\vspace*{2\baselineskip}

\section{Non-thermal dark matter: non-SUSY candidates}
\label{Sec:NonSUSY}
\subsection{Pseudo Nambu-Goldstone boson as DM}
\label{Subsec:psNGBDM}

Spontaneously broken global symmetries give rise to massless Nambu Goldstone (NGB) bosons.
The couplings of the NGB to the SM fields are suppressed by the spontaneous-symmetry breaking (SSB) scale. When the global symmetry is  broken explicitly, then the NGB become massive and this pseudo NGB (pNGB) can become in many circumstances a natural candidate for DM.
Some well-known examples include: axions, Majorons, familons,  branons, dilatons and more!
The properties of pNGB DM are derived from both the SSB sector and explicit symmetry breaking sector
so that the inferred relic density is highly model-dependent.

\subsubsection{The axion}

The axion --  which may arise as the pNGB of spontaneously broken PQ symmetry~\cite{Kim87,KimRMP10} or
from the fundamental antisymmetric tensor field $B_{MN}$ in string
theory~\cite{Witten84,ChoiKimFa,Witten85,ChoiKimStAxs,Kim88,ChoiKW97,Svrcek06,KimIWJE06} --
was previously discussed in Subsec.~\ref{Subsec:ThCollective}.

\subsubsection{Majorons}

The Majoron $J$ is the pNGB of  lepton number symmetry broken spontaneously by the VEV $v_J$
of a singlet scalar which couples to the Majorana mass term.
The Majoron  becomes massive due to the lepton number breaking terms and this massive pseudo-Majoron
has been studied as a DM candidate~\cite{GSV,ABST,RBS,BV,Lattanzi:2007ux,BJ,GMS}.

In an explicit model~\cite{BV}, the Majoron arises from a complex scalar field $\sigma$ which carries lepton number charge.
Below the symmetry breaking scale $v_J$, then $\sigma=(v_J+\rho)/\sqrt{2}\cdot\exp (iJ/v_J)$, where the real scalar
field $\rho$ mixes with the standard Higgs giving two Higgs scalars $h_1$ and $h_2$.
Additional singly- and doubly-charged Higgs fields $\eta^{+}$ and $\chi^{++}$ are introduced as well.
These latter fields couple directly to leptons.
The keV scale Majoron $J$ is then able to decay via loop-suppressed processes into
$\gamma\gamma$ or $\nu\bar{\nu}$ but with an age longer than that of the Universe.
The $J$ particle can be produced thermally if the charged scalars have mass below the $v_J$ scale
(in which case the $\eta^+$ and $\chi^{++}$ exist in thermal equilibrium with SM particles
and $J$ interacts with the $\eta^+$s and $\chi^{++}$s),
or non-thermally if the potential energy of the $\sigma$ field is converted to $\rho$s (or $h_{1,2}$s) which then later decay to $JJ$.
A third possibility is production of cosmic Majoron strings which then evaporate into Majorons.
The model introduces invisible decays of the Higgs scalars into $JJ$ which
ought to dominate the branching fractions. This scenario seems rather unlikely in light of the recent
Higgs discovery at LHC where the newfound $h$ decays with consistency into SM modes.
Further model-building may rescue this scenario. As an observable consequence, it is suggested to look
for $X$-ray lines arising from $J\to \gamma\gamma$ decays, with $E_{\gamma}\sim m_J/2$.

\subsubsection{Branons}

In models with extra spacetime dimensions, the brane where our world is located can move and fluctuate
along the extra dimensions. In this case, apart from the KK particle of the graviton in the bulk,
there appears a new degree of freedom which parametrizes the position of the visible brane in the extra dimensions. When the metric is not warped along the extra dimensions, the traverse brane fluctuations,
branons, can be parametrized by Goldstone boson associated to the spontaneous breaking of the extra-space translational symmetry and becomes massless~\cite{Dobado01}.
As an example, for a single extra dimension of a circle with brane parametrization $Y^M=(x^\mu,Y(x))$,
the metric induced on the brane is
\dis{
g_{\mu\nu} = \partial_\mu Y^M\partial_\nu Y^NG_{MN}.
}
With metric $G_{MN}=(\tilde{g}_{\mu\nu},-1)$, the brane action can be written as
\dis{
S_B=-f^4 \int_{M_4} d^4x \sqrt{g} \simeq -f^4  \int_{M_4} d^4x \sqrt{\tilde g}+ \frac{f^4}{2}
\int_{M_4}  d^4x \sqrt{\tilde g}\,\tilde{g}^{\mu\nu} \partial_\mu Y\partial_\nu Y,
}
where $f$ is the brane tension. This $Y$ is parametrized by the Goldstone boson.
The warp factor breaks the translational symmetry explicitly and generates a mass $m_b$ for the branons which is given by the bulk Riemann tensor evaluated at the brane position.
Parity on the brane then requires branons to couple as pairs to SM particles and then implies branon stability so that they may serve as a DM candidate~\cite{Cembranos:2003mr}.

In a more general setup, the brane action is obtained up to quadratic terms as~\cite{ACDM,Cembranos04}
\dis{
S_B =& \int_{M_4}d^4x \sqrt{\tilde g}\left[ \frac12 (\tilde{g}^{\mu\nu} \partial_\mu \pi^\alpha \partial_\nu \pi^\beta  - M^2_{\alpha \beta} \pi^\alpha\pi^\beta )  \right.\\
&+ \left.\frac{1}{8f^4}(4\partial_\mu \pi^\alpha \partial_\nu \pi^\beta - M^2_{\alpha \beta} \pi^\alpha\pi^\beta \tilde{g}_{\mu\nu} ) T^{\mu\nu}_{SM}  \right],
}
where the branon fields $\pi^\alpha(x)= f^2 \delta^\alpha_mY^m(x)$ and their mass matrix  is
$M^2_{\alpha\beta} = \tilde{g}^{\mu\nu}R_{\mu\alpha\nu\beta}|_{y=0}$.
The branons interact with SM particles through their energy-momentum tensor with a coupling suppressed by the
inverse of the brane tension scale $f$~\cite{Sundrum99br,Bando99,Dobado01,Cembranos02}.
In the case where $f$ and $m_b$ are $\sim m_{weak}$, then the branons behave as WIMPs and would give rise to CDM as
usual via the WIMP ``miracle'' scenario.
Branons may interact with nuclei thus giving direct WIMP detection signals, and they may annihilate
into SM particles giving indirect cosmic antimatter and gamma ray signatures.
The distinguishing characteristic of branons from other WIMPs ({\it e.g.} neutralinos of SUSY or KK photons from UED models)
is that they could be produced directly at colliders giving rise to monojet or monophoton signatures,
but without the additional cascade decay signatures expected from models like SUSY or UED.

\subsection{Sterile neutrinos as DM}

Perhaps the simplest model which gives rise to DM is the SM when
augmented by gauge singlet right-handed (sterile) neutrinos.
Such neutrinos are necessary in any case to explain the vast array of data on neutrino oscillations.
In this case, there is a Dirac mass $m_D$ which arises from the Yukawa coupling of active neutrinos
with their right-handed counterparts along with a Majorana mass  $M$ which is allowed for gauge singlets. The light active neutrinos gain a ``see-saw''mass $m_\nu \sim m_D^2/M$.
In GUT theories, one might expect the Dirac masses to be comparable to other quark and lepton masses;
however, in the see-saw case, a huge value of $M$ (which might arise from a GUT scale physics) acts
to suppress the active neutrino masses in accord with data.
But by abandoning prejudice from GUT theories, the Majorana mass could be any scale from eV up to $\MG$.
For sterile neutrinos to function as DM, a value $M\sim $keV is required.
Some recent reviews are available in Ref's.~\cite{Kusenko:2009up,Boyarsky:2009ix,Boyarsky:2012rt,Abazajian:2012ys}.

For keV-scale sterile neutrinos, they interact only weakly with matter through the mass mixing
with active neutrinos so their interaction strength is suppressed by the Fermi constant
and the mixing angle, $\theta \sim m_D/M$.
Due to their tiny interactions, the sterile neutrinos are never in the thermal equilibrium, but
nonetheless they can be produced (by several mechanisms) to give a sufficient amount of DM.
Dodelson and Widrow~\cite{Dodelson:1993je} proposed early on that the
right abundance of sterile neutrinos can be produced in the early Universe via
oscillation from active neutrinos.
Their original calculation was subsequently refined with further corrections and including
the degrees of freedom present around QCD phase
transition~\cite{Abazajian:2001nj,Abazajian:2001vt,Dolgov:2000ew,Asaka:2006rw,Asaka:2006nq,Abazajian:2005gj}.
The production rate is given by~\cite{Dodelson:1993je}
\dis{
\Gamma = \frac12\sin^22\theta_M\, \frac{7\pi}{24}\, G_F^2T^4E,
}
where the mixing angle  $ \sin 2 \theta_M$ is temperature dependent  due to matter effects
in the plasma of finite temperature and density and expressed as~\cite{NR88}
\dis{
\sin 2 \theta_M \simeq \frac{\sin 2\theta}{1+0.08\bfrac{T}{100\mev}^6 \bfrac{\kev^2}{\delta m^2} }.
}
Here $\theta$ is the mixing angle in the vacuum where $0.08$ is used for $\nu_\mu$ and $\nu_\tau$ mixing and is replaced by $0.27$ for mixing with $\nu_e$~\cite{Dolgov:2000ew}.

Due to the temperature dependence of the mixing angle and expansion rate, the maximum production rate
to Hubble expansion takes place at a temperature~\cite{Barbieri:1989ti,Enqvist:1990dq}
\dis{
T\simeq 0.1\gev \bfrac{M_s}{\mev}^{1/3}.
}
The present relic density is estimated to be
\dis{
\Omega_s \simeq 0.2 \bfrac{\sin^2\theta}{3\times 10^{-9}} \bfrac{m_s}{3\kev}^{1.8}.
}
Thus, if the mass of the sterile neutrino is around keV, then they can become a realistic candidate
for ``warm dark matter". It is emphasized that this mass range also works to explain pulsar ``kicks''
from supernova explosions~\cite{kicks}.

 Sterile neutrinos can decay to three neutrinos via $Z$-exchange and to $\gamma\nu$
via a $W$-boson loop~\cite{dicusprl77,dicusapj78,Dolgov:2000ew,Abazajian:2001vt,Pal:1981rm}.
The lifetime of sterile neutrino DM via three body decay gives a constraint on the mixing angle.
The decay into photon gives an even stronger bound on the mixing from $X$-ray observations as~\cite{Boyarsky:2005us}
\dis{
\sin^22\theta < 2.5 \times 10^{-18} \bfrac{0.86 \mev}{M_s}.
}
The original Dodelson-Widrow scenario seems at present in conflict with $X$-ray observations
as well as from Lyman-$\alpha$ forest and large scale structure data.

In further refinements, it was found that in the presence of a non-zero lepton asymmetry,
the sterile neutrinos can be produced non-thermally by a resonant transformation~\cite{Shi:1998km}.
This resonance enhances the production of lower energy sterile neutrinos and thus the resulting spectrum is non-thermal and non-relativistic  so that they make CDM and thus
avoid the constraints from Lyman-$\alpha$.
The relic density from this mechanism via the oscillation between $\nu_s$ and $\nu_\alpha$ is~\cite{Shi:1998km}
\dis{
\Omega_s^{\rm res}\simeq \bfrac{m_s}{343\ev}\bfrac{h}{0.5}^{-2}\bfrac{2L_{\nu_\alpha} + \sum_{\beta\neq \alpha}L_{\nu_\beta} }{0.1},
}
where $L_{\nu_\alpha}\equiv n_{\nu_\alpha}/n_\gamma$ is the lepton asymmetry in the active neutrino $\nu_\alpha$.

Additional sterile neutrino production mechanisms may include: 1. decays of
exotic singlet Higgs bosons into $\nu_s$s at $T\agt 100$ GeV which leads to CDM due to redshifting~\cite{kicks}, or 2. production via couplings to the inflaton~\cite{Shaposhnikov:2006xi} or radion~\cite{Kadota:2007mv} fields, which may lead to either warm or cold DM. There exist recent claims for an $X$-ray excess at around 3.5 keV which
can be interpreted in terms of sterile neutrino DM~\cite{Bulbul:2014sua,Boyarsky:2014jta}.
Sterile neutrinos may also accommodate the observed flux of 511 keV photons
from the galactic bulge~\cite{Khalil:2008kp}.

\subsection{Minimal DM}

In the Minimal Dark Matter scenario (MDM)~\cite{Cirelli:2005uq}, one extends the SM by simply adding
additional scalar or spin-1/2 $n$-tuplets of $SU(2)_L$ which may also carry weak hypercharge.
Stability of the lightest neutral member of the multiplet is guaranteed by gauge symmetry and
renormalizability; then, the minimality of the model gives definite predictions
depending only on the mass $M$ of the new matter states since the interactions are determined by gauge symmetry.
The un-hypercharged elements of the new multiplets obey direct detection constraints.
The fully successful MDM candidates include a fermionic 5-plet  with hypercharege $Y=0$.
The phenomenological predictions and direct and indirect DM detections are well summarized
in~\cite{Cirelli:2009uv}.

Since the MDM has electroweak interactions, the relic density can be naturally  obtained via thermal freeze-out as for WIMPs~\cite{Cirelli:2007xd} where co-annihilation with slightly heavier elements of
the multiplets ameliorates the result.
The annihilation  of MDM  occurs dominantly through $s$--wave with subdominant effect of $p$--wave.
When the mass of MDM $M\gtrsim M_V/\alpha$, the non-perturbative electroweak
Sommerfeld enhancement~\cite{Sommerfeld31,hisano05}
can play an important role.
Thus the correct relic density requires the mass of MDM
\dis{
M=(9.6\pm 0.2)\tev.
}

However, in the case of hypercharged MDM, one generates a large direct detection cross section
due to the tree-level $Z$-boson mediated interactions so that the allowed MDM mass is required
to be larger than $\mdm \gtrsim (2Y)^2 \times 3 \times 10^7\gev$ from Xenon 100 experiment~\cite{XENON100:2012}.
This is in contradiction with the above constraint on the mass from correct relic density.
To avoid this problem, it has been suggested that the dark matter may be produced during the reheating
process similar to Wimpzillas which are discussed in a following subsection~\cite{Feldstein13}.

\subsection{Primordial black hole DM}

When primordial density perturbations are large --  of order unity on the scale of the cosmological horizon
in the early Universe --  then Primordial Black Holes (PBHs) can form~\cite{Hawking:1971ei}.
The lifetime of a PBH is connected to its mass due to Hawking radiation~\cite{Hawking:1974rv} as
\dis{
\tau_{\rm PBH} \simeq 10^{64}\bfrac{M_{\rm PBH}}{M_{\odot}}^3  {\rm yr},
}
where $M_\odot \simeq 2\times 10^{33} {\rm gram}$ is Sun's mass.
In this case, PBHs with mass greater than $10^{15}$ gram can survive evaporation
and thus can be a candidate for DM. If the PBHs are formed early enough in the early Universe,
then they may also escape BBN constraints on the baryon-to-photon number ratio.

The mass of a PBH  is related to the mass contained in the horizon when the high density perturbation collapses.
Then, the mass of the PBH and  its comoving scale has a relation
\dis{
M_{\rm PBH} \simeq \left.\frac{4\pi}{3}(H^{-1})^3 \rho \right|_{k=aH}=  1.4\times 10^{13} \Msun \bfrac{g_*}{100}^{-1/6}\bfrac{k}{\mpc^{-1}}^{-2},
}
where $\rho$ is the density at the time of horizon entry,
$k$ is the comoving wave number of the corresponding Hubble horizon size and $g_*$ is the degrees of freedom. Therefore,  $M_{\rm PBH}> 10^{15} {\rm g} \simeq 10^{-18}\Msun$ leads to $k< 10^{15} \mpc^{-1}$ which corresponds
to a temperature $T<10^9\gev$.

The present contribution of PBHs to the matter density is given by~\cite{Bringmann:2001yp,Blais:2002nd}
\dis{
\Omega_{\rm PBH} (M) h^2= 6.35\times 10^{16} \beta(M) \bfrac{10^{15} {\rm g}}{M}^{1/2},
}
where $\beta$ is the probability that a region of comoving size  $R$ which corresponds to the mass scale $M$ will collapse into a PBH.

The large primordial density perturbations can be generated during inflation.
However, the power law spectrum from inflation gives too small a PBH density or else it is constrained by the photons evaporated~\cite{Page:1976wx}.
Therefore, a spectrum with special feature at some characteristic scale has been suggested to be a
dominant DM component~\cite{Blais:2002nd}.
The PBH can be formed also from phase transitions~\cite{Khlopov:1980mg,Jedamzik:1996mr},
collapse of the string loops, bubble collisions, or collapse of domain walls~\cite{Carr:2005zd}.

Some recent claims exist~\cite{Capela:2013yf,Pani:2014rca} which seem to exclude PBHs as DM
due to the possibility that they would bind with and subsequently swallow neutron stars although
these studies have been called into question due to overly optimistic assumptions
about the abundance of DM in globular clusters or on the PBH-neutron star capture rate.
Other studies using data from the Kepler satellite have looked for planetary level
microlensing effects and found no candidate events.
This has allowed a substantial range of $M_{PBH}\sim (10^{-9}-10^{-7})\Msun$ to be excluded for
a large enough PBH halo fraction~\cite{Griest:2013aaa}.

In the case where PBHs co-exist with WIMP DM, then it is claimed that the PBH will form an
ultra-compact mini halo of DM around itself which will serve as an intense $\gamma$-ray point source
due to WIMP-WIMP annihilations~\cite{Beacom}.
Constraints from $\gamma$-ray searches seem to rule out this possibility.

\subsection{Supermassive DM: Wimpzilla }\label{sec:wimpzilla}

Wimpzillas are very massive particles that cannot be produced thermally, since their mass $M_{\rm wz}$ is much larger than
the reheating temperature itself.
They however can be produced with mass of order the inflaton mass in the transition between inflation and a matter or radiation-dominated Universe due to non-adiabatic expansion by  classical gravitational effects~\cite{Ford:1986sy,Yajnik:1990un}.
Wimpzillas with mass in the range $0.04 \lesssim M_{\rm wz}/H_I \lesssim 2 $ (with the Hubble parameter at the end of inflation $H_I\simeq m_\phi\simeq 10^{13}\gev$)  can have the right relic density in the present Universe provided they are stable, independent of their couplings~\cite{Chung:1998zb,Kuzmin:1998uv}. They can be produced non-thermally by non-perturbative quantum effects in  preheating or by the collisions of vacuum bubbles in a first-order phase transition~\cite{Chung:1998ua}.

Supermassive particles can be produced also during reheating after inflation~\cite{Chung:1998rq}.
Since the maximum temperature after inflation is much higher than the reheating temperature, the
wimpzillas can be produced  thermally from scatterings during the reheating process.
In the slow reheating process, the analytic estimate for the relic density is~\cite{Chung:1998rq}
\dis{
\Omega h^2 \simeq M_{\rm wz}^2 \VEV{\sigma v} \bfrac{g_*}{200}^{-3/2} \bfrac{2000 \treh}{M_{\rm wz}}^7.
}

\subsection{Kaluza-Klein DM}

In models of universal extra dimensions (UED)~\cite{Appelquist:2000nn},
it is assumed that the SM particles exist in
an extra-dimensional universe, but that the extra dimensions are compactified on some sort of topological manifold such as
an orbifold~\cite{IKNQ,Kawamura00}.
Orbifolding eliminates unwanted ``wrong helicity'' modes, leaving just the chiral SM as the low energy
effective theory. In UED models, the SM particles exist as the $n=0$ Kaluza-Klein (KK) modes,
along with an infinite tower of their KK excitations $n=1,2,3,\cdots$.
The KK-excitations, basically heavier copies of the SM particles with the same spin and couplings,
have mass $m_{KK}^n\sim n/R$, where $R$ is the compactification radius.
LHC searches now constrain $R^{-1}\agt 1$ TeV.

In UED models,  KK-parity (which arises from the geometrical symmetries of the compactification)
is conserved wherein all $n=1$ (excited state) particles decay to other $n=1$ particles so that the lightest $n=1$ particle is absolutely stable, and denoted as the LKP, or lightest KK excitation/particle. Thus, the LKP becomes a good candidate for dark matter~\cite{Kolb:1983fm,Dienes:1998vg}. Possibilities for LKP include the $n=1$ KK photon, neutrino or graviton with mass $m_{LKP}\sim 1/R$.

The photon or neutrino LKPs enjoy electroweak interactions with SM particles and hence become WIMPs.
Their relic density is determined  thermally via freeze-out from the thermal plasma and it is found
that KK photons/neutrinos with mass $\sim 1$ TeV scale can account for the correct relic
density~\cite{Cheng:2002iz,Servant:2002aq,Kakizaki:2005en,Kakizaki:2005uy}.
Including co-annihilation processes, then even lighter LKPs around $0.6-0.9$ TeV are required.
Projections for direct LKP-photon-nucleon spin-independent scattering cross sections still lie about an order of magnitude below recent LUX limits~\cite{oai:arXiv.org:hep-ph/0209262}.
Even so, since one also expects a variety of $n=1$ KK quark and lepton states
around the 1 TeV scale, then-- in light of recent null results from LHC searches for new physics--
the simplest UED DM scenarios seem increasingly unlikely.

If instead the KK- graviton is the LKP, it interacts very weakly with SM particles and becomes an E-WIMP or super-WIMP~\cite{Feng:2003nr} and can be produced from the decays of heavier KK excitations which would be present in the thermal plasma (care being taken to avoid BBN constraints).
Alternatively, if the KK-graviton is not LKP, then it can still be produced thermally but may contribute via decay to the non-thermal production of the LKP WIMPs~\cite{Shah:2006gs}.
The phenomenology of KK DM including discussion of direct and indirect searches and physics at colliders is reviewed in~\cite{Hooper:2007qk}.

\subsection{Strongly interacting massive particles (SIMPs)}\label{subsub:SIDM}

An alternative class of dark matter particles are known as SIMPs, for
strongly interacting massive particles~\cite{Starkman:1990nj,Nardi:1990ku,Mohapatra:1997sc,Mohapatra:1999gg}.
SIMPs might arise from gauge theories containing exotic stable heavy quarks,
where the new heavy quarks are produced in the early Universe but later bind
with lighter quarks to form neutral, massive strongly interacting particles.
These types of SIMPs could be extremely massive, ranging far beyond the TeV scale.

SIMPs may also arise in supersymmetric models where the gluino is the lightest
SUSY particle~\cite{Baer:1998pg}. Then gluinos produced thermally or non-thermally in the early Universe would bind with gluons to make a neutral gluino-balls which could comprise the dark matter.
Gluino-balls would be expected around the TeV scale.
These types of SIMPs could give rise to exotic collider signatures
such as charged stable tracks which flip charge as they propagate.

Recently, a new SIMP paradigm was proposed~\cite{Hochberg14}
where DM arises from a secluded dark sector which is thermalized with the SM after re-heating.
These SIMPs could exist in the MeV-GeV range.

The variety of production mechanisms and huge span of mass possibilities
leads to great uncertainty in the SIMP production rate in the early Universe.
However, since SIMPs are strongly interacting, it is often expected that they might bind to nuclei
and so can be sought after in exotic, massive nuclei searches.
Indeed, heavy relic techni-baryon DM particles seem ruled out by this
approach~\cite{Chivukula:1989qb}.

\subsection{Dynamical dark matter (DDM)}

Dynamical dark matter (DDM)~\cite{Dienes:2011ja,Dienes:2011sa} is an entirely
new DM proposal in which
the dark matter consists of a (possibly vast) ensemble of fields
$\phi_i$, $i=1-N$. Instead of the DM fields $\phi_i$ being stable,
they are now unstable, but if production rates for the $\phi_i$
are balanced against decay rates, then there can still
exist at the present time one or more of the $\phi_i$ which constitute
the current dark matter abundance while at the same time satisfying experimental and
observational constraints.
An example is given for the case where the $\phi_i$ fields are
scalar fields which are produced via coherent oscillations. Since
each field $\phi_i$ has a different mass $m_i$ associated with it, then
its oscillation temperature (or time) $T_i$ will
differ from the other scalar fields, and the decay widths $\Gamma_i$
(and hence lifetimes) will also differ. In this case, at very early times
the equation of state of the $\phi_i$ fields which have not yet begun to oscillate
will be that of dark energy. Once the fields begin to oscillate, their
equation of state turns to that of CDM. Since a whole host of DM fields
$\phi_i$ are present, then the equation of state varies as a function of time.
Indeed the DM abundance will vary, and the constituency of the DM will vary
as successive fields $\phi_i$, or a collection, dominate the DM abundance.

As phenomenological consequences, it is envisioned that some of the
constituent $\phi_i$ quanta could be produced at colliders in association with
color carrying quanta associated with the dark sector. The colored quanta
$\psi_j$ could then decay into jets plus quanta of the $\phi_i$ fields
giving rise to distinctive dijet invariant mass distributions \cite{Dienes:2012yz}.
Alternatively, if the $\phi_i$ quanta are weakly interacting, then they
may collide with target nuclei in WIMP direct detection
experiments again with characteristic recoil energy spectra of the
target nuclei \cite{Dienes:2012cf}.

While DDM has been advocated as a general framework for
understanding DM in a new way, specific models of DDM have been proposed.
An intriguing possibility is that DDM can be realized by incorporating an
axion or axion-like particle along with an infinite tower of its Kaluza-Klein
excitations~\cite{Dienes:2012jb}.

\subsection{Chaplygin gas}

A perfect fluid with an  equation of state given by
\dis{
p=-\frac{A}{\rho},
}
is known as a Chaplygin gas.
Here, $p$ and $\rho$ are pressure and energy density respectively with $\rho>0$ and $A$ is a positive constant. The Chaplygin equation of state can be obtained from the Nambu-Goto action for $d$-branes
moving in a ($d$+2)- dimensional space-time in the lightcone parametrization.
This gas can be used to account for DM and DE simultaneously~\cite{Kamenshchik:2001cp},
and  has been later connected to M-theory and brane models~\cite{ViollierR02,BilicN04,KimViollier1,KimViollier2}.

In the FRW cosmology, the equation of state gives rise to a solution
\dis{
\rho = \sqrt{A+\frac{B}{a^6}},
}
where $a$ is the scale factor and $B$ is a constant of integration.
In the limit of small $a$, then the energy density behaves as matter:
\dis{
\rho \sim \frac{\sqrt{B}}{a^3}
}
while for large $a$ then the energy density
\dis{
\rho \sim  \sqrt{A},\ \ \ {\rm while}\ \ \ p\sim -\sqrt{A},
}
\ie a cosmological constant with value $\Lambda =\sqrt{A}$.

The generalized Chaplygin gas model was further considered in~\cite{Bento:2002ps}.
These models seem to have tension with both structure formation~\cite{Sandvik:2002jz} and
CMB~\cite{Carturan:2002si,Amendola:2003bz}. Thus, modified models have also been considered.

\subsection{Mirror-matter DM}

An intriguing assumption is that the universe is actually parity symmetric at
its fundamental level, and that the gauge group is then given by
$G\times G^\prime$ where $G=$SU(3)$_c\times$SU(2)$_L\times$U(1)$_Y$ and
$G^\prime\equiv $SU(3)$_c^\prime\times $SU(2)$_R\times $U(1)$_Y^\prime$.
Such a gauge group might be inspired by superstring models based on
the  E$_8\times $E$_8^\prime$ gauge group.

In such a case, the Universe consists of our own observable sector $O$
along with a mirror sector $M$ with very similar physics \cite{rfoot04}:
the mirror world contains quarks, leptons, gauge bosons, Higgs bosons etc.
The two sectors communicate with each other gravitationally, but also
possibly via $U(1)$ kinetic mixing or Higgs sector mixing. In such a world,
the DM could consist of mirror
baryons \cite{Berezhiani:2000gw,Hodges:1993yb},
which in turn might build up mirror planets and stars.
The BBN temperature $T_{BBN}^\prime$ of the mirror world $M$ is
taken to be somewhat lower than our own $T_{BBN}$ in order to avoid constraints
on dark radiation/ the apparent number of neutrinos.

Thus, in this scenario, the mirror baryons could compose the DM.
Mirror nuclei might interact with WIMP detectors via kinetic mixing, and have
been proposed to explain \cite{rfoot04,rfoot10,rfoot11,rfoot12,rfoot14} the variety of low mass WIMP anomalies in {\it e.g.}
DAMA/LIBRA \cite{Bernabei:2008yi,Bernabei:2010mq}, CoGeNT \cite{Aalseth:2010vx,Aalseth:2011wp},
CRESST \cite{Angloher:2011uu} and CDMS-Si \cite{Agnese:2013rvf}.

\subsection{Self-interacting dark matter (SI\,DM)}

While the $\Lambda$CDM model agrees well with observations on
cosmological scales, it apparently fails to agree with observations on
galactic or galactic cluster scales. This is the so-called ``cusp/core''
problem: simulations of CDM models predict a dense core to galaxies along with
large numbers of sub-galaxies in galactic clusters, while observations
favor a less dense galactic core with fewer sub-galaxies within clusters.

To address this problem, a number of modified DM candidates have been proposed:
self-interacting (SI)\footnote{
To distinguish it from SIMP of Subsec.
\ref{subsub:SIDM}, we separate SI from DM.}
DM \cite{Spergel:1999mh}, fuzzy DM \cite{Hu:2000ke}
and warm DM \cite{Bode:2000gq} (WDM).
The latter WDM seems inconsistent with Lyman $\alpha$ observations \cite{Viel:2005qj,VillaescusaNavarro:2010qy}
while fuzzy DM lacks a predictive model.
Self-interacting DM stands for a large class of DM models
wherein the particles are essentially CDM with suppressed annihilation cross
sections, but their self-scattering cross sections remain large enough
that as they collect in the galactic core, they can scatter one-with-another
to heat up the core so that their pressure extends it and reduces the central
density. Models of self-interacting DM include WIMPy models with large
self-interactions \cite{Kaplinghat:2013kqa}, SI asymmetric DM \cite{Petraki:2014uza}, SI Q-balls \cite{Kusenko:2001vu}
or just new particles scattering via new, perhaps hidden/dark sector
forces. Recent proposals include SI-atomic DM \cite{CyrRacine:2012fz} and
WIMP-like particles interacting via new massive or massless gauge
bosons \cite{Tulin:2012wi,Boddy:2014yra} or scalar bosons via a Yukawa-type
interaction \cite{Buckley:2009in,Loeb:2010gj}.

\subsection{Dark matter from $Q$-balls}

Supersymmetric models contain an assortment of scalar fields
carrying both lepton and baryon number-- the squarks and sleptons.
In addition, the SUSY scalar potential contains a variety of
``flat directions'' at the exact SUSY level.
These ingredients allow for
the possibility of what is known as Affleck-Dine (AD)
baryogenesis \cite{Affleck:1984fy,Dine:1995kz,Dine:2003ax}.
In the AD baryogenesis, the scalar potential is lifted by SUSY breaking terms and by
non-renormalizable operators- but these allow for baryon- and lepton-number
carrying condensate fields (fields with large VEVs) to form.
Similar to the axion or inflaton fields, the AD condensate will be
stable at high temperatures, but then begin to oscillate below the
oscillation temperature $T_{AD}$ as non-relativistic matter.
The presence of $B$ and $CP$ violation in SUSY breaking terms, coupled with
out-of-equilibrium decay of the condensate, can generate a sufficient baryon
number asymmetry (although exact rates are very model-dependent).

Along with generating the baryon asymmetry, the AD condensate can also produce
DM \cite{Kusenko:1997si}.
Analysis of inhomogeneities in the AD condensate show that
it should break up into non-topological solitons known as $Q$-balls,
which can contain large baryon number and mass \cite{Kusenko:1997vi}.
In some SUSY breaking
models, such as gauge-mediation, the $Q$-balls are expected to be stable, and
they themselves can form the DM.
In these cases, it is expected that $Q$-balls could produce signals in
large-size proton decay detectors such as SuperK \cite{Kusenko:1997vp}
(although at present no signals have been found).
In models such as gravity-mediation, the $Q$-balls decay to
fermionic SUSY particles and thus can
non-thermally augment the abundance of LSPs \cite{Fowlie:2012im}.

\vspace*{2\baselineskip}

\section{Conclusion} \label{Sec:Conclusion}

Recent astrophysical observations from the Planck
experiment~\cite{Ade:2013zuv} have only strengthened the view that
non-luminous DM~\cite{ZwickyF33} comprises the bulk of matter in our
Universe.  For many years, the paradigm of thermally-produced
WIMP-only CDM has held sway, and has motivated a panoply of
underground, space-based and terrestrial search experiments focussed on
either verifying its existence or ruling out the idea of WIMP dark matter.
Aside from a variety of seemingly
inconsistent anomalies, no clear evidence for WIMPs has yet
surfaced~\cite{Drees:2012ji}.  The situation has only been exacerbated
by recent null searches for WIMPs at Xenon100 and LUX, and by null
results from collider searches for new matter states at LHC8.  We
point out here that -- in spite of much hype over the WIMP ``miracle'' scenario --
thermally produced WIMP-only DM, while still possible, is under
considerable  pressure from a combination of null results from collider and direct/indirect
DM search experiments to reproduce the measured DM abundance.
This certainly holds true in calculable models such as
unified SUSY (with the notable exception of the ~1~TeV higgsino~\cite{Roszkowski:2009sm})
and UED, especially after LHC constraints are applied.
Such WIMP-only scenarios also tend to
neglect the presence of massive gravitinos or KK gravitons which may
modify the expected DM abundance or affect light element abundances from BBN.
Finally, the WIMP-only picture neglects the fundamental pathology of
QCD -- namely the strong CP problem -- and its solution via
introduction of the axion. Inclusion of the axion into WIMP models changes the DM abundance calculation in intricate ways as detailed in Sec. \ref{Sec:SUSY}.

This situation invites a broad review of
many DM candidates which can be produced both thermally and
non-thermally in the Universe.  In this spirit, we have reviewed DM
candidates which may be present now and which would be absolutely
stable or almost stable with lifetime larger than the age of Universe,
$t_U=4.3\times 10^{17}\,{\rm s}$.  We began with two theoretical
frameworks leading to DM: the bosonic coherent motion and also the
lightest particle among some charges of a discrete group such as among
the $\Z_2$ odd particles.

For BCM DM to live long enough, its lifetime must be very long,
typically many orders of magnitude longer than $t_U$.  Therefore, the
interactions leading to the decay of BCM must be highly suppressed
which invites an interpretation as pseudo-Goldstone bosons.  But the
seed global symmetry is not respected in gravity and in string theory.
However, discrete symmetries are not spoiled by gravity if certain
anomaly conditions are
satisfied~\cite{Krauss89,KimPLB13,KimNilles14,KimJKPS14}.  So, the
most plausible BCM candidate derived from a discrete symmetry is the
so-called invisible axion~\cite{KimPRL13}.  Other BCM candidates --
including saxions, inflatons, moduli and ALPs -- may not be DM
candidates but may well contribute in several important ways to DM
production in the early Universe.  Among these, saxions are reviewed
here at some length.  ALPs are sketched much more briefly because of
the unknown relation between their mass and coupling.

Among the $\Z_2$-odd DM candidates, the LSP remains most popular since
it emerges as a necessary consequence of the SUSY solution to the
gauge hierarchy problem.  But when combined with the invisible axion
solution to the strong CP problem, then the existence of both the
axino and saxion is inevitable. We reviewed at some length
the role of axinos in cosmology depending on whether they constitutes all or just a
fraction of the total DM density, or instead the case where they constitute
mother particles which might decay into DM.
In the first case, one might ultimately expect a
detection of axions,  which is possible for its contribution to CDM down to 10\,\% \cite{Yannis14},
along with ultimate discovery of
SUSY at collider experiments.  In the second case, one might
ultimately expect a discovery of both an axion and a WIMP particle.
Which of these dominates the DM abundance is model dependent, but
scenarios exist where they could occur with comparable amounts.

In recent years, a long list of DM candidates have been proposed, many
of them designed to explain various tenuous astrophysical anomalies.
To broaden the perspective, we have also reviewed a selection of
additional thermal and non-thermal DM candidates.  With so many DM
candidates available now, the option of multiple DM components is a
scenario in the cosmic evolution that is certainly worth
investigating.

As reviewed here, there exist numerous DM candidates which may fill
the halos of galaxies, including our own Milky Way.  It remains an
experimental task to identify one or more types of DM species in
either space-based or terrestrial experiments.  It is hoped that this
review presents some updated perspective on some of the possibilities
provided by non-thermal DM candidates.

\vspace*{2\baselineskip}

\section*{Acknowledgments}
J.E.K. thanks W. Buchm\"uller, K. Choi, L. Covi, V. Dombcke,
P. Gondolo, I.-W. Kim, D. J. E. Marsh, G. Raffelt, and Y. Semertzidis
for useful discussions, and Nordita (during ``What is the Dark Matter"
workshop, 05--30 May 2014), ICTP, MPI at Munich, and CERN Th Division,
for their hospitality, where parts of this review were written.  In
addition, we thank K. J. Bae for great assistance in preparation of
this manuscript.  H. B. was supported by the US DoE grant
No. DE-FG02-04ER41305.  K.-Y.C. was supported by the Basic Science
Research Program through the National Research Foundation of Korea
(NRF) funded by the Ministry of Education, Science and Technology
Grant No. 2011-0011083.  K.-Y.C. acknowledges the Max Planck Society
(MPG), the Korea Ministry of Education, Science and Technology (MEST),
Gyeongsangbuk-Do and Pohang City for the support of the Independent
Junior Research Group at the Asia Pacific Center for Theoretical
Physics (APCTP).  J.E.K. is supported in part by the National Research
Foundation (NRF) grant funded by the Korean Government (MEST)
(No. 2005-0093841) and by the IBS (IBS-R017-D1-2014-a00).
L.R. is supported by the Welcome Programme of the Foundation for
Polish Science and in part by the STFC consortium grant of
Lancaster, Manchester, and Sheffield Universities.


\end{widetext}

\end{document}